\documentclass[preprint,nofootinbib,floatfix,a4paper,prd,superscriptaddress,aps]{revtex4-1}   	% use "amsart" instead of "article" for AMSLaTeX format
\usepackage{geometry}                		% See geometry.pdf to learn the layout options. There are lots.
\usepackage[colorlinks,citecolor=blue]{hyperref}

\usepackage{amsmath}
\usepackage{mathrsfs}
\usepackage{upgreek}
\usepackage{amssymb}
\usepackage{graphicx,subfigure}
\usepackage{tikz}
\usepackage[justification=raggedright]{caption}
\usepackage{hyperref}
\usepackage{subfigure}
\usepackage{indentfirst}
\usepackage{autobreak}
\usepackage{extdash}
\usetikzlibrary{3d,calc}
\usepackage{tikz-3dplot}
\usetikzlibrary{spy}
\usetikzlibrary{decorations.fractals,spy}
\usepackage{makecell}
\usepackage{diagbox}
\usepackage{multirow}
\usepackage{ulem}

\usepackage{orcidlink}

\usepackage[colorlinks,citecolor=blue]{hyperref}
\usepackage{amsmath}
\usepackage{mathrsfs}
\usepackage{bbm}
\usepackage{amsfonts}
\usepackage{amssymb}
\usepackage{latexsym}
\usepackage{graphicx}
\usepackage[english]{babel}
\usepackage{multirow}
\usepackage{float}
\usepackage{url}
\usepackage{slashed}
\usepackage{xcolor} 
\usepackage{tabularx}
\usepackage{orcidlink}
\usepackage{appendix}
\usepackage{ulem}

\makeatletter
\@ifundefined{lesssim}{}{}
\@ifundefined{gtrsim}{}{}
\def\vereq#1#2{\lower3pt\vbox{\baselineskip1.5pt \lineskip1.5pt
\ialign{$\m@th#1\hfill##\hfil$\crcr#2\crcr\sim\crcr}}}
\makeatother

\newcommand{\XJ}{\affiliation{\small School of Physics, Xinjiang University, Urumqi 830046, China}}
\newcommand{\ZT}{\affiliation{\small School of Physics and Information Engineer, Zhaotong University, Zhaotong 657000, China}}
\newcommand{\CQ}{\affiliation{\small Chongqing Institute of East China Normal University, Chongqing 401120, China}}
\newcommand{\ECNU}{\affiliation{\small Department of Physics, East China Normal University, Shanghai 200241, China}}
\newcommand{\AS}{\affiliation{\small Institute of Physics, Academia Sinica, Nangang, Taipei 11529, Taiwan}}

\newcommand{\be}{\begin{equation}}
\newcommand{\ee}{\end{equation}}
\newcommand{\ba}{\begin{array}}
\newcommand{\ea}{\end{array}}
\newcommand{\bea}{\begin{eqnarray}}
\newcommand{\eea}{\end{eqnarray}}

\usepackage[utf8]{inputenc}
\usepackage[T1]{fontenc}
\begin{document}

\title{
 Line Operators in the Gauged Two-Higgs-Doublet Model
}

\author{Rui Yang}
\XJ

\author{Jiaming Guo}
\XJ \ZT \CQ

\author{Linghai Li}
\ECNU

\author{Xun Xue\footnote{Corresponding author}}
\email{xxue@phy.ecnu.edu.cn}
\ZT \XJ \CQ \ECNU 

\author{Tzu-Chiang Yuan\orcidlink{0000-0001-8546-5031}}
\email{tcyuan@phys.sinica.edu.tw}
\AS

\begin{abstract}

We investigate the global structure of the Gauged Two-Higgs-Doublet Model (G2HDM), a framework that extends the Standard Model by introducing a dark sector governed by the gauge symmetry $U(1)_X \times SU(2)_H$. The full gauge symmetry of the theory, including the visible sector, is given by the universal covering group $\tilde{G} = U(1)_Y \times SU(2)_L \times SU(3)_C \times U(1)_X \times SU(2)_H$. However, the true gauge group may instead be a quotient $G = \tilde{G} / \Gamma$, where $\Gamma$ is the center of $\tilde{G}$ or a subgroup thereof, leading to different global structures that cannot be distinguished by local experiments.
We explore the physical implications of these global structures, analyzing their effects on Wilson, 't Hooft, and dyonic line operators, as well as the periodicity of the CP-violating $\theta$-angles associated with each group factor. Furthermore, we determine the minimal electric and magnetic charges that arise after electroweak symmetry breaking, highlighting their dependence on the choice of $\Gamma$. These findings provide a systematic characterization of the G2HDM's global properties and their potential phenomenological consequences.

\end{abstract}

\maketitle

\section{Introduction}\label{sec:intro}

The discovery of the 125 GeV Higgs boson in 2012 at CERN's Large Hadron Collider (LHC) was a landmark achievement that not only confirmed the accuracy of the Standard Model (SM) but also solidified its central role in modern particle physics. The SM is based on the gauge group $\tilde{G} = U(1)_Y \times SU(2)_L \times SU(3)_C$, which has been extensively tested in local experiments, demonstrating remarkable agreement with experimental data. However, the global structure of the gauge group remains an open question. The physically realized gauge group could be a quotient group~\cite{ORaifeartaigh:1986agb,Hucks:1990nw} 
$G = \tilde{G}/\Gamma$, where $\Gamma$ is the discrete center of $\tilde{G}$ or a subgroup of it. 
For the SM, $\Gamma$ can be either $\boldsymbol{1}$, $\boldsymbol{Z}_2$, $\boldsymbol{Z}_3$ or $\boldsymbol{Z}_6$.
Since both $\tilde{G}$ and $G$ share the same Lie algebra, they exhibit identical local dynamics. Consequently, any differences between these gauge groups would manifest only in global properties, which could have profound implications for topics such as anomaly cancellation, topological effects, and the behavior of non-perturbative objects like monopoles and instantons. Understanding the global structure of the SM gauge group is therefore crucial for fully grasping the theoretical underpinnings of fundamental interactions.

Generalized symmetries~\cite{Brennan:2023mmt,Apruzzi:2023uma,Bhardwaj:2023wzd} have undergone remarkable developments in recent years, significantly enhancing our understanding of symmetry in theoretical physics. These extensions of conventional symmetries go beyond traditional point-particle interactions, encompassing symmetries that act on extended objects and more intricate structures. They include higher-form symmetries~\cite{Gomes:2023ahz,Hsin:2018vcg,Alonso:2024pmq}, non-invertible symmetries~\cite{Choi:2022jqy,Yokokura:2022alv}, and other advanced frameworks, broadening the scope of symmetry-based approaches in modern physics.

In gauge theories, line operators~\cite{Tong:2017oea,Aharony:2013hda,Garner:2023pmt} and $\theta$-angles~\cite{Aharony:2013hda,Hsin:2020nts,Choi:2023pdp} are indispensable tools for exploring non-perturbative phenomena and topological structures. Foundational studies~\cite{Kapustin:2005py,Henningson:2006hp} introduced Wilson~\cite{Wilson:1974sk} and 't Hooft~\cite{tHooft:1977nqb} loops, which provided critical insights into electromagnetic duality~\cite{Lizzi:1997ss}. These operators, functioning as probes into gauge field dynamics, have enabled a deeper investigation into key concepts such as screening, confinement and duality, enriching our understanding of Yang-Mills gauge theories.

Global structures of the SM had been analyzed from the viewpoint of line operators in~\cite{Tong:2017oea}. 
In this paper, we investigate the line operators in a particle model known as the Gauged Two-Higgs-Doublet Model (G2HDM)~\cite{Huang:2015wts,Arhrib:2018sbz,Huang:2019obt,Ramos:2021txu,Ramos:2021omo}, a theory that incorporates various dark matter candidates. The corresponding gauge group is given by
\begin{equation}
\label{eq:G2HDMcoveringgroup}
\tilde{G}=U{{\left( 1 \right)}_{Y}}\times SU{{\left( 2 \right)}_{L}}\times SU{{\left( 3 \right)}_{C}}\times U{{\left( 1 \right)}_{X}
\times SU{{\left( 2 \right)}_{H}} } \; . 
\end{equation}
Here, $\tilde{G}$ represents the universal covering of the gauge group. The minimal matter content of the model~\cite{Ramos:2021txu,Ramos:2021omo} is tabulated in Table~\ref{tab:mattercontent} for convenience~\footnote{For phenomenology purpose, a Stueckelberg field $S$ with mass $M_X$ for the $U(1)_X$ was also introduced in~\cite{Huang:2015wts,Arhrib:2018sbz,Huang:2019obt,Ramos:2021txu,Ramos:2021omo} to provide gauge invariant mass for the physical dark photon. We will omit this issue here since it is irrelevant to what we are discussing in what follows.}. A subtle ambiguity arises from the quotient of the discrete center group $\Gamma$ and its subgroups, yielding $G = \tilde{G}/\Gamma$. The reason why $G$ can legitimately take this form is that both $G$ and $\tilde{G}$ share the same dynamics and cannot be distinguished by experiments performed locally. We note that $\tilde{G}$ contains two $U(1)$s, two $SU(2)$s and various $\Gamma$ can be constructed. Therefore the general gauge group of G2HDM can be written as
\be
\label{eq:G2HDMQuotientGroup1}
G=\frac{U{{\left( 1 \right)}_{Y}}\times SU{{\left( 2 \right)}_{L}}\times SU{{\left( 3 \right)}_{C}}\times U{{\left( 1 \right)}_{X}}\times SU{{\left( 2 \right)}_{H}}}{\Gamma } \; ,
\ee
with $\Gamma$ can take on one of the following forms~\footnote{ Note that while the center of $SU(2) \times SU(2)$ is $\boldsymbol{Z_2} \times \boldsymbol{Z}_2 \not \cong \boldsymbol{Z}_4$, the center of $SU(2) \times SU(3)$ is $\boldsymbol{Z}_2 \times \boldsymbol{Z}_3 \cong \boldsymbol{Z}_6$. This is in accordance with the algebra version of the Chinese Remainder Theorem, which states that when $m$ and $n$ are two distinct positive coprime integers, $\boldsymbol{Z}_m \times \boldsymbol{Z}_n \cong \boldsymbol{Z}_{m n}$.}: 
$\boldsymbol{1}$, $\boldsymbol{Z}_2$ (2 choices), 
$\boldsymbol{Z}_3$, $\boldsymbol{Z}_6$ (2 choices), 
$\boldsymbol{Z}_2 \times \boldsymbol{Z}_2$, 
$\boldsymbol{Z}_2 \times \boldsymbol{Z}_3$ (2 choices), $\boldsymbol{Z}_2 \times \boldsymbol{Z}_6$ (2 choices).
Alternatively, it can be expressed as
\be
\label{eq:G2HDMQuotientGroup2}
G=\frac{U{{\left( 1 \right)}_{V}}\times SU{{\left( 2 \right)}_{L}}\times SU{{\left( 3 \right)}_{C}}\times U{{\left( 1 \right)}_{A}}\times SU{{\left( 2 \right)}_{H}}}{\Gamma } \; . 
\ee
Here, $U(1)_V \times U(1)_A$ represents a mixture of SM hypercharge $U(1)_Y$  and dark hypercharge $U(1)_X$,
forming two independent $U(1)$s that we will elaborate on in further detail in Section~\ref{sec:LO-G2HDM}. 
This mixing allows for a broader range of interactions between the visible and dark sectors, consistent with the extended symmetry structure of the model. Furthermore, the matter content of the minimal G2HDM, as presented in Table~\ref{tab:mattercontent}, remains invariant under the appropriate transformations governed by the discrete group $\Gamma$. This invariance ensures the theoretical consistency of the framework and aligns with the symmetry constraints imposed by the extended gauge structure.

The main focus of this work is to extend the SM results in~\cite{Tong:2017oea}, obtaining the spectra of line operators and the periodicity of various $CP$ violating $\theta$-angles in the minimal G2HDM gauge group specified by either Eq.~(\ref{eq:G2HDMQuotientGroup1}) or (\ref{eq:G2HDMQuotientGroup2}) with appropriate choices of $
\Gamma$. In Section~\ref{sec:G2HDM}, we provide a concise review of the model and discuss the possible structures of its quotient groups. 
Section~\ref{sec:lineoperators} reviews some general features of line operators, exploring the spectra 
of the line operators and periodicity of the $\theta$-angles for $SU(N)$,  $SU(N)/\boldsymbol{Z}_N$ and $(U(1) 
\times SU(N))/\boldsymbol{Z}_N$, as well as the distinct behaviors of $\theta$-angles in these three theories. These results of the line operators and $\theta$-angles are then applied to the G2HDM in Sections~\ref{sec:LO-G2HDM} and \ref{sec:theta} respectively. Specific values of the $\theta$-angles leading to $CP$ invariant theories are also given.
Section~\ref{sec:symmetrybreaking} examines the effects of spontaneous symmetry breaking by the Higgs condensations on the spectra of the line operators. In particular, the minimal electric and magnetic charges are deduced for different choices of $\Gamma$ for several examples of the quotients. Impact from the symmetry breaking to the $\theta$-terms is also discussed. Section~\ref{sec:summary} summarizes our paper. Witten effect~\cite{Witten:1979ey} exhibited in the Abelian lines for $(U(1) \times SU(2))/\boldsymbol{Z}_2$, $(U(1) \times SU(3))/\boldsymbol{Z}_3$ and $(U(1) \times SU(2) \times SU(3))/\boldsymbol{Z}_6$ are shown in Appendix~\ref{sec:appendix}.

\small
\begin{table}[htbp!]
	\begin{center}
		\caption{Matter content in the minimal G2HDM. Physical SM hypercharge $Y$ and dark hypercharge $X$ are given by $Y={\widetilde Y}/6$ 
        and $X={\widetilde X}/2$.}
        \label{tab:mattercontent}
		\begin{tabular}{|c|c|c|c|c|c|c||c|c|}
			
			\hline
			\text{Matter Fields} & $SU(2)_L$ & $SU(3)_C$ & $SU(2)_H$ & $U(1)_{\widetilde Y}$ & $U(1)_{\widetilde X}$ & $h$-parity & $\; B \; $ & $\; L \;$ \\ 
			\hline
			$Q_L=(u_L, d_L)^{\rm T}$         &   $\bf 2$ & $\bf 3$ & $\bf 1$ & $1$ & $0$ & $(+,+)^{\rm T}$ & $\frac{1}{3}$ & $0$ \\ 
			$U_R=(u_R, u_R^H)^{\rm T}$       &   $\bf 1$ & $\bf 3$ & $\bf 2$ & $4$ & $-1$ & $(+,-)^{\rm T}$ &  $\frac{1}{3}$ & $0$ \\ 
			$D_R=(d_R^H, d_R)^{\rm T}$       &   $\bf 1$ & $\bf 3$ & $\bf 2$ & $-2$ & $1$ & $(-,+)^{\rm T}$ & $\frac{1}{3}$ & $0$ \\ 
			\hline
			$u_L^H$ &   $\bf 1$ & $\bf 3$ & $\bf 1$ & $4$ & $-2$ & $-$ & $\frac{1}{3}$ & $0$ \\ 
			$d_L^H$ &   $\bf 1$ & $\bf 3$ & $\bf 1$ & $-2$ & $+2$ & $-$ & $\frac{1}{3}$ & $0$ \\ 
			\hline
			$L_L=(\nu_L, e_L)^{\rm T}$ &   $\bf 2$ & $\bf 1$ & $\bf 1$ & $-3$ & $0$ & $(+,+)^{\rm T}$  & $0$ & $1$ \\ 
			$N_R=(\nu_R, \nu_R^H)^{\rm T}$   &   $\bf 1$ & $\bf 1$ & $\bf 2$ & $0$ & $-1$ & $(+,-)^{\rm T}$ & $0$ & $1$ \\ 
			$E_R=(e_R^H, e_R)^{\rm T}$ & $\bf 1$ & $\bf 1$ & $\bf 2$ & $-6$ & $1$ & $(-,+)^{\rm T}$ & $0$ & $1$ \\ 
			\hline
			$\nu_L^H$  & $\bf 1$ & $\bf 1$ & $\bf 1$ & $0$ & $-2$ & $-$ & $0$ & $1$ \\ 
			$e_L^H$   &   $\bf 1$ & $\bf 1$ & $\bf 1$ & $-6$ & $+2$ & $-$ & $0$ & $1$ \\
			\hline\hline
			$H=(H_1, H_2)^{\rm T}$ &   $\bf 2$ & $\bf 1$ & $\bf 2$ & $3$ & $-1$ & $(+,-)^{\rm T}$ & 0 & 0 \\ 
			$\Phi_H=(\Phi_1,\Phi_2)^{\rm T}$ &   $\bf 1$ & $\bf 1$ & $\bf 2$ & $0$ & $+1$ & $(-,+)^{\rm T}$ & 0 & 0  \\ 
			\hline   
		\end{tabular}
	\end{center}
\end{table}
\normalsize

\section{G2HDM - A Concise Review}
\label{sec:G2HDM}

The dark sector of the minimal G2HDM is based on a hidden gauge group $U(1)_{X} \times SU(2)_{H}$, which is a dark replica of the electroweak SM with its minimal matter content shown in Table~\ref{tab:mattercontent}. A distinguishing feature of G2HDM 
is to group together the two Higgs doublets $H_1$ and $H_2$ in the popular Inert Two-Higgs-Doublet Model (I2HDM) of dark matter, forming a two-dimensional spinor representation of a hidden $SU(2)_H$ group. This is denoted by $H$ in Table~\ref{tab:mattercontent}. Another hidden doublet $\Phi_H$ is also necessary to introduce in order to provide masses to the hidden particles in the model. This aspect will be elaborated further in Section~\ref{sec:symmetrybreaking} when we discuss symmetry breaking effects. 
One may specify explicitly all the quantum numbers of a particle $P$ by using the notation $P(R_{SU(2)_L},R_{SU(2)_H},R_{SU(3)_C)})_{q,h}$. 
Here $R_{G}$ denotes the particle representation under the group $G$; and the subscripts $q$ and $h$ represent the hypercharge under $U(1)_{\widetilde Y}$ and dark hypercharge under $U(1)_{\widetilde X}$, respectively. The normalization for the Abelian group we adopted here is: $q={\widetilde Y}=6Y$ and $h={\widetilde{X}}=2X$. This implies that all particles listed in Table~\ref{tab:mattercontent} for the minimal G2HDM have integral values of $q$ and $h$.~\footnote{
We note that the $U(1)_{\tilde X}$ quantum numbers presented in Table~\ref{tab:mattercontent} are slightly modified compared to those used in previous works~\cite{Ramos:2021txu,Ramos:2021omo}. These changes affect neither the cancellations of gauge and gravitational anomalies discussed below nor the previously phenomenological studies.}
Thus, $H=H(\boldsymbol2,\boldsymbol2,\boldsymbol1)_{3,-1}$ is the bi-doublet, $\Phi_H = H(\boldsymbol1,\boldsymbol2,\boldsymbol1)_{0,1}$ is the hidden doublet, {\it etc.} 

Another distinctive feature of the G2HDM model is the presence of an accidental discrete symmetry, known as  $h$-parity, under which all SM particles are  $h$-parity even. This symmetry emerges naturally within the model and, if unbroken, ensures the stability of the lightest  $h$-parity odd particle. Consequently, the lightest  $h$-parity odd particle can serve as a viable dark matter candidate. Depending on the parameter space, this candidate may be a hidden scalar  $H_2^0$ (the neutral component of  the inert doublet $H_2$), a hidden Dirac fermion  $\nu^H$, or a hidden gauge boson  $W^{\prime}$. This $h$-parity offers a compelling explanation for the stability of dark matter within the G2HDM framework. The  $h$-parity assignments for the matter content in minimal G2HDM are summarized in the seventh column in Table~\ref{tab:mattercontent}. For the hidden gauge bosons, it is notable that while  $W^{\prime}$ is  $h$-parity odd, the dark photon and dark  $Z$  boson are  $h$-parity even. Recent studies of dark matter phenomenology in minimal G2HDM can be found in~\cite{Ramos:2021txu,Ramos:2021omo,Chen:2019pnt,Ramsey-Musolf:2024zex}.

With the quantum number assignments for the fermion content in the minimal G2HDM, it is straightforward to verify that the model is free from both gauge and gravitational anomalies, mixed or unmixed. Additionally, since the model includes 12 left-handed doublets of $SU(2)_L$ and 24 right-handed doublets of $SU(2)_H$, it is also free from Witten's global $SU(2)$ anomaly~\cite{Witten:1982fp}, just like SM.

For the global baryon and lepton number assignments of the fermions in the minimal G2HDM, as provided in the last two columns of Table~\ref{tab:mattercontent}, it can be readily confirmed that while $B$ and $L$ are anomalous, $B-L$ is also anomaly-free per generation in G2HDM.
\be
\begin{aligned} 
\label{eq:G2HDMBminusL}
\sum{LSU(2)_L^2} =\sum{BSU(2)_L^2}=+1 \;, & \quad\quad \sum{L{\widetilde Y}^2}=\sum{B{\widetilde Y}^2}=-18\;, \\
\sum{L SU(2)_H^2} = \sum{B SU(2)_H^2} = -2 \; , & \quad\quad  \sum{L {\tilde X}^2} = \sum{B {\tilde X}^2} = +4 \; , \\
& \quad\quad \sum{L\tilde{X}\tilde{Y}} 
  = \sum{B\tilde{X}\tilde{Y}}= 0\;. 
\end{aligned}
\ee
We note that the anomalies of $LSU(2)_L^2$, $BSU(2)_L^2$, $L{\widetilde Y}^2$, and $B{\widetilde Y}^2$ per generation have the same values in both the SM and minimal G2HDM.

Besides the three SM $\theta$-angles discussed in previous section, we have two more $\theta$-angles, $\theta_{2H}$ and $\tilde \theta_{X}$ for $SU(2)_H$ and $U(1)_{\widetilde X}$ respectively. Under the global lepton number $U(1)_L$ (or global baryon number $U(1)_B$) rotation with angle $\alpha$, according to the analysis of the mixed anomalies in Eq.~(\ref{eq:G2HDMBminusL}), we have \be
\begin{aligned}
\theta_{2L}\rightarrow\theta_{2L} +\alpha \; , &\quad \tilde{\theta}_Y\rightarrow\tilde{\theta}_Y - 18\alpha \; ; \\
\theta_{2H}\rightarrow\theta_{2H} - 2\alpha \; , &\quad \tilde{\theta}_X\rightarrow\tilde{\theta}_X +4\alpha \; .
\end{aligned}
\ee
Therefore, in this model, the two $SU(2)$ $\theta$-angles, $\theta_{2L}$ and $\theta_{2H}$ are unphysical and can always be rotated away by chiral rotation; while two linear combinations can be defined: $\tilde{\theta}_Y + 18\theta_{2L}$ and $\tilde{\theta}_X + 2\theta_{2H}$, both of which are physical.
Following~\cite{Tong:2017oea}, in our analysis in Section~\ref{sec:theta} for the $\theta$-angles in G2HDM, we will ignore the effects of the global mixed anomalies from the lepton number and baryon number currents to the two angles $\theta_{2L}$ and $\theta_{2H}$.

Since the covering group $\tilde G$ in Eq.~(\ref{eq:G2HDMcoveringgroup}) has a total of 5 factors, there are many ways to do the partitions for quotienting. Thus it is useful to classify its quotient group $G$ given by Eq.~(\ref{eq:G2HDMQuotientGroup1}) (or~(\ref{eq:G2HDMQuotientGroup2})) into various patterns.
We will focus on the following three specific partitions:
\be
\begin{aligned}
\label{eq:quotientpatterns}
(A) &    \quad \quad \quad
    G= \frac{U\left( 1 \right)\times SU(2) \times SU(3) }{\Gamma_p}\times \frac{U\left( 1 \right)\times SU(2)}{\Gamma_m} \;  , \\
(B) &    \quad \quad \quad 
    G=\frac{U{{\left( 1 \right)}}\times SU{{\left( 3 \right)}}}{\Gamma_p}\times \frac{ U{{\left( 1 \right)}} \times SU{{\left( 2 \right)}}}{\Gamma_m}\times \frac{SU{{\left( 2 \right)}}}{\Gamma_n} \; , \\
(C) &  \quad \quad \quad  
    G=\frac{U{{\left( 1 \right)}}\times SU{{\left( 2 \right)}}}{\Gamma_p}\times \frac{ U{{\left( 1 \right)}} \times SU{{\left( 2 \right)}}}{\Gamma_m}\times \frac{SU{{\left( 3 \right)}}}{\Gamma_n} \; .
\end{aligned}
\ee

For case (A), $G$ has 8 possibilities:\\ 
$\frac{U{{\left( 1 \right)}_{Y}}\times SU{{\left( 2 \right)}_{L}}\times SU{{\left( 3 \right)}_{C}}}{\Gamma_p}\times \frac{ U{{\left( 1 \right)}_{X}} \times SU{{\left( 2 \right)}_{H}}}{\Gamma_m}$,
$\frac{U{{\left( 1 \right)}_{X}}\times SU{{\left( 2 \right)}_{L}}\times SU{{\left( 3 \right)}_{C}}}{\Gamma_p}\times \frac{ U{{\left( 1 \right)}_{Y}} \times SU{{\left( 2 \right)}_{H}}}{\Gamma_m}$, \\
$\frac{U{{\left( 1 \right)}_{Y}}\times SU{{\left( 2 \right)}_{H}}\times SU{{\left( 3 \right)}_{C}}}{\Gamma_p}\times \frac{U{{\left( 1 \right)}_{X}} \times SU{{\left( 2 \right)}_{L}}}{\Gamma_m}$,
$\frac{U{{\left( 1 \right)}_{X}}\times SU{{\left( 2 \right)}_{H}}\times SU{{\left( 3 \right)}_{C}}}{\Gamma_p}\times \frac{ U{{\left( 1 \right)}_{Y}} \times SU{{\left( 2 \right)}_{L}}}{\Gamma_m}$, plus four other terms with 
$Y \to V$ and $X \to A$. 

For case (B), $G$ has 8 possibilities:\\
$\frac{U{{\left( 1 \right)}_{X}}\times SU{{\left( 3 \right)}_{C}}}{\Gamma_p}\times \frac{ U{{\left( 1 \right)}_{Y}} \times SU{{\left( 2 \right)}_{H}}}{\Gamma_m}\times \frac{SU{{\left( 2 \right)}_{L}}}{\Gamma_n}$, 
$\frac{U{{\left( 1 \right)}_{Y}}\times SU{{\left( 3 \right)}_{C}}}{\Gamma_p}\times \frac{ U{{\left( 1 \right)}_{X}} \times SU{{\left( 2 \right)}_{H}}}{\Gamma_m}\times \frac{SU{{\left( 2 \right)}_{L}}} {\Gamma_n}$, \\
$\frac{U{{\left( 1 \right)}_{X}}\times SU{{\left( 3 \right)}_{C}}}{\Gamma_p}\times \frac{ U{{\left( 1 \right)}_{Y}} \times SU{{\left( 2 \right)}_{L}}}{\Gamma_m}\times \frac{SU{{\left( 2 \right)}_{H}}}{\Gamma_n}$, 
$\frac{U{{\left( 1 \right)}_{Y}}\times SU{{\left( 3 \right)}_{C}}}{\Gamma_p}\times \frac{ U{{\left( 1 \right)}_{X}} \times SU{{\left( 2 \right)}_{L}}}{\Gamma_m}\times \frac{SU{{\left( 2 \right)}_{H}}} {\Gamma_n}$,
plus four other terms with
$Y \to V$ and $X \to A$. 

For case (C), $G$ has 4 possibilities: \\
$\frac{U{{\left( 1 \right)}_{Y}}\times SU{{\left( 2 \right)}_{H}}}{\Gamma_p}\times \frac{ U{{\left( 1 \right)}_{X}} \times SU{{\left( 2 \right)}_{L}}}{\Gamma_m}\times \frac{SU{{\left( 3 \right)}_{C}}}{\Gamma_n}$, 
$\frac{U{{\left( 1 \right)}_{X}}\times SU{{\left( 2 \right)}_{H}}}{\Gamma_p}\times \frac{ U{{\left( 1 \right)}_{Y}} \times SU{{\left( 2 \right)}_{L}}}{\Gamma_m}\times \frac{SU{{\left( 3 \right)}_{C}}}{\Gamma_n}$, 
plus two other terms with $Y \to V$ and $X \to A$. 

Together, we have 20 possible quotient groups $G$ for the 3 cases of (A), (B), and (C).
Furthermore, we can even allow one or both $U(1)$ factors to lie outside the center quotient structure, as seen in cases like $G=U(1)\times \frac{SU(2)\times SU(3)}{\Gamma_p}\times \frac{U(1) \times SU(2)}{\Gamma_m}$, $U(1)\times U(1) \times \frac{SU(2)}{\Gamma_p}\times \frac{SU(2)\times SU(3)}{\Gamma_m}$, and similar patterns. However, these constructions are relatively straightforward and do not introduce any new features beyond what we will be exploring in the following sections for the previous three cases. As such, we will not delve into these cases further and will instead focus on the 3 representative cases of (A), (B), and (C).

\section{ Line Operators  }\label{sec:lineoperators}

For a simple $SU(N)$ group, its center is a discrete cyclic group $\boldsymbol{Z}_N$. For example, the center of $SU(3)$ is $\boldsymbol{Z}_3$, while the center  of the semi-simple group $SU(2) \times SU(3)$ is $\boldsymbol{Z}_6$. 
When considering a quotient group $G = \tilde{G}/{\Gamma}$, where $\Gamma$ is a discrete subgroup of the center of the universal covering group  $\tilde{G}$ (with $\Gamma = \boldsymbol{1}$ corresponding to the universal cover itself), the local properties of the quotient group remain unchanged. Fortunately, the global symmetries differ under different associated gauge groups~\cite{ORaifeartaigh:1986agb,Hucks:1990nw}. For example, as mentioned earlier the $\Gamma$ in the SM $U(1)_Y \times SU(2)_L \times SU(3)_C$ can be either $\boldsymbol{1}$, $\boldsymbol{Z}_2$, $\boldsymbol{Z}_3$ or $\boldsymbol{Z}_6$, and the global properties are all different in each case~\cite{Tong:2017oea}.

A Wilson line, a special configuration of Wilson loop~\cite{Wilson:1974sk}, is a static operator extending along the time direction, representing a electrically charged object with infinite mass sitting at the origin. A non-Abelian Wilson line operator, defined in terms of the Lie algebra $\boldsymbol{\mathrm{g}}$ of a gauge group $G$, is expressed as
\be
W_R [\mathcal{C}] = {\rm Tr}_{R}  P \exp\left(i\int_{\mathcal{C}} dt A_{0}\right) \; ,
\ee
where $A_0 = A_0^a T^a(R)$ with $T^a(R)$ the generators of $\boldsymbol{\mathbf g}$ in a representation $R$ of $G$, $\mathcal{C}$ is a line along the time direction,
and $P$ denotes path ordering. Wilson line operators exist for any representation $R$ and, as elements of the gauge symmetry, must remain invariant under any choice of $\Gamma$. Wilson lines are denoted as $\Lambda_e/{W}$, where $\Lambda_e$ is the weight lattice of $\boldsymbol{\mathrm{g}}$, and $W$ is the Weyl group. A purely electric line is characterized by the label $(\lambda^e, 0)$, where $\lambda^e \in \Lambda_e$ with the equivalent relation $\lambda^e \sim w \lambda^e$, where $w \in W$.

A 't Hooft line, a special configuration of 't Hooft loop~\cite{tHooft:1977nqb}, which represents an infinitely heavy magnetically charged particle sitting at the origin, is expressed as
\be
T_{R^*} [\mathcal{C}] = {\rm Tr}_{{R}^{*}}  P \exp\left(i\int_{\mathcal{C}} d t A^{*}_{0}\right) \; ,
\ee
where $R^{*}$ is the representation associated with the dual Lie algebra 
$\boldsymbol{\mathrm{g}}^*$ of 
$\boldsymbol{\mathrm{g}}$, corresponding to the gauge group $G$. 't Hooft lines are represented as 
$\Lambda_{m}/{W}$, where 
$\Lambda_{m}$ is the weight lattice of the dual $\boldsymbol{\mathrm{g}}^*$, or equivalently, the dual of root lattice of $\boldsymbol{\mathrm{g}}$. A purely magnetic line is labeled by 
$(0, \lambda^m )$, where $\lambda^m \in \Lambda_m$ with $\lambda^m \sim w \lambda^m$ and $w \in W$.

For a quotient group $G = \tilde{G}/{\Gamma}$, the Wilson and 't Hooft lines form subsets of the corresponding Wilson and 't Hooft lines associated with $\tilde{G}$. For each $G$, these lines exhibit partially distinct lattices, depending on the choice of $\Gamma$. The lines of interest are those that demonstrate different behavior under varying $\Gamma$. Specifically, the distinct lattices are given by ${\Lambda_e}/{\Lambda_r} = \boldsymbol{Z}_N$ for Wilson lines and ${\Lambda_m}/{\Lambda_{cr}} = \boldsymbol{Z}_N$ for 't Hooft lines. Here, $\Lambda_r$ and $\Lambda_{cr}$ represent the root and co-root lattices of the Lie algebra $\boldsymbol{\mathrm{g}}$, respectively.

Dyons, classical configurations which carry both electric and magnetic charges, can arise in both Abelian~\cite{Schwinger:1966nj,Schwinger:1968rq,Zwanziger:1968ams,Zwanziger:1968rs} and non-Abelian~\cite{Julia:1975ff} gauge theories.
When quantized, these dyons become quantum states carrying both electric and magnetic charges.
More generally, there exist Wilson-'t Hooft non-Abelian dyonic lines, which can be labeled as
\be
(z^e, z^m) \in \Gamma \times \Gamma \; .
\ee
The pairs $(z^e, z^m) \sim (w z^e, w z^m)$ with $w \in W$ form a class, with the center imposing invariance, reflecting constraints on the charges. As noted earlier, only $z^e$ elements invariant under transformations by elements of $\Gamma$ are allowed. The existence and structure of magnetic charges are further constrained by the following generalized Dirac quantization condition, which must be satisfied: If two non-Abelian dyonic lines, $(z^e, z^m)$ and $(z^{e}{'}, z^{m}{'})$, are present, they can coexist only if the following condition is met~\cite{Corrigan:1976wk}:
\begin{equation}\label{dyonicQC}
z^e z^{m}{'} - z^{e}{'} z^m = 0 \pmod{N} \; .
\end{equation}

In summary, Wilson and 't Hooft lines correspond to heavy electric and magnetic insertions in space, respectively. Once $\boldsymbol{Z}_{N}$ is determined for the quotient group, Wilson lines can be directly constructed, while the generalized Dirac quantization condition~\cite{Corrigan:1976wk} places restrictions on both the magnetic and dyonic lines. In general, if the conditions permit, a larger value of $N$ allows the existence of a greater number of magnetic lines.

The introduction of the $\theta$-term highlights the crucial role of $\theta$-angles~\cite{Aharony:2013hda,Hsin:2020nts,Choi:2023pdp} as topological parameters in shaping the structure of the quantum vacuum. This is particularly significant in the context of strong interactions in QCD, where the $\theta$-angle is closely tied to $CP$ violation and poses the so-called strong $CP$ problem, related to the neutron electric dipole moment. The Peccei-Quinn mechanism~\cite{Peccei:1977hh,Peccei:1977ur} proposed addressing this issue by introducing the axion particle~\cite{Weinberg:1977ma,Wilczek:1977pj}, allowing for the dynamical adjustment of the $\theta$-angle and stabilizing the vacuum state of the system.

Moreover, the global structure of the gauge group affects the periodicity of the $\theta$-angle~\cite{Tong:2017oea}. In systems such as $SU(N)$ and $SU(N)/\boldsymbol{Z}_N$, the global structure of the gauge group determines the allowed range of $\theta$-value, influencing the complexity of the vacuum state and the behavior of line operators. For the gauge group $SU(N)$, the periodicity of the $\theta$-angle is given by $\theta \in [0, 2\pi)$. However, for $SU(N)/\boldsymbol{Z}_N$, the presence of the center group extends the periodicity of the $\theta$-angle to
\be
\theta \in [0, 2\pi N) \; .
\ee
This extension of the $\theta$-angle's periodicity is linked to the theory of $SU(N)/\boldsymbol{Z}_N$, which allows instantons to carry a fractional charge of $\frac{1}{N}$ ~\cite{Tong2018,Jackiw:1976pf,Callan:1976je}.

The broader range of $\theta$-value enables the system to explore different topological sectors~\cite{Longo:2003wy}, thereby distinguishing gauge theories with distinct physical properties. The existence of these unique topological sectors enriches the theoretical framework and provides a broader context for studying gauge theories with varying topological characteristics.

The Witten effect~\cite{Witten:1979ey} informs us that when the $\theta$-angle is shifted by $\theta \to \theta + 2\pi$, the pairs of minimal lattices $(z^e, z^m)$ transform as $(z^e, z^m) \to (z^e + z^m, z^m)$. 
Thus, purely magnetic lines can turn into dyonic lines via the Witten effect.
The $\theta$-angle can also be interpreted as additional solutions to Eq.~\eqref{dyonicQC}. Specifically, for $G = SU(3)/\boldsymbol{Z}_3$, when $\theta$ undergoes a rotation $\theta \to \theta + 2\pi n$ with $n=0,1,2$, the lattices transform as $(z^e, z^m) \to (z^e + nz^m, z^m)$. 
This relationship is clearly illustrated in Figs.~\ref{fig:SU(2)theta2pi} and \ref{fig:SU(3)theta2&4pi} for the $SU(2)$ and $SU(3)$ respectively, where the blue boxes represent the smallest classes of this theory, and the green circles denote the allowed lattices~\cite{Tong:2017oea}. 

\begin{figure}[htbp!]
	
	\subfigure[$\, SU(2)$]{
		\begin{tikzpicture}[scale=0.75]
		\draw[->](-1.5,0)--(2.5,0)node[below]{$z_2^e$};
		\draw[->](0,-1.5)--(0,2.5)node[left]{$z_2^m$};
		
		\filldraw[opacity=0.25] [fill=blue,draw=black](-0.5,-0.5) rectangle (1.5,1.5);
		
		\foreach \x in {-1,0,1,2} {
			\foreach \y in {-1,1,2}{
				\filldraw [fill=white,draw=black] (\x,\y) circle(0.25);		}	}
		
		\foreach \x in {-1,0,1,2} {
			\foreach \y in {0,2}{
				\filldraw [fill=green,draw=black] (\x,\y) circle(0.25);		}	}
	\end{tikzpicture}		
	}
	\hspace{5pt}
	\subfigure[$\, SU(2)/\boldsymbol{Z}_{2}\;$ at $\theta=0$]{
		\begin{tikzpicture}[scale=0.75]
		\draw[->](-1.5,0)--(2.5,0)node[below]{$z_2^e$};
		\draw[->](0,-1.5)--(0,2.5)node[left]{$z_2^m$};
		
		\filldraw[opacity=0.25] [fill=blue,draw=black](-0.5,-0.5) rectangle (1.5,1.5);
		
		\foreach \x in {-1,1,2} {
			\foreach \y in {-1,0,1,2}{
				\filldraw [fill=white,draw=black] (\x,\y) circle(0.25);		}	}
		
		\foreach \x in {0,2} {
			\foreach \y in {-1,0,1,2}{
				\filldraw [fill=green,draw=black] (\x,\y) circle(0.25);		}	}
	\end{tikzpicture}
	}
	\hspace{5pt}
	\subfigure[$\,SU(2)/\boldsymbol{Z}_{2}\;$ at $\theta=2\pi$]{
		\begin{tikzpicture}[scale=0.75]		
		\draw[->](-1.5,0)--(2.5,0)node[below]{$z_2^e$};
		\draw[->](0,-1.5)--(0,2.5)node[left]{$z_2^m$};
		
		\filldraw[opacity=0.25] [fill=blue,draw=black](-0.5,-0.5) rectangle (1.5,1.5);
		
		\foreach \x in {-1,0,1,2} {
			\foreach \y in {-1,0,1,2}{
				\filldraw [fill=white,draw=black] (\x,\y) circle(0.25);		}	}
		
		\foreach \x in {-1,1} {
			\foreach \y in {-1,1}{
				\filldraw [fill=green,draw=black] (\x,\y) circle(0.25);		}	}
		\foreach \x in {0,2} {
			\foreach \y in {0,2}{
				\filldraw [fill=green,draw=black] (\x,\y) circle(0.25);		}	}
	\end{tikzpicture}
	}
	\caption{The spectra of line operators for $SU(2)$ in (a) and $SU(2)/\boldsymbol{Z}_2$ at $\theta = 0$ (b), $2\pi$ (c), respectively. Last case of (c) represent the Witten effect. 
    See~\cite{Tong:2017oea} for more detailed discussions.
    }
	\label{fig:SU(2)theta2pi}
\end{figure}

\begin{figure}[htbp!]
	
	\subfigure[$\, SU(3)$]{
		\begin{tikzpicture}[scale=0.5]
			\draw[->](-2.5,0)--(3.5,0)node[below]{$z_3^e$};
			\draw[->](0,-2.5)--(0,3.5)node[left]{$z_3^m$};
			
			\filldraw[opacity=0.25] [fill=blue,draw=black](-0.5,-0.5) rectangle (2.5,2.5);
			
			\foreach \x in {-2,-1,0,1,2,3} {
				\foreach \y in {-2,-1,1,2}{
					\filldraw [fill=white,draw=black] (\x,\y) circle(0.25);		}	}
			
			\foreach \x in {-2,-1,0,1,2,3} {
				\foreach \y in {0,3}{
					\filldraw [fill=green,draw=black] (\x,\y) circle(0.25);		}	}
		\end{tikzpicture}
	}
	\hspace{5pt}
	\subfigure[$\, SU(3)/\boldsymbol{Z}_{3}\;$ at $\theta=0$]{
		\begin{tikzpicture}[scale=0.5]		
			\draw[->](-2.5,0)--(3.5,0)node[below]{$z_3^e$};
			\draw[->](0,-2.5)--(0,3.5)node[left]{$z_3^m$};
			
			\filldraw[opacity=0.25] [fill=blue,draw=black](-0.5,-0.5) rectangle (2.5,2.5);
			
			\foreach \x in {-2,-1,1,2} {
				\foreach \y in {-2,-1,0,1,2,3}{
					\filldraw [fill=white,draw=black] (\x,\y) circle(0.25);		}	}
			
			\foreach \x in {0,3} {
				\foreach \y in {-2,-1,0,1,2,3}{
					\filldraw [fill=green,draw=black] (\x,\y) circle(0.25);		}	}
		\end{tikzpicture}
	}
	\hspace{5pt}
	\subfigure[$\,SU(3)/\boldsymbol{Z}_{3}\;$ at $\theta=2\pi$]{
		\begin{tikzpicture}[scale=0.5]		
			\draw[->](-2.5,0)--(3.5,0)node[below]{$z_3^e$};
			\draw[->](0,-2.5)--(0,3.5)node[left]{$z_3^m$};
			
			\filldraw[opacity=0.25] [fill=blue,draw=black](-0.5,-0.5) rectangle (2.5,2.5);
			
			\foreach \x in {-2,-1,0,1,2,3} {
				\foreach \y in {-2,-1,0,1,2,3}{
					\filldraw [fill=white,draw=black] (\x,\y) circle(0.25);		}	}
			
			\foreach \x in {-2,-1,0,1,2,3} {
				\filldraw [fill=green,draw=black] (\x,\x) circle(0.25);     	}
			
			\foreach \x in {1,2,3} {
				\filldraw [fill=green,draw=black] (\x,\x-3) circle(0.25);     	}
			
			\foreach \x in {-2,-1,0} {
				\filldraw [fill=green,draw=black] (\x,\x+3) circle(0.25);     	}
		\end{tikzpicture}
	}
	\hspace{5pt}
	\subfigure[$\,SU(3)/\boldsymbol{Z}_{3}\;$ at $\theta=4\pi$]{
		\begin{tikzpicture}[scale=0.5]		
			\draw[->](-2.5,0)--(3.5,0)node[below]{$z_3^e$};
			\draw[->](0,-2.5)--(0,3.5)node[left]{$z_3^m$};
			
			\filldraw[opacity=0.25] [fill=blue,draw=black](-0.5,-0.5) rectangle (2.5,2.5);
			
			\foreach \x in {-2,-1,0,1,2,3} {
				\foreach \y in {-2,-1,0,1,2,3}{
					\filldraw [fill=white,draw=black] (\x,\y) circle(0.25);		}	}
			
			\foreach \x in {-2,-1,0,1,2} {
				\filldraw [fill=green,draw=black] (\x,-\x) circle(0.25);     	}
			
			\foreach \x in {0,1,2,3} {
				\filldraw [fill=green,draw=black] (\x,-\x+3) circle(0.25);     	}
			
			\foreach \x in {-2,-1} {
				\filldraw [fill=green,draw=black] (\x,-\x-3) circle(0.25);     	}
			
			\filldraw [fill=green,draw=black] (3,3) circle(0.25);
		\end{tikzpicture}
	}
	\caption{The spectra of line operators for $SU(3)$ in (a) and $SU(3)/\boldsymbol{Z}_3$ at $\theta = 0$ (b), $2\pi$ (c), and $4\pi$ (d), respectively. Last two cases of (c) and (d) represent the Witten effect. See~\cite{Tong:2017oea} for more detailed discussions.
    }
    \label{fig:SU(3)theta2&4pi}
\end{figure}

In the case of $U(N) \cong {(U(1) \times SU(N))}/{\boldsymbol{Z}_N}$ theory, the $U(1)$ factor $\theta$-angle, denoted as $\tilde \theta$, extends to $[0, 2\pi N^2)$, instead of the usual $[0, 2\pi)$. Meanwhile, the periodicity of the $SU(N)$ $\theta$-angle, denoted as $\theta_N$, remains unchanged.
This extended range introduces a richer structure in terms of vacuum states and the behavior of line operators, thereby distinguishing different gauge theories based on the allowed values of $\theta$-angle and their physical implications. This effect is particularly pronounced in the magnetic sector through the Witten effect~\cite{Tong:2017oea}, as we will explore further in Section~\ref{sec:theta}.

In Table~\ref{tab:periodicitySUN}, we summarize the periodicity of the $CP$-violating $\theta$-angles for $SU(N)$, $SU(N)/\boldsymbol{Z}_N$ and $( U(1) \times SU(N) ) / \boldsymbol{Z}_N \cong U(N)$ obtained in~\cite{Tong:2017oea}.

\begin{table}[htbp!]
	\begin{center}
		\caption{Periodicity of the $CP$-violating $\theta$-angles for $SU(N)$, $SU(N)/\boldsymbol{Z}_N$ and $U(N) \cong ( U(1) \times SU(N) )/\boldsymbol{Z}_N$.}
        \label{tab:periodicitySUN}
		\begin{tabular}{|c|c|c|c|}
			\hline
			& $SU(N)$ & $SU(N)/\boldsymbol{Z}_N$ & $ \, U(N) \cong (U(1) \times SU(N))/\boldsymbol{Z}_N$ \, \\ 
            \hline
            \, \text{Periodicity} \, & \, $\theta_N \in [0, 2 \pi)$ \, & \, $\theta_N \in [0, 2 \pi N)  $ \, & \, $ {\tilde \theta} \in [0, 2 \pi N^2), \theta_N \in [0, 2 \pi)$ \, \\
			\hline
		\end{tabular}
	\end{center}
\end{table}

In the SM, the $\theta$-angle for $SU(3)_C$, denoted as ${\theta_3}$, plays a central role in phenomenology. The extremely small value of ${\theta_3}$ (${ \theta_3 \leq 10^{-10}}$), inferred from the non-observation of a neutron electric dipole moment, represents a major unresolved issue in particle physics.
For the $\theta$-angle of $SU(2)_L$, denoted as $\theta_{2L}$, the presence of an anomalous global $B + L$ symmetry allows it to be rotated away by chiral rotation. However, as Tong pointed out in~\cite{Tong:2017oea}, there is a caveat warrants a clarification: There is actually a linear combination of $\theta_{2L}$ and $\tilde{\theta}_Y$ (the $\theta$-angle of $U(1)_{\widetilde Y}$ with the physical hypercharge $Y=\widetilde{Y}/6$) in the SM that cannot be rotated away. This is due to the fact that the global baryon and lepton number symmetry currents, when coupled to the gauge fields of $SU(2)_L $ and $U(1)_{\widetilde Y}$, suffer from mixed anomalies per generation as
\be
\label{SMBminusL}
\sum{LSU(2)_L^2}=\sum{BSU(2)_L^2}=+1 \;,\quad\quad \sum{L{\widetilde Y}^2}=\sum{B{\widetilde Y}^2}=-18\;. 
\ee
Nevertheless, $B-L$ remains anomaly-free in the SM.
Consequently, under a global lepton number $U(1)_L$ (or baryon number $U(1)_B$) transformation with parameter $\alpha$, the $\theta_{2L}$ and $\tilde{\theta}_Y$ angles transform as follows:
\be
\theta_{2L}\rightarrow\theta_{2L} + \, \alpha \; ,\quad \quad \tilde{\theta}_Y\rightarrow\tilde{\theta}_Y - 18\alpha \; . 
\ee
Thus, while $\theta_{2L}$ can be rotated away by appropriate $\alpha$, there is always one physical combined $\theta$-angle $\tilde{\theta}_Y+18\theta_{2L}$ remains for the unbroken electromagnetism $U_{\rm em}(1)$~\cite{Tong:2017oea}.

\section{Line Operators in G2HDM}
\label{sec:LO-G2HDM}

We will now explore the line operators for some examples of the three representative cases of quotients discussed in Section~\ref{sec:G2HDM}.

We begin by establishing some common notations.
The center of the two $SU(2)_{L}$ and $SU(2)_{H}$ will be denoted as $\boldsymbol{Z}_{2L}$ and  $\boldsymbol{Z}_{2H}$ respectively, while the center of $SU(3)_C$ is simply denoted as $\boldsymbol{Z}_{3}$.
Let $z_{2}^{e}$ represents the center weight lattice of $SU{{\left( 2 \right)}_{L}}$, $x_{2}^{e}$ for $SU{{\left( 2 \right)}_{H}}$, and $z_{3}^{e}$ for $SU{{\left( 3 \right)}_{C}}$, with the symbols $z_{2}^{m}$, $x_{2}^{m}$, and $z_{3}^{m}$ denote their root lattices, respectively. 
The SM and dark hypercharges of $U{{\left( 1 \right)}_{Y}}$ and $U{{\left( 1 \right)}_{X}}$ are denoted by $q=\widetilde Y = 6 Y$ with magnetic charge $g$ and $h=\widetilde X = 2X$ with magnetic charge $k$, respectively. As noted earlier, we adopt the convention that both $q,h \in \boldsymbol{Z}$. 
To avoid word clutter, we'll use ``electric charge'' and ``magnetic charge'' to refer to either the SM or dark versions of these charges. The context will make it clear which one is being referred to.

First, consider the following quotient  
\be
\label{eq:CaseAex1}
G=\frac{U{{\left( 1 \right)}_{Y}}\times SU{{\left( 2 \right)}_{L}}\times SU{{\left( 3 \right)}_{C}}}{\Gamma_p}\times \frac{U{{\left( 1 \right)}_{X}} \times SU{{\left( 2 \right)}_{H}}}{\Gamma_m}\;,
\ee
which is one of the eight possibilities of case (A).
The possible $\Gamma$ for quotienting is therefore
\be
\Gamma= \{ \boldsymbol{1} \times \boldsymbol{1},
{\boldsymbol{Z}_{2L}} \times \boldsymbol{1},
\boldsymbol{1} \times {\boldsymbol{Z}_{2H}},
{\boldsymbol{Z}_{3}} \times \boldsymbol{1}, {\boldsymbol{Z}_{2L}}\times {\boldsymbol{Z}_{2H}},{\boldsymbol{Z}_{6L}} \times \boldsymbol{1}, 
{\boldsymbol{Z}_{3}} \times {\boldsymbol{Z}_{2H}},{\boldsymbol{Z}_{6L}} \times {\boldsymbol{Z}_{2H}} \} \; .
\label{eq:ex1centers}
\ee
The first quotient generator is associated with the centers of $SU{{\left( 2 \right)}_{L}}$ and $SU\left( 3 \right)_C$, combined with an appropriate $U{{\left( 1 \right)}_{Y}}$ rotation, while the second is linked to the center of $SU{{\left( 2 \right)}_{H}}$, accompanied by a $U{{\left( 1 \right)}_{X}}$ factor. These quotients are generated by
\be
\begin{aligned}
	& \xi ={{e}^{2i\pi \frac{q}{6}}}\otimes \eta \otimes \omega ={{e}^{2i\pi \frac{q}{6}}}{{e}^{i\pi z_{2}^{e}}}{{e}^{\frac{2}{3}i\pi z_{3}^{e}}} \; , \\ 
	& \chi ={{e}^{2i\pi \frac{h}{2}}}\otimes \rho ={{e}^{2i\pi \frac{h}{2}}}{{e}^{i\pi x_{2}^{e}}} \; .\\ 
\end{aligned} 
\ee
Here, $\eta$, $\rho$ and $\omega$ belong to the centers of $SU{{\left( 2 \right)}_{L}}$, $SU\left( 2 \right)_H$, and $SU{{\left( 3 \right)}_{C}}$, respectively, obeying ${{\eta }^{2}}=1$, ${{\rho }^{2}}=1$, and ${{\omega }^{3}}=1$. As mentioned before, $q$ is the $U{{\left( 1 \right)}_{\widetilde Y}}$ charge and $h$ is the $U{{\left( 1 \right)}_{\widetilde X}}$ charge. 
For example, the 
center group 
${\boldsymbol{Z}_{6L}} \times \boldsymbol{1}$ is generated by $\xi \times \chi^2$,
${\boldsymbol{Z}_{3}}\times {\boldsymbol{Z}_{2H}}$ by ${{\xi }^{2}}\times \chi $, and ${\boldsymbol{Z}_{2L}}\times \boldsymbol{1}$ by ${{\xi }^{3}}\times {{\chi }^{2}}$, {\it etc.}

As in the SM case~\cite{Tong:2017oea}, the generalized Dirac quantization condition for the gauge group of G2HDM is most easily obtained by studying purely electric or magnetic lines. 
Thus, there are two types of line operators to be considered, two Wilson lines for each quotient factor in Eq.~(\ref{eq:ex1centers}) referred to as Wilson lines $\xi$ and $\chi$, named after their respective quotient generators, and their magnetic dual of 't Hooft lines. They are labeled by three electrical charges $\left( z_{2}^{e},z_{3}^{e},q \right)$ and three magnetic charges $\left( z_{2}^{m},z_{3}^{m},g \right)$, and the other is labeled by two electrical charges $\left( x_{2}^{e},h \right)$ and two magnetic charges $\left( x_{2}^{m},k \right)$. These two sets of charges are linked together via the generalized Dirac quantization condition (GDQC), which is given by~\cite{Corrigan:1976wk}
\begin{equation}
\label{eq:DiracCondition0}
	{{e}^{-\frac{2i\pi q}{6}6g}}{{e}^{i\pi z_{2}^{e}z_{2}^{m}}}{{e}^{\frac{2}{3}i\pi z_{3}^{e}z_{3}^{m}}}{{e}^{i\pi x_{2}^{e}x_{2}^{m}}}{{e}^{-\frac{2i\pi h}{2}2k}}=1 \; .
\end{equation}
Or, equivalently, in terms of electric and magnetic charges
\begin{equation}
\label{eq:DiracCondition}
-6gq+3z_{2}^{e}z_{2}^{m}+2z_{3}^{e}z_{3}^{m}+3x_{2}^{e}x_{2}^{m}-6kh=0\pmod 6 \; .
\end{equation}

For each choice of $\Gamma$ in Eq.~(\ref{eq:ex1centers}), we have to solve the spectra of the line operators in sequence.

$\Gamma=\boldsymbol{1} \times \boldsymbol{1} $: With no quotients, both Wilson lines $\xi$ and $\chi$ are unrestricted. Wilson lines $\xi$ with electric charges ${z}_{2}^{e}=0,1$, ${z}_{3}^{e}=0,1,2$, and ${q\in{\boldsymbol{Z}}}$, are all allowed. Similarly, Wilson lines $\chi$ have electric charges ${x}_{2}^{e}=0,1$ and ${h\in{\boldsymbol{Z}}}$. The GDQC dictates that magnetic charges can only take values $g\in{\boldsymbol{Z}}$ with ${z}_{2}^{m}=0 \,\bmod{\,2}$, ${z}_{3}^{m}=0\,\bmod{\,3}$ and $k\in{\boldsymbol{Z}}$ with ${x}_{2}^{m}=0\,\bmod{\,2}$, respectively. 
This non-Abelian portion of the resulting spectrum corresponding to $SU(2)_L \times SU(3)_C \times SU(2)_H$ is shown in Fig.~\ref{fig:LOex1case1} with the green circles representing the allowed charges.
The left side of Fig.~\ref{fig:LOex1case1} coincides with the SM result~\cite{Tong:2017oea}, as it should be!
Additionally, Abelian lines with $(q,g)=\{(0,1), (1,0)\}$ and $(h,k)=\{(0,1),(1,0)\}$ can be added to generate any other electric and magnetic lattices. 

%Ex 1 case 1
\begin{figure}[htbp!]
	\centering
	\begin{tikzpicture}[scale=0.75]
	\draw[thick][->](-1,0.6)--(9,0.6)node[below]{$z_3^e$};
	\draw[thick][->](0.6,-1)--(0.6,9)node[left]{$z_3^m$};
	\foreach \x in {0,3,6}{
		\foreach \y in {0,3,6}{
			\draw [thick] [->] (\x-0.5,\y)--(\x+2,\y) node [below] {$z_2^e$};		}	}
	\foreach \y in {0,3,6}{
		\foreach \x in {0,3,6}{
			\draw [thick] [->] (\x,\y-0.5)--(\x,\y+2) node [left] {$z_2^m$};		}	}
	
	\foreach \x in {0,1.2,3,4.2,6,7.2} {
		\foreach \y in {0,1.2,3,4.2,6,7.2}{
			\filldraw [fill=white,draw=black] (\x,\y) circle(0.3);		}	}
	
	\foreach \x in {0,1.2,3,4.2,6,7.2} {
		\filldraw [fill=green,draw=black] (\x,0) circle(0.3);		}
	\foreach \x in {0,1.2,3,4.2,6,7.2} {\node at (\x,0) [below=10pt] {$q=0$};	}
	\node at (0,0) [left=10pt] {$g=0$};	
	
	\draw [thick] [->] (11,0.6)--(13.5,0.6) node [below] {$x_2^e$};
	\draw [thick] [->] (11.5,0.1)--(11.5,2.6) node [left] {$x_2^m$};
	\foreach \x in {11.5,12.7} {
		\foreach \y in {0.6,1.8} {
			\filldraw [fill=white,draw=black] (\x,\y) circle(0.3);		}	}
	\foreach \x in {11.5,12.7} {
		\filldraw [fill=green,draw=black] (\x,0.6) circle(0.3);		}
	\node at (11.5,0.6) [left=10pt] {$k=0$};
	\foreach \x in {11.5,12.7} {
		\node at (\x,0.6) [below=10pt] {$h=0$};		}
\end{tikzpicture}
	\caption{$\Gamma = \boldsymbol{1} \times \boldsymbol{1}$ for $G=\frac{U{{\left( 1 \right)}_{Y}}\times SU{{\left( 2 \right)}_{L}}\times SU{{\left( 3 \right)}_{C}}}{\Gamma_p}\times \frac{U{{\left( 1 \right)}_{X}} \times SU{{\left( 2 \right)}_{H}}}{\Gamma_m}$. Abelian lines generated by $(q,g)=\{(0,1),(1,0)\}$ and 
    $(h,k)=\{(0,1),(1,0)\}$ can be included.}
	\label{fig:LOex1case1}
\end{figure}

$\Gamma={\boldsymbol{Z}_{2L}} \times \boldsymbol{1} $: Wilson lines $\chi$ are unrestricted, thus as in previous case, the electric charges ${h\in{\boldsymbol{Z}}}$ with ${x}_{2}^{e}=0,1$ and magnetic charges $k\in \boldsymbol{Z}$ with $x_2^m=0\, \bmod{2}$ are all allowed. 
On the other hand, the Wilson lines $\xi$, invariant under $\xi^3$, are restricted by ${q={z}_{2}^{e}\bmod2}$ with ${z}_{3}^{e}=0,1,2$ in both cases. The GDQC requires magnetic charges $g$ to take $g=0$ when ${z}_{2}^{m}=0\bmod2$ and $g=\frac{1}{2}$ when ${z}_{2}^{m}=1\bmod2$, while ${z}_{3}^{m}=0,1,2\bmod3$. 
The resulting spectrum of these line operators corresponds to $U(2)_{L} \times SU(3)_C \times U(1)_X \times SU(2)_{H} $ and is shown in Fig.~\ref{fig:LOex1case2}.
Again, the left side of Fig.~\ref{fig:LOex1case2} agrees with the SM result.
Additional Abelian lines can be generated by $(q,g)=\{(0,1),(2,0)\}$ and $(h,k)=\{(0,1),(1,0)\}$. 

%Ex 1 case 2
\begin{figure}[htbp!]
	\centering
\begin{tikzpicture}[scale=0.75]
	\draw[thick][->](-1,0.6)--(9,0.6)node[below]{$z_3^e$};
	\draw[thick][->](0.6,-1)--(0.6,9)node[left]{$z_3^m$};
	\foreach \x in {0,3,6}{
		\foreach \y in {0,3,6}{
			\draw [thick] [->] (\x-0.5,\y)--(\x+2,\y) node [below] {$z_2^e$};		}	}
	\foreach \y in {0,3,6}{
		\foreach \x in {0,3,6}{
			\draw [thick] [->] (\x,\y-0.5)--(\x,\y+2) node [left] {$z_2^m$};		}	}
	
	\foreach \x in {0,1.2,3,4.2,6,7.2} {
		\foreach \y in {0,1.2,3,4.2,6,7.2}{
			\filldraw [fill=white,draw=black] (\x,\y) circle(0.3);		}	}
	
	\foreach \x in {0,1.2,3,4.2,6,7.2} {
		\foreach \y in {0,1.2}{
			\filldraw [fill=green,draw=black] (\x,\y) circle(0.3);		}   }
	\foreach \x in {0,3,6} {\node at (\x,0) [below=10pt] {$q=0$};	}
	\foreach \x in {1.2,4.2,7.2} {\node at (\x,0) [below=10pt] {$q=1$};	}
	\node at (0,0) [left=10pt] {$g=0$};
	\node at (0,1.2) [left=10pt] {$g={1}/{2}\;$};
	
	\draw [thick] [->] (11,0.6)--(13.5,0.6) node [below] {$x_2^e$};
	\draw [thick] [->] (11.5,0.1)--(11.5,2.6) node [left] {$x_2^m$};
	\foreach \x in {11.5,12.7} {
		\foreach \y in {0.6,1.8} {
			\filldraw [fill=white,draw=black] (\x,\y) circle(0.3);		}	}
	\foreach \x in {11.5,12.7} {
		\filldraw [fill=green,draw=black] (\x,0.6) circle(0.3);		}
	\node at (11.5,0.6) [left=10pt] {$k=0$};
	\foreach \x in {11.5,12.7} {
		\node at (\x,0.6) [below=10pt] {$h=0$};		}
\end{tikzpicture}
	\caption{$\Gamma = \boldsymbol{Z}_{2L} \times \boldsymbol{1} $ for $G=\frac{U{{\left( 1 \right)}_{Y}}\times SU{{\left( 2 \right)}_{L}}\times SU{{\left( 3 \right)}_{C}}}{\Gamma_p}\times \frac{U{{\left( 1 \right)}_{X}} \times SU{{\left( 2 \right)}_{H}}}{\Gamma_m}$. Abelian lines generated by $(q,g)=\{(0,1),(2,0)\}$ and 
    $(h,k)=\{(0,1),(1,0)\}$ can be included.}
	\label{fig:LOex1case2}
\end{figure}

$\Gamma={ \boldsymbol{1}  \times \boldsymbol{Z}_{2H}}$: Wilson lines $\xi$ are unrestricted, with $q\in \boldsymbol{Z}$ and $z_2^e=0,1$, $z_3^e=0,1,2$. The corresponding magnetic charges are ${g\in{\boldsymbol{Z}}}$, ${z}_{2}^{m}=0 \, \bmod \, 2$ and ${z}_{3}^{m}=0 \, \bmod \, 3$. On the other hand, Wilson lines $\chi$, invariant under $\chi$, are restricted by ${h={x}_{2}^{e}\bmod2}$. The GDQC demands magnetic charges $k$ to satisfy ${2k={x}_{2}^{m}\bmod2}$. The spectrum representing $SU(2)_L \times SU(3)_C \times U(2)_H$ is shown in Fig.~\ref{fig:LOex1case3}.
Additional Abelian lines are generated by $(q,g)=\{(0,1),(1,0)\}$ and $(h,k)=\{(0,1),(2,0)\}$. 

%Ex 1 case 3
\begin{figure}[htbp!]
	\centering
\begin{tikzpicture}[scale=0.75]
	\draw[thick][->](-1,0.6)--(9,0.6)node[below]{$z_3^e$};
	\draw[thick][->](0.6,-1)--(0.6,9)node[left]{$z_3^m$};
	\foreach \x in {0,3,6}{
		\foreach \y in {0,3,6}{
			\draw [thick] [->] (\x-0.5,\y)--(\x+2,\y) node [below] {$z_2^e$};		}	}
	\foreach \y in {0,3,6}{
		\foreach \x in {0,3,6}{
			\draw [thick] [->] (\x,\y-0.5)--(\x,\y+2) node [left] {$z_2^m$};		}	}
	
	\foreach \x in {0,1.2,3,4.2,6,7.2} {
		\foreach \y in {0,1.2,3,4.2,6,7.2}{
			\filldraw [fill=white,draw=black] (\x,\y) circle(0.3);		}	}
	
	\foreach \x in {0,1.2,3,4.2,6,7.2} {
		\filldraw [fill=green,draw=black] (\x,0) circle(0.3);		}
	\foreach \x in {0,1.2,3,4.2,6,7.2} {\node at (\x,0) [below=10pt] {$q=0$};	}
	\node at (0,0) [left=10pt] {$g=0$};
	
	\draw [thick] [->] (11,0.6)--(13.5,0.6) node [below] {$x_2^e$};
	\draw [thick] [->] (11.5,0.1)--(11.5,2.6) node [left] {$x_2^m$};
	\foreach \x in {11.5,12.7} {
		\foreach \y in {0.6,1.8} {
			\filldraw [fill=green,draw=black] (\x,\y) circle(0.3);		}	}
	\foreach \x in {11.5,12.7} {
		\filldraw [fill=green,draw=black] (\x,0.6) circle(0.3);		}
	\node at (11.5,0.6) [left=10pt] {$k=0$};
	\node at (11.5,0.6) [below=10pt] {$h=0$};
	\node at (11.5,1.8) [left=10pt] {$k={1}/{2}\;$};
	\node at (12.7,0.6) [below=10pt] {$h=1$};
\end{tikzpicture}
	\caption{$\Gamma =  \boldsymbol{1}  \times \boldsymbol{Z}_{2H}$ for $G=\frac{U{{\left( 1 \right)}_{Y}}\times SU{{\left( 2 \right)}_{L}}\times SU{{\left( 3 \right)}_{C}}}{\Gamma_p}\times \frac{U{{\left( 1 \right)}_{X}} \times SU{{\left( 2 \right)}_{H}}}{\Gamma_m}$. Abelian lines generated by $(q,g)=\{(0,1),(1,0)\}$ and 
    $(h,k)=\{(0,1),(2,0)\}$ can be included.}
	\label{fig:LOex1case3}
\end{figure}

$\Gamma={\boldsymbol{Z}_{3}} \times \boldsymbol{1} $: Wilson lines $\chi$ are unrestricted. However, Wilson lines $\xi$ must be invariant under ${\xi}^{2}$ and are restricted by ${q={z}_{3}^{e}\bmod3}$. Each of these cases has ${z}_{2}^{e}=0,1$. The GDQC, now allowing solely $SU(3)_C$ magnetic charges, requires $3g = z_3^m \bmod3$. While ${z}_{2}^{m}=0\bmod2$, there are no $SU(2)_L$ magnetic charges. The resulting spectrum of line operators corresponds to $SU(2)_{L} \times U(3)_C \times U(1)_X \times SU(2)_{H}$ and is shown in Fig.~\ref{fig:LOex1case4}. 
This also agrees with SM result.
The Abelian lines generated by $(q,g)=\{(0,1),(3,0)\}$ and $(h,k)=\{(0,1),(1,0)\}$ can be included. 

%Ex 1 case 4
\begin{figure}[htbp!]
	\centering
\begin{tikzpicture}[scale=0.75]
	\draw[thick][->](-1,0.6)--(9,0.6)node[below]{$z_3^e$};
	\draw[thick][->](0.6,-1)--(0.6,9)node[left]{$z_3^m$};
	\foreach \x in {0,3,6}{
		\foreach \y in {0,3,6}{
			\draw [thick] [->] (\x-0.5,\y)--(\x+2,\y) node [below] {$z_2^e$};		}	}
	\foreach \y in {0,3,6}{
		\foreach \x in {0,3,6}{
			\draw [thick] [->] (\x,\y-0.5)--(\x,\y+2) node [left] {$z_2^m$};		}	}
	
	\foreach \x in {0,1.2,3,4.2,6,7.2} {
		\foreach \y in {0,1.2,3,4.2,6,7.2}{
			\filldraw [fill=white,draw=black] (\x,\y) circle(0.3);		}	}
	
	\foreach \x in {0,1.2,3,4.2,6,7.2} {
		\foreach \y in {0,3,6} {
			\filldraw [fill=green,draw=black] (\x,\y) circle(0.3);		}	}
	\foreach \x in {0,1.2} {\node at (\x,0) [below=10pt] {$q=0$};	}
	\foreach \x in {3,4.2} {\node at (\x,0) [below=10pt] {$q=1$};	}
	\foreach \x in {6,7.2} {\node at (\x,0) [below=10pt] {$q=2$};	}
	\node at (0,0) [left=10pt] {$g=0$};
	\node at (0,3) [left=10pt] {$g={1}/{3}\;$};
	\node at (0,6) [left=10pt] {$g={2}/{3}\;$};
	
	\draw [thick] [->] (11,0.6)--(13.5,0.6) node [below] {$x_2^e$};
	\draw [thick] [->] (11.5,0.1)--(11.5,2.6) node [left] {$x_2^m$};
	\foreach \x in {11.5,12.7} {
		\foreach \y in {0.6,1.8} {
			\filldraw [fill=white,draw=black] (\x,\y) circle(0.3);		}	}
	\foreach \x in {11.5,12.7} {
		\filldraw [fill=green,draw=black] (\x,0.6) circle(0.3);		}
	\node at (11.5,0.6) [left=10pt] {$k=0$};
	\foreach \x in {11.5,12.7} {
		\node at (\x,0.6) [below=10pt] {$h=0$};		}
\end{tikzpicture}
	\caption{$\Gamma = \boldsymbol{Z}_{3} \times \boldsymbol{1} $  for $G=\frac{U{{\left( 1 \right)}_{Y}}\times SU{{\left( 2 \right)}_{L}}\times SU{{\left( 3 \right)}_{C}}}{\Gamma_p}\times \frac{U{{\left( 1 \right)}_{X}} \times SU{{\left( 2 \right)}_{H}}}{\Gamma_m}$. Abelian lines generated by $(q,g)=\{(0,1),(3,0)\}$ and 
    $(h,k)=\{(0,1),(1,0)\}$ can be included.}
	\label{fig:LOex1case4}
\end{figure}

$\Gamma ={\boldsymbol{Z}_{2L}} \times {\boldsymbol{Z}_{2H}}$: The behavior of Wilson lines $\xi$ and their corresponding magnetic charges follows the pattern seen in the case of $\Gamma ={\boldsymbol{Z}_{2L}} \times \boldsymbol{1} $, while $\chi$ follows the pattern of the $\Gamma = \boldsymbol{1} \times {\boldsymbol{Z}_{2H}}$ case. The resulting spectrum corresponding to $U(2)_L \times SU(3)_C \times U(2)_H$ is shown in Fig.~\ref{fig:LOex1case5}. Additional Abelian lines with the generators $(q,g)=\{(0,1),(2,0)\}$ and $(h,k)=\{(0,1),(2,0)\}$ can be included.

%Ex 1 case 5
\begin{figure}[htbp!]
	\centering
\begin{tikzpicture}[scale=0.75]
	\draw[thick][->](-1,0.6)--(9,0.6)node[below]{$z_3^e$};
	\draw[thick][->](0.6,-1)--(0.6,9)node[left]{$z_3^m$};
	\foreach \x in {0,3,6}{
		\foreach \y in {0,3,6}{
			\draw [thick] [->] (\x-0.5,\y)--(\x+2,\y) node [below] {$z_2^e$};		}	}
	\foreach \y in {0,3,6}{
		\foreach \x in {0,3,6}{
			\draw [thick] [->] (\x,\y-0.5)--(\x,\y+2) node [left] {$z_2^m$};		}	}
	
	\foreach \x in {0,1.2,3,4.2,6,7.2} {
		\foreach \y in {0,1.2,3,4.2,6,7.2}{
			\filldraw [fill=white,draw=black] (\x,\y) circle(0.3);		}	}
	
	\foreach \x in {0,1.2,3,4.2,6,7.2} {
		\foreach \y in {0,1.2}{
			\filldraw [fill=green,draw=black] (\x,\y) circle(0.3);		}   }
	\foreach \x in {0,3,6} {\node at (\x,0) [below=10pt] {$q=0$};	}
	\foreach \x in {1.2,4.2,7.2} {\node at (\x,0) [below=10pt] {$q=1$};	}
	\node at (0,0) [left=10pt] {$g=0$};
	\node at (0,1.2) [left=10pt] {$g={1}/{2}\;$};
	
	\draw [thick] [->] (11,0.6)--(13.5,0.6) node [below] {$x_2^e$};
	\draw [thick] [->] (11.5,0.1)--(11.5,2.6) node [left] {$x_2^m$};
	\foreach \x in {11.5,12.7} {
		\foreach \y in {0.6,1.8} {
			\filldraw [fill=green,draw=black] (\x,\y) circle(0.3);		}	}
	\foreach \x in {11.5,12.7} {
		\filldraw [fill=green,draw=black] (\x,0.6) circle(0.3);		}
	\node at (11.5,0.6) [left=10pt] {$k=0$};
	\node at (11.5,0.6) [below=10pt] {$h=0$};
	\node at (11.5,1.8) [left=10pt] {$k={1}/{2}\;$};
	\node at (12.7,0.6) [below=10pt] {$h=1$};
\end{tikzpicture}
	\caption{$\Gamma ={\boldsymbol{Z}_{2L}}\times {\boldsymbol{Z}_{2H}}$ for $G=\frac{U{{\left( 1 \right)}_{Y}}\times SU{{\left( 2 \right)}_{L}}\times SU{{\left( 3 \right)}_{C}}}{\Gamma_p}\times \frac{U{{\left( 1 \right)}_{X}} \times SU{{\left( 2 \right)}_{H}}}{\Gamma_m}$. Abelian lines generated by $(q,g)=\{(0,1),(2,0)\}$ and 
    $(h,k)=\{(0,1),(2,0)\}$ can be included.}
	\label{fig:LOex1case5}
\end{figure}

$\Gamma={\boldsymbol{Z}_{6L}} \times \boldsymbol{1} $: While Wilson lines $\chi$ are unrestricted, Wilson lines $\xi$ should be invariant under generator $\xi$. This implies that the Abelian electric charges satisfy ${q={3{z}_{2}^{e}-2{z}_{3}^{e}}\bmod6}$. 
Just like in the SM case~\cite{Tong:2017oea}, all the particles in the minimal G2HDM satisfy this relation.
In this scenario, the spectrum of line operators becomes more abundant, particularly the 't Hooft lines, which can take $SU(2)_L \times {SU(3)_C}$ magnetic charges $6g={3{z}_{2}^{m}+2{z}_{3}^{m}}\bmod6$. The resulting spectrum corresponding to $S(U(2)_L \times U(3)_C) \times SU(2)_H\times U(1)_X$~\footnote{Following Hucks~\cite{Hucks:1990nw}, we define $ S(U(2) \times U(3)) = ( U(1) \times SU(2) \times SU(3) ) / Z_6 $. } is shown in Fig.~\ref{fig:LOex1case6} with its left side agreeing with SM. Additional Abelian lines are generated by $(q,g)=\{(0,1),(6,0)\}$ and $(h,k)=\{(0,1),(1,0)\}$. 

%Ex 1 case 6
\begin{figure}[htbp!]
	\centering
\begin{tikzpicture}[scale=0.75]
	\draw[thick][->](-1,0.6)--(9,0.6)node[below]{$z_3^e$};
	\draw[thick][->](0.6,-1)--(0.6,9)node[left]{$z_3^m$};
	\foreach \x in {0,3,6}{
		\foreach \y in {0,3,6}{
			\draw [thick] [->] (\x-0.5,\y)--(\x+2,\y) node [below] {$z_2^e$};		}	}
	\foreach \y in {0,3,6}{
		\foreach \x in {0,3,6}{
			\draw [thick] [->] (\x,\y-0.5)--(\x,\y+2) node [left] {$z_2^m$};		}	}
	
	\foreach \x in {0,1.2,3,4.2,6,7.2} {
		\foreach \y in {0,1.2,3,4.2,6,7.2}{
			\filldraw [fill=green,draw=black] (\x,\y) circle(0.3);		}	}
	
	\node at (0,0) [below=10pt] {$q=0$};
	\node at (1.2,0) [below=10pt] {$q=3$};
	\node at (3,0) [below=10pt] {$q=4$};
	\node at (4.2,0) [below=10pt] {$q=1$};
	\node at (6,0) [below=10pt] {$q=2$};
	\node at (7.2,0) [below=10pt] {$q=5$};
	\node at (0,0) [left=10pt] {$g=0$};
	\node at (0,1.2) [left=10pt] {$g={1}/{2}\;$};	
	\node at (0,3) [left=10pt] {$g={1}/{3}\;$};	
	\node at (0,4.2) [left=10pt] {$g={5}/{6}\;$};	
	\node at (0,6) [left=10pt] {$g={2}/{3}\;$};	
	\node at (0,7.2) [left=10pt] {$g={1}/{6}\;$};	
	
	\draw [thick] [->] (11,0.6)--(13.5,0.6) node [below] {$x_2^e$};
	\draw [thick] [->] (11.5,0.1)--(11.5,2.6) node [left] {$x_2^m$};
	\foreach \x in {11.5,12.7} {
		\foreach \y in {0.6,1.8} {
			\filldraw [fill=white,draw=black] (\x,\y) circle(0.3);		}	}
	\foreach \x in {11.5,12.7} {
		\filldraw [fill=green,draw=black] (\x,0.6) circle(0.3);		}
	\node at (11.5,0.6) [left=10pt] {$k=0$};
	\foreach \x in {11.5,12.7} {
		\node at (\x,0.6) [below=10pt] {$h=0$};		}
\end{tikzpicture}
	\caption{$\Gamma = \boldsymbol{Z}_{6L} \times \boldsymbol{1} $ for $G=\frac{U{{\left( 1 \right)}_{Y}}\times SU{{\left( 2 \right)}_{L}}\times SU{{\left( 3 \right)}_{C}}}{\Gamma_p}\times \frac{U{{\left( 1 \right)}_{X}} \times SU{{\left( 2 \right)}_{H}}}{\Gamma_m}$. Abelian lines generated by $(q,g)=\{(0,1),(6,0)\}$ and 
    $(h,k)=\{(0,1),(1,0)\}$ can be included.}
	\label{fig:LOex1case6}
\end{figure}

$\Gamma ={\boldsymbol{Z}_{3}}\times {\boldsymbol{Z}_{2H}}$: The behavior of Wilson lines $\xi$ and their corresponding magnetic charges follows the pattern seen in the case of $\Gamma ={\boldsymbol{Z}_{3}} \times \boldsymbol{1}$, while $\chi$ follows the pattern of the $\Gamma =\boldsymbol{1} \times {\boldsymbol{Z}_{2H}}$ case. 
The resulting spectrum corresponding to $SU(2)_L \times U(3)_C \times U(2)_H$ is shown in Fig.~\ref{fig:LOex1case7}. Additional Abelian lines with the following generators $(q,g)=\{(0,1),(3,0)\}$ and $(h,k)=\{(0,1),(2,0)\}$ can be included.

%Ex 1 case 7
\begin{figure}[htbp!]
	\centering
    	\begin{tikzpicture}[scale=0.75]
	\draw[thick][->](-1,0.6)--(9,0.6)node[below]{$z_3^e$};
	\draw[thick][->](0.6,-1)--(0.6,9)node[left]{$z_3^m$};
	\foreach \x in {0,3,6}{
		\foreach \y in {0,3,6}{
			\draw [thick] [->] (\x-0.5,\y)--(\x+2,\y) node [below] {$z_2^e$};		}	}
	\foreach \y in {0,3,6}{
		\foreach \x in {0,3,6}{
			\draw [thick] [->] (\x,\y-0.5)--(\x,\y+2) node [left] {$z_2^m$};		}	}
	
	\foreach \x in {0,1.2,3,4.2,6,7.2} {
		\foreach \y in {0,1.2,3,4.2,6,7.2}{
			\filldraw [fill=white,draw=black] (\x,\y) circle(0.3);		}	}
	
	\foreach \x in {0,1.2,3,4.2,6,7.2} {
		\foreach \y in {0,3,6} {
			\filldraw [fill=green,draw=black] (\x,\y) circle(0.3);		}	}
	\foreach \x in {0,1.2} {\node at (\x,0) [below=10pt] {$q=0$};	}
	\foreach \x in {3,4.2} {\node at (\x,0) [below=10pt] {$q=1$};	}
	\foreach \x in {6,7.2} {\node at (\x,0) [below=10pt] {$q=2$};	}
	\node at (0,0) [left=10pt] {$g=0$};
	\node at (0,3) [left=10pt] {$g={1}/{3}\;$};
	\node at (0,6) [left=10pt] {$g={2}/{3}\;$};
	
	\draw [thick] [->] (11,0.6)--(13.5,0.6) node [below] {$x_2^e$};
	\draw [thick] [->] (11.5,0.1)--(11.5,2.6) node [left] {$x_2^m$};
	\foreach \x in {11.5,12.7} {
		\foreach \y in {0.6,1.8} {
			\filldraw [fill=green,draw=black] (\x,\y) circle(0.3);		}	}
	\foreach \x in {11.5,12.7} {
		\filldraw [fill=green,draw=black] (\x,0.6) circle(0.3);		}
	\node at (11.5,0.6) [left=10pt] {$k=0$};
	\node at (11.5,0.6) [below=10pt] {$h=0$};
	\node at (11.5,1.8) [left=10pt] {$k={1}/{2}\;$};
	\node at (12.7,0.6) [below=10pt] {$h=1$};

\end{tikzpicture}
	\caption{$\Gamma ={\boldsymbol{Z}_{3}}\times {\boldsymbol{Z}_{2H}}$ for $G=\frac{U{{\left( 1 \right)}_{Y}}\times SU{{\left( 2 \right)}_{L}}\times SU{{\left( 3 \right)}_{C}}}{\Gamma_p}\times \frac{U{{\left( 1 \right)}_{X}} \times SU{{\left( 2 \right)}_{H}}}{\Gamma_m}$. Abelian lines generated by $(q,g)=\{(0,1),(3,0)\}$ and 
    $(h,k)=\{(0,1),(2,0)\}$ can be included.}
	\label{fig:LOex1case7}
\end{figure}

$\Gamma ={\boldsymbol{Z}_{6L}}\times {\boldsymbol{Z}_{2H}}$: The behavior of Wilson lines $\xi$ and their corresponding magnetic charges follows the pattern seen in the $\Gamma ={\boldsymbol{Z}_{6L}} \times \boldsymbol{1}$ case, while $\chi$ follows the pattern of the $\Gamma =\boldsymbol{1} \times {\boldsymbol{Z}_{2H}}$ case. 
The resulting spectrum corresponding to $S(U(2)_L \times U(3)_C) \times U(2)_H$ is shown in Fig.~\ref{fig:LOex1case8}.
In addition, Abelian lines with the following generators $(q,g)=\{(0,1),(6,0)\}$ and $(h,k)=\{(0,1),(2,0)\}$ can be included.

%Ex 1 case 8
\begin{figure}[htbp!]
	\centering
    \begin{tikzpicture}[scale=0.75]
	\draw[thick][->](-1,0.6)--(9,0.6)node[below]{$z_3^e$};
	\draw[thick][->](0.6,-1)--(0.6,9)node[left]{$z_3^m$};
	\foreach \x in {0,3,6}{
		\foreach \y in {0,3,6}{
			\draw [thick] [->] (\x-0.5,\y)--(\x+2,\y) node [below] {$z_2^e$};		}	}
	\foreach \y in {0,3,6}{
		\foreach \x in {0,3,6}{
			\draw [thick] [->] (\x,\y-0.5)--(\x,\y+2) node [left] {$z_2^m$};		}	}
	
	\foreach \x in {0,1.2,3,4.2,6,7.2} {
		\foreach \y in {0,1.2,3,4.2,6,7.2}{
			\filldraw [fill=green,draw=black] (\x,\y) circle(0.3);		}	}
	
	\node at (0,0) [below=10pt] {$q=0$};
	\node at (1.2,0) [below=10pt] {$q=3$};
	\node at (3,0) [below=10pt] {$q=4$};
	\node at (4.2,0) [below=10pt] {$q=1$};
	\node at (6,0) [below=10pt] {$q=2$};
	\node at (7.2,0) [below=10pt] {$q=5$};
	\node at (0,0) [left=10pt] {$g=0$};
	\node at (0,1.2) [left=10pt] {$g={1}/{2}\;$};	
	\node at (0,3) [left=10pt] {$g={1}/{3}\;$};	
	\node at (0,4.2) [left=10pt] {$g={5}/{6}\;$};	
	\node at (0,6) [left=10pt] {$g={2}/{3}\;$};	
	\node at (0,7.2) [left=10pt] {$g={1}/{6}\;$};	
	
	\draw [thick] [->] (11,0.6)--(13.5,0.6) node [below] {$x_2^e$};
	\draw [thick] [->] (11.5,0.1)--(11.5,2.6) node [left] {$x_2^m$};
	\foreach \x in {11.5,12.7} {
		\foreach \y in {0.6,1.8} {
			\filldraw [fill=green,draw=black] (\x,\y) circle(0.3);		}	}
	\foreach \x in {11.5,12.7} {
		\filldraw [fill=green,draw=black] (\x,0.6) circle(0.3);		}
	\node at (11.5,0.6) [left=10pt] {$k=0$};
	\node at (11.5,0.6) [below=10pt] {$h=0$};
	\node at (11.5,1.8) [left=10pt] {$k={1}/{2}\;$};
	\node at (12.7,0.6) [below=10pt] {$h=1$};

\end{tikzpicture}
	\caption{$\Gamma ={\boldsymbol{Z}_{6L}}\times {\boldsymbol{Z}_{2H}}$ for $G=\frac{U{{\left( 1 \right)}_{Y}}\times SU{{\left( 2 \right)}_{L}}\times SU{{\left( 3 \right)}_{C}}}{\Gamma_p}\times \frac{U{{\left( 1 \right)}_{X}} \times SU{{\left( 2 \right)}_{H}}}{\Gamma_m}$. Abelian lines generated by $(q,g)=\{(0,1),(6,0)\}$ and 
    $(h,k)=\{(0,1),(2,0)\}$ can be included.}
	\label{fig:LOex1case8}
\end{figure}

Let's consider yet another one of the 8 possibilities of case (A):
\be
\label{eq:ex2}
 G=\frac{U{{\left( 1 \right)}_{X}}\times SU{{\left( 2 \right)}_{H}}\times SU{{\left( 3 \right)}_{C}}}{\Gamma_p}\times \frac{ U{{\left( 1 \right)}_{Y}} \times SU{{\left( 2 \right)}_{L}} }{\Gamma_m} \; .
\ee
This pattern is similar to Eq.~(\ref{eq:CaseAex1}) with the SM $U(1)_Y \times SU(2)_L$ and its dark replica $U(1)_X \times SU(2)_H$ in the two quotient factors interchanged, so the possible $\Gamma$ is similar to Eq.~(\ref{eq:ex1centers}) with $L \leftrightarrow H$, namely,
\be
\Gamma= \{ \boldsymbol{1} \times \boldsymbol{1},
\boldsymbol{1} \times {\boldsymbol{Z}_{2L}},
{\boldsymbol{Z}_{2H}} \times \boldsymbol{1},
{\boldsymbol{Z}_{3}} \times \boldsymbol{1}, {\boldsymbol{Z}_{2H}}\times {\boldsymbol{Z}_{2L}},{\boldsymbol{Z}_{6H}} \times \boldsymbol{1}, 
{\boldsymbol{Z}_{3}} \times {\boldsymbol{Z}_{2L}},{\boldsymbol{Z}_{6H}} \times {\boldsymbol{Z}_{2L}} \} \; .
\label{eq:ex2centers}
\ee
The two discrete cyclic groups in Eq.~(\ref{eq:ex2centers}) correspond to generators:
\be
\begin{aligned}
	& \xi ={{e}^{2i\pi \frac{h}{6}}}\otimes \rho \otimes \omega ={{e}^{2i\pi \frac{h}{6}}}{{e}^{i\pi x_{2}^{e}}}{{e}^{\frac{2}{3}i\pi z_{3}^{e}}} \; , \\ 
	& \chi ={{e}^{2i\pi \frac{q}{2}}}\otimes \eta ={{e}^{2i\pi \frac{q}{2}}}{{e}^{i\pi z_{2}^{e}}}  \; .\\ 
\end{aligned}
\ee
Although the generators have changed, the GDQC remains the same as Eq.~(\ref{eq:DiracCondition}).

The analysis is completely mirroring the previous example. The resulting spectra of the line operators for each element of $\Gamma$ in Eq.~(\ref{eq:ex2centers}) are shown in Figs.~(\ref{fig:LOex2case1}), (\ref{fig:LOex2case2}), (\ref{fig:LOex2case3}), (\ref{fig:LOex2case4}), (\ref{fig:LOex2case5}), (\ref{fig:LOex2case6}), (\ref{fig:LOex2case7}), and (\ref{fig:LOex2case8}) respectively. 
We will not bother the details here.

%Ex 2 case 1
\begin{figure}[htbp!]
	\centering
\begin{tikzpicture}[scale=0.75]
	\draw[thick][->](-1,0.6)--(9,0.6)node[below]{$z_3^e$};
	\draw[thick][->](0.6,-1)--(0.6,9)node[left]{$z_3^m$};
	\foreach \x in {0,3,6}{
		\foreach \y in {0,3,6}{
			\draw [thick] [->] (\x-0.5,\y)--(\x+2,\y) node [below] {$x_2^e$};		}	}
	\foreach \y in {0,3,6}{
		\foreach \x in {0,3,6}{
			\draw [thick] [->] (\x,\y-0.5)--(\x,\y+2) node [left] {$x_2^m$};		}	}
	
	\foreach \x in {0,1.2,3,4.2,6,7.2} {
		\foreach \y in {0,1.2,3,4.2,6,7.2}{
			\filldraw [fill=white,draw=black] (\x,\y) circle(0.3);		}	}
	
	\foreach \x in {0,1.2,3,4.2,6,7.2} {
		\filldraw [fill=green,draw=black] (\x,0) circle(0.3);		}
	\foreach \x in {0,1.2,3,4.2,6,7.2} {\node at (\x,0) [below=10pt] {$h=0$};	}
	\node at (0,0) [left=10pt] {$k=0$};	
	
	\draw [thick] [->] (11,0.6)--(13.5,0.6) node [below] {$z_2^e$};
	\draw [thick] [->] (11.5,0.1)--(11.5,2.6) node [left] {$z_2^m$};
	\foreach \x in {11.5,12.7} {
		\foreach \y in {0.6,1.8} {
			\filldraw [fill=white,draw=black] (\x,\y) circle(0.3);		}	}
	\foreach \x in {11.5,12.7} {
		\filldraw [fill=green,draw=black] (\x,0.6) circle(0.3);		}
	\node at (11.5,0.6) [left=10pt] {$g=0$};
	\foreach \x in {11.5,12.7} {
		\node at (\x,0.6) [below=10pt] {$q=0$};		}
\end{tikzpicture}
	\caption{$\Gamma = \boldsymbol{1} \times \boldsymbol{1}$ for  $G=\frac{U{{\left( 1 \right)}_{X}}\times SU{{\left( 2 \right)}_{H}}\times SU{{\left( 3 \right)}_{C}}}{\Gamma_p}\times \frac{ U{{\left( 1 \right)}_{Y}} \times SU{{\left( 2 \right)}_{L}} }{\Gamma_m}$. Abelian lines generated by $(q,g)=\{(0,1),(1,0)\}$ and 
    $(h,k)=\{(0,1),(1,0)\}$ can be included.}
	\label{fig:LOex2case1}
\end{figure}

%Ex 2 case 2
\begin{figure}[htbp!]
	\centering
	\begin{tikzpicture}[scale=0.75]
		\draw[thick][->](-1,0.6)--(9,0.6)node[below]{$z_3^e$};
		\draw[thick][->](0.6,-1)--(0.6,9)node[left]{$z_3^m$};
		\foreach \x in {0,3,6}{
			\foreach \y in {0,3,6}{
				\draw [thick] [->] (\x-0.5,\y)--(\x+2,\y) node [below] {$x_2^e$};		}	}
		\foreach \y in {0,3,6}{
			\foreach \x in {0,3,6}{
				\draw [thick] [->] (\x,\y-0.5)--(\x,\y+2) node [left] {$x_2^m$};		}	}
		
		\foreach \x in {0,1.2,3,4.2,6,7.2} {
			\foreach \y in {0,1.2,3,4.2,6,7.2}{
				\filldraw [fill=white,draw=black] (\x,\y) circle(0.3);		}	}
		
		\foreach \x in {0,1.2,3,4.2,6,7.2} {
			\filldraw [fill=green,draw=black] (\x,0) circle(0.3);		}
		\foreach \x in {0,1.2,3,4.2,6,7.2} {\node at (\x,0) [below=10pt] {$h=0$};	}
		\node at (0,0) [left=10pt] {$k=0$};
		
		\draw [thick] [->] (11,0.6)--(13.5,0.6) node [below] {$z_2^e$};
		\draw [thick] [->] (11.5,0.1)--(11.5,2.6) node [left] {$z_2^m$};
		\foreach \x in {11.5,12.7} {
			\foreach \y in {0.6,1.8} {
				\filldraw [fill=green,draw=black] (\x,\y) circle(0.3);		}	}
		\foreach \x in {11.5,12.7} {
			\filldraw [fill=green,draw=black] (\x,0.6) circle(0.3);		}
		\node at (11.5,0.6) [left=10pt] {$g=0$};
		\node at (11.5,0.6) [below=10pt] {$q=0$};
		\node at (11.5,1.8) [left=10pt] {$g={1}/{2}\;$};
		\node at (12.7,0.6) [below=10pt] {$q=1$};
	\end{tikzpicture}
	\caption{$\Gamma = \boldsymbol{1} \times \boldsymbol{Z}_{2L}$ for  $G=\frac{U{{\left( 1 \right)}_{X}}\times SU{{\left( 2 \right)}_{H}}\times SU{{\left( 3 \right)}_{C}}}{\Gamma_p}\times \frac{ U{{\left( 1 \right)}_{Y}} \times SU{{\left( 2 \right)}_{L}} }{\Gamma_m}$. Abelian lines generated by $(q,g)=\{(0,1),(2,0)\}$ and 
    $(h,k)=\{(0,1),(1,0)\}$ can be included.}
	\label{fig:LOex2case2}
\end{figure}

%Ex 2 case 3
\begin{figure}[htbp!]
	\centering
	\begin{tikzpicture}[scale=0.75]
		\draw[thick][->](-1,0.6)--(9,0.6)node[below]{$z_3^e$};
		\draw[thick][->](0.6,-1)--(0.6,9)node[left]{$z_3^m$};
		\foreach \x in {0,3,6}{
			\foreach \y in {0,3,6}{
				\draw [thick] [->] (\x-0.5,\y)--(\x+2,\y) node [below] {$x_2^e$};		}	}
		\foreach \y in {0,3,6}{
			\foreach \x in {0,3,6}{
				\draw [thick] [->] (\x,\y-0.5)--(\x,\y+2) node [left] {$x_2^m$};		}	}
		
		\foreach \x in {0,1.2,3,4.2,6,7.2} {
			\foreach \y in {0,1.2,3,4.2,6,7.2}{
				\filldraw [fill=white,draw=black] (\x,\y) circle(0.3);		}	}
		
		\foreach \x in {0,1.2,3,4.2,6,7.2} {
			\foreach \y in {0,1.2}{
				\filldraw [fill=green,draw=black] (\x,\y) circle(0.3);		}   }
		\foreach \x in {0,3,6} {\node at (\x,0) [below=10pt] {$h=0$};	}
		\foreach \x in {1.2,4.2,7.2} {\node at (\x,0) [below=10pt] {$h=1$};	}
		\node at (0,0) [left=10pt] {$k=0$};
		\node at (0,1.2) [left=10pt] {$k={1}/{2}\;$};
		
		\draw [thick] [->] (11,0.6)--(13.5,0.6) node [below] {$z_2^e$};
		\draw [thick] [->] (11.5,0.1)--(11.5,2.6) node [left] {$z_2^m$};
		\foreach \x in {11.5,12.7} {
			\foreach \y in {0.6,1.8} {
				\filldraw [fill=white,draw=black] (\x,\y) circle(0.3);		}	}
		\foreach \x in {11.5,12.7} {
			\filldraw [fill=green,draw=black] (\x,0.6) circle(0.3);		}
		\node at (11.5,0.6) [left=10pt] {$g=0$};
		\foreach \x in {11.5,12.7} {
			\node at (\x,0.6) [below=10pt] {$q=0$};		}
	\end{tikzpicture}
	\caption{$\Gamma = \boldsymbol{Z}_{2H} \times \boldsymbol{1}$ for  $G=\frac{U{{\left( 1 \right)}_{X}}\times SU{{\left( 2 \right)}_{H}}\times SU{{\left( 3 \right)}_{C}}}{\Gamma_p}\times \frac{ U{{\left( 1 \right)}_{Y}} \times SU{{\left( 2 \right)}_{L}} }{\Gamma_m}$. Abelian lines generated by $(q,g)=\{(0,1),(1,0)\}$ and 
    $(h,k)=\{(0,1),(2,0)\}$ can be included.}
	\label{fig:LOex2case3}
\end{figure}

%Ex 2 case 4
\begin{figure}[htbp!]
	\centering
\begin{tikzpicture}[scale=0.75]
	\draw[thick][->](-1,0.6)--(9,0.6)node[below]{$z_3^e$};
	\draw[thick][->](0.6,-1)--(0.6,9)node[left]{$z_3^m$};
	\foreach \x in {0,3,6}{
		\foreach \y in {0,3,6}{
			\draw [thick] [->] (\x-0.5,\y)--(\x+2,\y) node [below] {$x_2^e$};		}	}
	\foreach \y in {0,3,6}{
		\foreach \x in {0,3,6}{
			\draw [thick] [->] (\x,\y-0.5)--(\x,\y+2) node [left] {$x_2^m$};		}	}
	
	\foreach \x in {0,1.2,3,4.2,6,7.2} {
		\foreach \y in {0,1.2,3,4.2,6,7.2}{
			\filldraw [fill=white,draw=black] (\x,\y) circle(0.3);		}	}
	
	\foreach \x in {0,1.2,3,4.2,6,7.2} {
		\foreach \y in {0,3,6} {
			\filldraw [fill=green,draw=black] (\x,\y) circle(0.3);		}	}
	\foreach \x in {0,1.2} {\node at (\x,0) [below=10pt] {$h=0$};	}
	\foreach \x in {3,4.2} {\node at (\x,0) [below=10pt] {$h=1$};	}
	\foreach \x in {6,7.2} {\node at (\x,0) [below=10pt] {$h=2$};	}
	\node at (0,0) [left=10pt] {$k=0$};
	\node at (0,3) [left=10pt] {$k={1}/{3}\;$};
	\node at (0,6) [left=10pt] {$k={2}/{3}\;$};
	
	\draw [thick] [->] (11,0.6)--(13.5,0.6) node [below] {$z_2^e$};
	\draw [thick] [->] (11.5,0.1)--(11.5,2.6) node [left] {$z_2^m$};
	\foreach \x in {11.5,12.7} {
		\foreach \y in {0.6,1.8} {
			\filldraw [fill=white,draw=black] (\x,\y) circle(0.3);		}	}
	\foreach \x in {11.5,12.7} {
		\filldraw [fill=green,draw=black] (\x,0.6) circle(0.3);		}
	\node at (11.5,0.6) [left=10pt] {$g=0$};
	\foreach \x in {11.5,12.7} {
		\node at (\x,0.6) [below=10pt] {$q=0$};		}
\end{tikzpicture}
	\caption{$\Gamma = \boldsymbol{Z}_{3} \times \boldsymbol{1}$ for  $G=\frac{U{{\left( 1 \right)}_{X}}\times SU{{\left( 2 \right)}_{H}}\times SU{{\left( 3 \right)}_{C}}}{\Gamma_p}\times \frac{ U{{\left( 1 \right)}_{Y}} \times SU{{\left( 2 \right)}_{L}} }{\Gamma_m}$. Abelian lines generated by $(q,g)=\{(0,1),(1,0)\}$ and 
    $(h,k)=\{(0,1),(3,0)\}$ can be included.}
	\label{fig:LOex2case4}
\end{figure}

%Ex 2 case 5
\begin{figure}[htbp!]
	\centering
\begin{tikzpicture}[scale=0.75]
	\draw[thick][->](-1,0.6)--(9,0.6)node[below]{$z_3^e$};
	\draw[thick][->](0.6,-1)--(0.6,9)node[left]{$z_3^m$};
	\foreach \x in {0,3,6}{
		\foreach \y in {0,3,6}{
			\draw [thick] [->] (\x-0.5,\y)--(\x+2,\y) node [below] {$x_2^e$};		}	}
	\foreach \y in {0,3,6}{
		\foreach \x in {0,3,6}{
			\draw [thick] [->] (\x,\y-0.5)--(\x,\y+2) node [left] {$x_2^m$};		}	}
	
	\foreach \x in {0,1.2,3,4.2,6,7.2} {
		\foreach \y in {0,1.2,3,4.2,6,7.2}{
			\filldraw [fill=white,draw=black] (\x,\y) circle(0.3);		}	}
	
	\foreach \x in {0,1.2,3,4.2,6,7.2} {
		\foreach \y in {0,1.2}{
			\filldraw [fill=green,draw=black] (\x,\y) circle(0.3);		}   }
	\foreach \x in {0,3,6} {\node at (\x,0) [below=10pt] {$h=0$};	}
	\foreach \x in {1.2,4.2,7.2} {\node at (\x,0) [below=10pt] {$h=1$};	}
	\node at (0,0) [left=10pt] {$k=0$};
	\node at (0,1.2) [left=10pt] {$k={1}/{2}\;$};
	
	\draw [thick] [->] (11,0.6)--(13.5,0.6) node [below] {$z_2^e$};
	\draw [thick] [->] (11.5,0.1)--(11.5,2.6) node [left] {$z_2^m$};
	\foreach \x in {11.5,12.7} {
		\foreach \y in {0.6,1.8} {
			\filldraw [fill=green,draw=black] (\x,\y) circle(0.3);		}	}
	\foreach \x in {11.5,12.7} {
		\filldraw [fill=green,draw=black] (\x,0.6) circle(0.3);		}
	\node at (11.5,0.6) [left=10pt] {$g=0$};
	\node at (11.5,0.6) [below=10pt] {$q=0$};
	\node at (11.5,1.8) [left=10pt] {$g={1}/{2}\;$};
	\node at (12.7,0.6) [below=10pt] {$q=1$};
\end{tikzpicture}
	\caption{$\Gamma = \boldsymbol{Z}_{2H} \times \boldsymbol{Z}_{2L}$ 
    for  $G=\frac{U{{\left( 1 \right)}_{X}}\times SU{{\left( 2 \right)}_{H}}\times SU{{\left( 3 \right)}_{C}}}{\Gamma_p}\times \frac{ U{{\left( 1 \right)}_{Y}} \times SU{{\left( 2 \right)}_{L}} }{\Gamma_m}$. Abelian lines generated by $(q,g)=\{(0,1),(2,0)\}$ and 
    $(h,k)=\{(0,1),(2,0)\}$ can be included.}
	\label{fig:LOex2case5}
\end{figure}

%Ex 2 case 6
\begin{figure}[htbp!]
	\centering
\begin{tikzpicture}[scale=0.75]
	\draw[thick][->](-1,0.6)--(9,0.6)node[below]{$z_3^e$};
	\draw[thick][->](0.6,-1)--(0.6,9)node[left]{$z_3^m$};
	\foreach \x in {0,3,6}{
		\foreach \y in {0,3,6}{
			\draw [thick] [->] (\x-0.5,\y)--(\x+2,\y) node [below] {$x_2^e$};		}	}
	\foreach \y in {0,3,6}{
		\foreach \x in {0,3,6}{
			\draw [thick] [->] (\x,\y-0.5)--(\x,\y+2) node [left] {$x_2^m$};		}	}
	
	\foreach \x in {0,1.2,3,4.2,6,7.2} {
		\foreach \y in {0,1.2,3,4.2,6,7.2}{
			\filldraw [fill=green,draw=black] (\x,\y) circle(0.3);		}	}
	
	\node at (0,0) [below=10pt] {$h=0$};
	\node at (1.2,0) [below=10pt] {$h=3$};
	\node at (3,0) [below=10pt] {$h=4$};
	\node at (4.2,0) [below=10pt] {$h=1$};
	\node at (6,0) [below=10pt] {$h=2$};
	\node at (7.2,0) [below=10pt] {$h=5$};
	\node at (0,0) [left=10pt] {$k=0$};
	\node at (0,1.2) [left=10pt] {$k={1}/{2}\;$};	
	\node at (0,3) [left=10pt] {$k={1}/{3}\;$};	
	\node at (0,4.2) [left=10pt] {$k={5}/{6}\;$};	
	\node at (0,6) [left=10pt] {$k={2}/{3}\;$};	
	\node at (0,7.2) [left=10pt] {$k={1}/{6}\;$};	
	
	\draw [thick] [->] (11,0.6)--(13.5,0.6) node [below] {$z_2^e$};
	\draw [thick] [->] (11.5,0.1)--(11.5,2.6) node [left] {$z_2^m$};
	\foreach \x in {11.5,12.7} {
		\foreach \y in {0.6,1.8} {
			\filldraw [fill=white,draw=black] (\x,\y) circle(0.3);		}	}
	\foreach \x in {11.5,12.7} {
		\filldraw [fill=green,draw=black] (\x,0.6) circle(0.3);		}
	\node at (11.5,0.6) [left=10pt] {$g=0$};
	\foreach \x in {11.5,12.7} {
		\node at (\x,0.6) [below=10pt] {$q=0$};		}
\end{tikzpicture}
	\caption{$\Gamma = \boldsymbol{Z}_{6H} \times \boldsymbol{1}$ for  $G=\frac{U{{\left( 1 \right)}_{X}}\times SU{{\left( 2 \right)}_{H}}\times SU{{\left( 3 \right)}_{C}}}{\Gamma_p}\times \frac{ U{{\left( 1 \right)}_{Y}} \times SU{{\left( 2 \right)}_{L}} }{\Gamma_m}$. Abelian lines generated by $(q,g)=\{(0,1),(1,0)\}$ and 
    $(h,k)=\{(0,1),(6,0)\}$ can be included.}
	\label{fig:LOex2case6}
\end{figure}

%Ex 2 case 7
\begin{figure}[htbp!]
	\centering
\begin{tikzpicture}[scale=0.75]
	\draw[thick][->](-1,0.6)--(9,0.6)node[below]{$z_3^e$};
	\draw[thick][->](0.6,-1)--(0.6,9)node[left]{$z_3^m$};
	\foreach \x in {0,3,6}{
		\foreach \y in {0,3,6}{
			\draw [thick] [->] (\x-0.5,\y)--(\x+2,\y) node [below] {$x_2^e$};		}	}
	\foreach \y in {0,3,6}{
		\foreach \x in {0,3,6}{
			\draw [thick] [->] (\x,\y-0.5)--(\x,\y+2) node [left] {$x_2^m$};		}	}
	
	\foreach \x in {0,1.2,3,4.2,6,7.2} {
		\foreach \y in {0,1.2,3,4.2,6,7.2}{
			\filldraw [fill=white,draw=black] (\x,\y) circle(0.3);		}	}
	
	\foreach \x in {0,1.2,3,4.2,6,7.2} {
		\foreach \y in {0,3,6} {
			\filldraw [fill=green,draw=black] (\x,\y) circle(0.3);		}	}
	\foreach \x in {0,1.2} {\node at (\x,0) [below=10pt] {$h=0$};	}
	\foreach \x in {3,4.2} {\node at (\x,0) [below=10pt] {$h=1$};	}
	\foreach \x in {6,7.2} {\node at (\x,0) [below=10pt] {$h=2$};	}
	\node at (0,0) [left=10pt] {$k=0$};
	\node at (0,3) [left=10pt] {$k={1}/{3}\;$};
	\node at (0,6) [left=10pt] {$k={2}/{3}\;$};
	
	\draw [thick] [->] (11,0.6)--(13.5,0.6) node [below] {$z_2^e$};
	\draw [thick] [->] (11.5,0.1)--(11.5,2.6) node [left] {$z_2^m$};
	\foreach \x in {11.5,12.7} {
		\foreach \y in {0.6,1.8} {
			\filldraw [fill=green,draw=black] (\x,\y) circle(0.3);		}	}
	\foreach \x in {11.5,12.7} {
		\filldraw [fill=green,draw=black] (\x,0.6) circle(0.3);		}
	\node at (11.5,0.6) [left=10pt] {$g=0$};
	\node at (11.5,0.6) [below=10pt] {$q=0$};
	\node at (11.5,1.8) [left=10pt] {$g={1}/{2}\;$};
	\node at (12.7,0.6) [below=10pt] {$q=1$};
\end{tikzpicture}
	\caption{$\Gamma = \boldsymbol{Z}_{3} \times \boldsymbol{Z}_{2L}$ for  $G=\frac{U{{\left( 1 \right)}_{X}}\times SU{{\left( 2 \right)}_{H}}\times SU{{\left( 3 \right)}_{C}}}{\Gamma_p}\times \frac{ U{{\left( 1 \right)}_{Y}} \times SU{{\left( 2 \right)}_{L}} }{\Gamma_m}$. Abelian lines generated by $(q,g)=\{(0,1),(2,0)\}$ and 
    $(h,k)=\{(0,1),(3,0)\}$ can be included.}
	\label{fig:LOex2case7}
\end{figure}

%Ex 2 case 8
\begin{figure}[htbp!]
	\centering
\begin{tikzpicture}[scale=0.75]
	\draw[thick][->](-1,0.6)--(9,0.6)node[below]{$z_3^e$};
	\draw[thick][->](0.6,-1)--(0.6,9)node[left]{$z_3^m$};
	\foreach \x in {0,3,6}{
		\foreach \y in {0,3,6}{
			\draw [thick] [->] (\x-0.5,\y)--(\x+2,\y) node [below] {$x_2^e$};		}	}
	\foreach \y in {0,3,6}{
		\foreach \x in {0,3,6}{
			\draw [thick] [->] (\x,\y-0.5)--(\x,\y+2) node [left] {$x_2^m$};		}	}
	
	\foreach \x in {0,1.2,3,4.2,6,7.2} {
		\foreach \y in {0,1.2,3,4.2,6,7.2}{
			\filldraw [fill=green,draw=black] (\x,\y) circle(0.3);		}	}
	
	\node at (0,0) [below=10pt] {$h=0$};
	\node at (1.2,0) [below=10pt] {$h=3$};
	\node at (3,0) [below=10pt] {$h=4$};
	\node at (4.2,0) [below=10pt] {$h=1$};
	\node at (6,0) [below=10pt] {$h=2$};
	\node at (7.2,0) [below=10pt] {$h=5$};
	\node at (0,0) [left=10pt] {$k=0$};
	\node at (0,1.2) [left=10pt] {$k={1}/{2}\;$};	
	\node at (0,3) [left=10pt] {$k={1}/{3}\;$};	
	\node at (0,4.2) [left=10pt] {$k={5}/{6}\;$};	
	\node at (0,6) [left=10pt] {$k={2}/{3}\;$};	
	\node at (0,7.2) [left=10pt] {$k={1}/{6}\;$};	
	
	\draw [thick] [->] (11,0.6)--(13.5,0.6) node [below] {$z_2^e$};
	\draw [thick] [->] (11.5,0.1)--(11.5,2.6) node [left] {$z_2^m$};
	\foreach \x in {11.5,12.7} {
		\foreach \y in {0.6,1.8} {
			\filldraw [fill=green,draw=black] (\x,\y) circle(0.3);		}	}
	\foreach \x in {11.5,12.7} {
		\filldraw [fill=green,draw=black] (\x,0.6) circle(0.3);		}
	\node at (11.5,0.6) [left=10pt] {$g=0$};
	\node at (11.5,0.6) [below=10pt] {$q=0$};
	\node at (11.5,1.8) [left=10pt] {$g={1}/{2}\;$};
	\node at (12.7,0.6) [below=10pt] {$q=1$};
\end{tikzpicture}
	\caption{$\Gamma = \boldsymbol{Z}_{6H} \times \boldsymbol{Z}_{2L}$ for $G=\frac{U{{\left( 1 \right)}_{X}}\times SU{{\left( 2 \right)}_{H}}\times SU{{\left( 3 \right)}_{C}}}{\Gamma_p}\times \frac{ U{{\left( 1 \right)}_{Y}} \times SU{{\left( 2 \right)}_{L}} }{\Gamma_m}$. Abelian lines generated by $(q,g)=\{(0,1),(2,0)\}$ and 
    $(h,k)=\{(0,1),(6,0)\}$ can be included.}
	\label{fig:LOex2case8}
\end{figure}

As our third example, consider a significantly different quotient,
\be
G=\frac{U{{\left( 1 \right)}_{V}}\times SU{{\left( 3 \right)}_{C}}}{\Gamma_p}\times \frac{ U{{\left( 1 \right)}_{A}} \times SU{{\left( 2 \right)}_{H}} }{\Gamma_m}\times \frac{SU{{\left( 2 \right)}_{L}}}{\Gamma_n} \; ,
\label{eq:ex3}
\ee
which is one of the 8 possibilities of case (B),
in which the original two $U(1)$ groups have been combined into $U(1)_V$ and $U(1)_A$. The SM hypercharge $\widetilde Y$ and dark hypercharge $\widetilde X$ now mix together. We assume the ``vector'' hypercharge $U(1)_V$ to be $q_+=q+h$ and the other ``axial'' hypercharge $U(1)_A$ as $q_-=q-h$. 

As before, we need to identify its centers and subgroups, and then determine their generators. The possible $\Gamma$ for quotienting in Eq.~(\ref{eq:ex3}) is 
\be
\label{eq:ex3centers}
\begin{aligned}
\Gamma = &\{ 
{\boldsymbol{1}} \times {\boldsymbol{1}} \times {\boldsymbol{1}},
{\boldsymbol{1}} \times {\boldsymbol{1}} \times {\boldsymbol{Z}_{2L}},
{\boldsymbol{1}} \times {\boldsymbol{Z}_{2H}} \times {\boldsymbol{1}}, {\boldsymbol{Z}_{3}} \times {\boldsymbol{1}} \times {\boldsymbol{1}}, \\
& \;\, {\boldsymbol{1}} \times {\boldsymbol{Z}_{2H}} \times{\boldsymbol{Z}_{2L}}, 
{\boldsymbol{Z}_{3}} \times \boldsymbol{1} \times {\boldsymbol{Z}_{2L}},
{\boldsymbol{Z}_{3}} \times {\boldsymbol{Z}_{2H}} \times \boldsymbol{1} ,
{\boldsymbol{Z}_{3}} \times {\boldsymbol{Z}_{2H}}\times {\boldsymbol{Z}_{2L}}
\} \; .
\end{aligned}
\ee
The generators for the Wilson lines are given by
\be
\begin{aligned}
	&\xi ={{e}^{\frac{2}{3}\pi i\left( q_+ \right)}}\otimes \omega ={{e}^{\frac{2}{3}\pi i\left( q_+ \right)}}{{e}^{\frac{2}{3}\pi iz_{3}^{e}}}  \;, \\
	&\chi ={{e}^{\pi i\left( q_- \right)}}\otimes \rho ={{e}^{\pi i\left( q_- \right)}}{{e}^{\pi ix_{2}^{e}}} \; , \\
	&\eta ={{e}^{\pi iz_{2}^{e}}} \; .
\end{aligned}
\ee
Since the $U(1)$ charges have changed, it is necessary to modify the GDQC as
\begin{equation}\label{eq:DiracConditionAlt0}
	{{e}^{-\frac{2i\pi q_+}{3}3g_+}}{{e}^{i\pi z_{2}^{e}z_{2}^{m}}}{{e}^{\frac{2}{3}i\pi z_{3}^{e}z_{3}^{m}}}{{e}^{i\pi x_{2}^{e}x_{2}^{m}}}{{e}^{-\frac{2i\pi q_-}{2}2g_-}}=1 \; ,
\end{equation}
with $g_\pm$ being the ``magnetic'' counterpart of $q_\pm$.
Or, equivalently,
\begin{equation}\label{eq:DiracConditionAlt}
-6g_+q_++3z_{2}^{e}z_{2}^{m}+2z_{3}^{e}z_{3}^{m}+3x_{2}^{e}x_{2}^{m}-6g_-q_-= 0 \pmod{6} \; .
\end{equation}

With this modified GDQC, we can repeat the same exercises as in the two previous examples to obtain the spectra of the line operators. The results are shown in Figs.~(\ref{fig:LOex3case1}), (\ref{fig:LOex3case2}), 
(\ref{fig:LOex3case3}), (\ref{fig:LOex3case4}), (\ref{fig:LOex3case5}), (\ref{fig:LOex3case6}), (\ref{fig:LOex3case7}) and (\ref{fig:LOex3case8}) for each element of $\Gamma$ in Eq.~(\ref{eq:ex3centers}), respectively.
Details will be omitted here. 
We just note that the spectra  change dramatically compared to the previous two examples.
% 

%Ex 3 case 1
\begin{figure}[htbp!]
	\centering
\begin{tikzpicture}
	\draw[thick][->](-0.5,0)--(3,0)node[below]{$z_3^e$};
	\draw[thick][->](0,-0.5)--(0,3)node[left]{$z_3^m$};
	\draw[thick][->](4.5,0)--(7,0)node[below]{$x_2^e$};
	\draw[thick][->](5,-0.5)--(5,2)node[left]{$x_2^m$};
	\draw[thick][->](8.5,0)--(11,0)node[below]{$z_2^m$};
	\draw[thick][->](9,-0.5)--(9,2)node[left]{$z_2^m$};	
	\foreach \x in {0,1.2,2.4}{
		\foreach \y in {0,1.2,2.4}{
			\filldraw[fill=white,draw=black] (\x,\y) circle(0.3);   }	}
	\foreach \x in {5,6.2,9,10.2}{
		\foreach \y in {0,1.2}{
			\filldraw[fill=white,draw=black] (\x,\y) circle(0.3);   }	}	
	\foreach \x in {0,1.2,2.4,5,6.2,9,10.2}{
		\filldraw[fill=green,draw=black] (\x,0) circle(0.3);   }
	\foreach \x in {5,6.2}{
		\node at (\x,0) [below=10pt] {$q_{-}=0$};   }
	\node at (5,0) [left=12pt] {$g_{-}=0$};	
	\foreach \x in {0,1.2,2.4}{
		\node at (\x,0) [below=10pt] {$q_{+}=0$};   }
	\node at (0,0) [left=12pt] {$g_{+}=0$};
\end{tikzpicture}
	\caption{$\Gamma = \boldsymbol{1} \times \boldsymbol{1} \times \boldsymbol{1}$ for $G=\frac{U{{\left( 1 \right)}_{V}}\times SU{{\left( 3 \right)}_{C}}}{\Gamma_p}\times \frac{ U{{\left( 1 \right)}_{A}} \times SU{{\left( 2 \right)}_{H}} }{\Gamma_m}\times \frac{SU{{\left( 2 \right)}_{L}}}{\Gamma_n}.$ Abelian lines generated by $(q_+,g_+)=\{(0,1),(1,0)\}$ and 
    $(q_-,g_-)=\{(0,1),(1,0)\}$ can be included.
}\label{fig:LOex3case1}
\end{figure}

%Ex 3 case 2
\begin{figure}[htbp!]
	\centering
	\begin{tikzpicture}
		\draw[thick][->](-0.5,0)--(3,0)node[below]{$z_3^e$};
		\draw[thick][->](0,-0.5)--(0,3)node[left]{$z_3^m$};
		\draw[thick][->](4.5,0)--(7,0)node[below]{$x_2^e$};
		\draw[thick][->](5,-0.5)--(5,2)node[left]{$x_2^m$};
		\draw[thick][->](8.5,0)--(11,0)node[below]{$z_2^m$};
		\draw[thick][->](9,-0.5)--(9,2)node[left]{$z_2^m$};
		
		\foreach \x in {0,1.2,2.4}{
			\foreach \y in {0,1.2,2.4}{
				\filldraw[fill=white,draw=black] (\x,\y) circle(0.3);   }	}
		\foreach \x in {5,6.2,9,10.2}{
			\foreach \y in {0,1.2}{
				\filldraw[fill=white,draw=black] (\x,\y) circle(0.3);   }	}
		
		\foreach \x in {0,1.2,2.4,5,6.2,9}{
			\filldraw[fill=green,draw=black] (\x,0) circle(0.3);   }
		
		\filldraw[fill=green,draw=black] (9,1.2) circle(0.3);
		
		\foreach \x in {5,6.2}{
			\node at (\x,0) [below=10pt] {$q_{-}=0$};   }
		\node at (5,0) [left=12pt] {$g_{-}=0$};

		\foreach \x in {0,1.2,2.4}{
			\node at (\x,0) [below=10pt] {$q_{+}=0$};   }
		\node at (0,0) [left=12pt] {$g_{+}=0$};

	\end{tikzpicture}
	\caption{$\Gamma = \boldsymbol{1} \times \boldsymbol{1} \times \boldsymbol{Z}_{2L}$ for  $G=\frac{U{{\left( 1 \right)}_{V}}\times SU{{\left( 3 \right)}_{C}}}{\Gamma_p}\times \frac{ U{{\left( 1 \right)}_{A}} \times SU{{\left( 2 \right)}_{H}} }{\Gamma_m}\times \frac{SU{{\left( 2 \right)}_{L}}}{\Gamma_n}.$  Abelian lines generated by $(q_+,g_+)=\{(0,1),(1,0)\}$ and 
    $(q_-,g_-)=\{(0,1),(1,0)\}$ can be included.
}
	\label{fig:LOex3case2}
\end{figure}

%Ex 3 case 3
\begin{figure}[htbp!]
	\centering
	\begin{tikzpicture}
	\draw[thick][->](-0.5,0)--(3,0)node[below]{$z_3^e$};
	\draw[thick][->](0,-0.5)--(0,3)node[left]{$z_3^m$};
	\draw[thick][->](4.5,0)--(7,0)node[below]{$x_2^e$};
	\draw[thick][->](5,-0.5)--(5,2)node[left]{$x_2^m$};
	\draw[thick][->](8.5,0)--(11,0)node[below]{$z_2^m$};
	\draw[thick][->](9,-0.5)--(9,2)node[left]{$z_2^m$};
	
	\foreach \x in {0,1.2,2.4}{
		\foreach \y in {0,1.2,2.4}{
			\filldraw[fill=white,draw=black] (\x,\y) circle(0.3);   }	}
	\foreach \x in {5,6.2,9,10.2}{
		\foreach \y in {0,1.2}{
			\filldraw[fill=white,draw=black] (\x,\y) circle(0.3);   }	}
	
	\foreach \x in {0,1.2,2.4,5,6.2,9,10.2}{
		\filldraw[fill=green,draw=black] (\x,0) circle(0.3);   }
	
	\foreach \x in {5,6.2}{
		\filldraw[fill=green,draw=black] (\x,1.2) circle(0.3);   }
	
	\node at (5,0) [below=10pt] {$q_{-}=0$};
	\node at (6.2,0) [below=10pt] {$q_{-}=1$};
	\node at (5,0) [left=12pt] {$g_{-}=0$};
	\node at (5,1.2) [left=10pt] {$g_{-}={1}/{2}\;$};
	
	\foreach \x in {0,1.2,2.4}{
		\node at (\x,0) [below=10pt] {$q_{+}=0$};   }
	\node at (0,0) [left=12pt] {$g_{+}=0$};

	\end{tikzpicture}
	\caption{$\Gamma = \boldsymbol{1} \times \boldsymbol{Z}_{2H} \times \boldsymbol{1} $ for  $G=\frac{U{{\left( 1 \right)}_{V}}\times SU{{\left( 3 \right)}_{C}}}{\Gamma_p}\times \frac{ U{{\left( 1 \right)}_{A}} \times SU{{\left( 2 \right)}_{H}} }{\Gamma_m}\times \frac{SU{{\left( 2 \right)}_{L}}}{\Gamma_n}.$  Abelian lines generated by $(q_+,g_+)=\{(0,1),(1,0)\}$ and 
    $(q_-,g_-)=\{(0,1),(2,0)\}$ can be included.}
	\label{fig:LOex3case3}
\end{figure}

%Ex 3 case 4
\begin{figure}[htbp!]
	\centering
	\begin{tikzpicture}
		\draw[thick][->](-0.5,0)--(3,0)node[below]{$z_3^e$};
		\draw[thick][->](0,-0.5)--(0,3)node[left]{$z_3^m$};
		\draw[thick][->](4.5,0)--(7,0)node[below]{$x_2^e$};
		\draw[thick][->](5,-0.5)--(5,2)node[left]{$x_2^m$};
		\draw[thick][->](8.5,0)--(11,0)node[below]{$z_2^m$};
		\draw[thick][->](9,-0.5)--(9,2)node[left]{$z_2^m$};
		
		\foreach \x in {0,1.2,2.4}{
			\foreach \y in {0,1.2,2.4}{
				\filldraw[fill=white,draw=black] (\x,\y) circle(0.3);   }	}
		\foreach \x in {5,6.2,9,10.2}{
			\foreach \y in {0,1.2}{
				\filldraw[fill=white,draw=black] (\x,\y) circle(0.3);   }	}
		
		\foreach \x in {0,1.2,2.4,5,6.2,9,10.2}{
			\filldraw[fill=green,draw=black] (\x,0) circle(0.3);   }		
		\foreach \y in {0,1.2,2.4}{
			\foreach \x in {0,1.2,2.4}{
				\filldraw[fill=green,draw=black] (\x,\y) circle(0.3);   }   }
		
		\foreach \x in {5,6.2}{
			\node at (\x,0) [below=10pt] {$q_{-}=0$};   }
		\node at (5,0) [left=12pt] {$g_{-}=0$};
			
		\node at (0,0) [below=10pt] {$q_{+}=0$};
		\node at (1.2,0) [below=10pt] {$q_{+}=1$};
		\node at (2.3,0) [below=10pt] {$q_{+}=2$};
		\node at (0,0) [left=12pt] {$g_{+}=0$};
		\node at (0,1.2) [left=10pt] {$g_{+}={2}/{3}\;$};
		\node at (0,2.4) [left=10pt] {$g_{+}={1}/{3}\;$};

	\end{tikzpicture}
	\caption{$\Gamma = \boldsymbol{Z}_{3} \times \boldsymbol{1} \times \boldsymbol{1}$ for  $G=\frac{U{{\left( 1 \right)}_{V}}\times SU{{\left( 3 \right)}_{C}}}{\Gamma_p}\times \frac{ U{{\left( 1 \right)}_{A}} \times SU{{\left( 2 \right)}_{H}} }{\Gamma_m}\times \frac{SU{{\left( 2 \right)}_{L}}}{\Gamma_n}.$  Abelian lines generated by $(q_+,g_+)=\{(0,1),(3,0)\}$ and 
    $(q_-,g_-)=\{(0,1),(1,0)\}$ can be included.}
	\label{fig:LOex3case4}
\end{figure}

%Ex 3 case 5
\begin{figure}[htbp!]
	\centering
\begin{tikzpicture}
	\draw[thick][->](-0.5,0)--(3,0)node[below]{$z_3^e$};
	\draw[thick][->](0,-0.5)--(0,3)node[left]{$z_3^m$};
	\draw[thick][->](4.5,0)--(7,0)node[below]{$x_2^e$};
	\draw[thick][->](5,-0.5)--(5,2)node[left]{$x_2^m$};
	\draw[thick][->](8.5,0)--(11,0)node[below]{$z_2^m$};
	\draw[thick][->](9,-0.5)--(9,2)node[left]{$z_2^m$};	
	\foreach \x in {0,1.2,2.4}{
		\foreach \y in {0,1.2,2.4}{
			\filldraw[fill=white,draw=black] (\x,\y) circle(0.3);   }	}
	\foreach \x in {5,6.2,9,10.2}{
		\foreach \y in {0,1.2}{
			\filldraw[fill=white,draw=black] (\x,\y) circle(0.3);   }	}	
	\foreach \x in {0,1.2,2.4,5,6.2,9}{
		\filldraw[fill=green,draw=black] (\x,0) circle(0.3);   }	
	\filldraw[fill=green,draw=black] (9,1.2) circle(0.3);	
	\foreach \x in {5,6.2}{
		\filldraw[fill=green,draw=black] (\x,1.2) circle(0.3);   }	
	\node at (5,0) [below=10pt] {$q_{-}=0$};
	\node at (6.2,0) [below=10pt] {$q_{-}=1$};
	\node at (5,0) [left=12pt] {$g_{-}=0$};
	\node at (5,1.2) [left=10pt] {$g_{-}={1}/{2}\;$};	
	\foreach \x in {0,1.2,2.4}{
		\node at (\x,0) [below=10pt] {$q_{+}=0$};   }
	\node at (0,0) [left=12pt] {$g_{+}=0$};		
	\foreach \x in {0,1.2,2.4}{
		\node at (\x,0) [below=10pt] {$q_{+}=0$};   }
	\node at (0,0) [left=12pt] {$g_{+}=0$};
\end{tikzpicture}
	\caption{$\Gamma = \boldsymbol{1} \times \boldsymbol{Z}_{2H} \times \boldsymbol{Z}_{2L}$ for  $G=\frac{U{{\left( 1 \right)}_{V}}\times SU{{\left( 3 \right)}_{C}}}{\Gamma_p}\times \frac{ U{{\left( 1 \right)}_{A}} \times SU{{\left( 2 \right)}_{H}} }{\Gamma_m}\times \frac{SU{{\left( 2 \right)}_{L}}}{\Gamma_n}.$  Abelian lines generated by $(q_+,g_+)=\{(0,1),(1,0)\}$ and 
    $(q_-,g_-)=\{(0,1),(2,0)\}$ can be included.}
	\label{fig:LOex3case5}
\end{figure}

%Ex 3 case 6
\begin{figure}[htbp!]
	\centering
\begin{tikzpicture}
	\draw[thick][->](-0.5,0)--(3,0)node[below]{$z_3^e$};
	\draw[thick][->](0,-0.5)--(0,3)node[left]{$z_3^m$};
	\draw[thick][->](4.5,0)--(7,0)node[below]{$x_2^e$};
	\draw[thick][->](5,-0.5)--(5,2)node[left]{$x_2^m$};
	\draw[thick][->](8.5,0)--(11,0)node[below]{$z_2^m$};
	\draw[thick][->](9,-0.5)--(9,2)node[left]{$z_2^m$};	
	\foreach \x in {0,1.2,2.4}{
		\foreach \y in {0,1.2,2.4}{
			\filldraw[fill=white,draw=black] (\x,\y) circle(0.3);   }	}
	\foreach \x in {5,6.2,9,10.2}{
		\foreach \y in {0,1.2}{
			\filldraw[fill=white,draw=black] (\x,\y) circle(0.3);   }	}	
	\foreach \x in {0,1.2,2.4,5,6.2,9}{
		\filldraw[fill=green,draw=black] (\x,0) circle(0.3);   }	
	\filldraw[fill=green,draw=black] (9,1.2) circle(0.3);	
	\foreach \y in {0,1.2,2.4}{
		\foreach \x in {0,1.2,2.4}{
			\filldraw[fill=green,draw=black] (\x,\y) circle(0.3);   }   }	
	\foreach \x in {5,6.2}{
		\node at (\x,0) [below=10pt] {$q_{-}=0$};   }
	\node at (5,0) [left=12pt] {$g_{-}=0$};	
	\node at (0,0) [below=10pt] {$q_{+}=0$};
	\node at (1.2,0) [below=10pt] {$q_{+}=1$};
	\node at (2.3,0) [below=10pt] {$q_{+}=2$};
	\node at (0,0) [left=12pt] {$g_{+}=0$};
	\node at (0,1.2) [left=10pt] {$g_{+}={2}/{3}\;$};
	\node at (0,2.4) [left=10pt] {$g_{+}={1}/{3}\;$};
\end{tikzpicture}
	\caption{$\Gamma={\boldsymbol{Z}_{3}} \times \boldsymbol{1} \times {\boldsymbol{Z}_{2L}}$ for  $G=\frac{U{{\left( 1 \right)}_{V}}\times SU{{\left( 3 \right)}_{C}}}{\Gamma_p}\times \frac{ U{{\left( 1 \right)}_{A}} \times SU{{\left( 2 \right)}_{H}} }{\Gamma_m}\times \frac{SU{{\left( 2 \right)}_{L}}}{\Gamma_n}.$  Abelian lines generated by $(q_+,g_+)=\{(0,1),(3,0)\}$ and 
    $(q_-,g_-)=\{(0,1),(1,0)\}$ can be included.}
	\label{fig:LOex3case6}
\end{figure}

%Ex 3 case 7
\begin{figure}[htbp!]
	\centering
\begin{tikzpicture}
	\draw[thick][->](-0.5,0)--(3,0)node[below]{$z_3^e$};
	\draw[thick][->](0,-0.5)--(0,3)node[left]{$z_3^m$};
	\draw[thick][->](4.5,0)--(7,0)node[below]{$x_2^e$};
	\draw[thick][->](5,-0.5)--(5,2)node[left]{$x_2^m$};
	\draw[thick][->](8.5,0)--(11,0)node[below]{$z_2^m$};
	\draw[thick][->](9,-0.5)--(9,2)node[left]{$z_2^m$};	
	\foreach \x in {0,1.2,2.4}{
		\foreach \y in {0,1.2,2.4}{
			\filldraw[fill=white,draw=black] (\x,\y) circle(0.3);   }	}
	\foreach \x in {5,6.2,9,10.2}{
		\foreach \y in {0,1.2}{
			\filldraw[fill=white,draw=black] (\x,\y) circle(0.3);   }	}	
	\foreach \x in {0,1.2,2.4,5,6.2,9,10.2}{
		\filldraw[fill=green,draw=black] (\x,0) circle(0.3);   }	
	\foreach \y in {0,1.2,2.4}{
		\foreach \x in {0,1.2,2.4}{
			\filldraw[fill=green,draw=black] (\x,\y) circle(0.3);   }   }	
	\foreach \x in {5,6.2}{
		\filldraw[fill=green,draw=black] (\x,1.2) circle(0.3);   }	
	\node at (5,0) [below=10pt] {$q_{-}=0$};
	\node at (6.2,0) [below=10pt] {$q_{-}=1$};
	\node at (5,0) [left=12pt] {$g_{-}=0$};
	\node at (5,1.2) [left=10pt] {$g_{-}={1}/{2}\;$};		
	\node at (0,0) [below=10pt] {$q_{+}=0$};
	\node at (1.2,0) [below=10pt] {$q_{+}=1$};
	\node at (2.3,0) [below=10pt] {$q_{+}=2$};
	\node at (0,0) [left=12pt] {$g_{+}=0$};
	\node at (0,1.2) [left=10pt] {$g_{+}={2}/{3}\;$};
	\node at (0,2.4) [left=10pt] {$g_{+}={1}/{3}\;$};
\end{tikzpicture}
	\caption{$\Gamma={\boldsymbol{Z}_{3}}  \times {\boldsymbol{Z}_{2H}} \times \boldsymbol{1}$ for  $G=\frac{U{{\left( 1 \right)}_{V}}\times SU{{\left( 3 \right)}_{C}}}{\Gamma_p}\times \frac{ U{{\left( 1 \right)}_{A}} \times SU{{\left( 2 \right)}_{H}} }{\Gamma_m}\times \frac{SU{{\left( 2 \right)}_{L}}}{\Gamma_n}.$  Abelian lines generated by $(q_+,g_+)=\{(0,1),(3,0)\}$ and 
    $(q_-,g_-)=\{(0,1),(2,0)\}$ can be included.}
	\label{fig:LOex3case7}
\end{figure}

%Ex 3 case 8
\begin{figure}[htbp!]
	\centering
\begin{tikzpicture}
	\draw[thick][->](-0.5,0)--(3,0)node[below]{$z_3^e$};
	\draw[thick][->](0,-0.5)--(0,3)node[left]{$z_3^m$};
	\draw[thick][->](4.5,0)--(7,0)node[below]{$x_2^e$};
	\draw[thick][->](5,-0.5)--(5,2)node[left]{$x_2^m$};
	\draw[thick][->](8.5,0)--(11,0)node[below]{$z_2^m$};
	\draw[thick][->](9,-0.5)--(9,2)node[left]{$z_2^m$};	
	\foreach \x in {0,1.2,2.4}{
		\foreach \y in {0,1.2,2.4}{
			\filldraw[fill=white,draw=black] (\x,\y) circle(0.3);   }	}
	\foreach \x in {5,6.2,9,10.2}{
		\foreach \y in {0,1.2}{
			\filldraw[fill=white,draw=black] (\x,\y) circle(0.3);   }	}	
	\foreach \x in {0,1.2,2.4,5,6.2,9}{
		\filldraw[fill=green,draw=black] (\x,0) circle(0.3);   }	
	\filldraw[fill=green,draw=black] (9,1.2) circle(0.3);	
	\foreach \y in {0,1.2,2.4}{
		\foreach \x in {0,1.2,2.4}{
			\filldraw[fill=green,draw=black] (\x,\y) circle(0.3);   }   }	
	\foreach \x in {5,6.2}{
		\filldraw[fill=green,draw=black] (\x,1.2) circle(0.3);   }	
	\node at (5,0) [below=10pt] {$q_{-}=0$};
	\node at (6.2,0) [below=10pt] {$q_{-}=1$};
	\node at (5,0) [left=12pt] {$g_{-}=0$};
	\node at (5,1.2) [left=10pt] {$g_{-}={1}/{2}\;$};		
	\node at (0,0) [below=10pt] {$q_{+}=0$};
	\node at (1.2,0) [below=10pt] {$q_{+}=1$};
	\node at (2.3,0) [below=10pt] {$q_{+}=2$};
	\node at (0,0) [left=12pt] {$g_{+}=0$};
	\node at (0,1.2) [left=10pt] {$g_{+}={2}/{3}\;$};
	\node at (0,2.4) [left=10pt] {$g_{+}={1}/{3}\;$};
\end{tikzpicture}
	\caption{$\Gamma={\boldsymbol{Z}_{3}}  \times {\boldsymbol{Z}_{2H}} \times \boldsymbol{Z}_{2L}$ for  $G=\frac{U{{\left( 1 \right)}_{V}}\times SU{{\left( 3 \right)}_{C}}}{\Gamma_p}\times \frac{ U{{\left( 1 \right)}_{A}} \times SU{{\left( 2 \right)}_{H}} }{\Gamma_m}\times \frac{SU{{\left( 2 \right)}_{L}}}{\Gamma_n}.$  Abelian lines generated by $(q_+,g_+)=\{(0,1),(3,0)\}$ and 
    $(q_-,g_-)=\{(0,1),(2,0)\}$ can be included.}
	\label{fig:LOex3case8}
\end{figure}

There are still 17 types of $G$ with different central quotient divisions that have not been discussed. However, we have already clarified the basic method and the remaining cases are quite similar to those above, so there is no need to present the extensive analysis here. For example, consider $G=\frac{U{{\left( 1 \right)}_{Y}}\times SU{{\left( 2 \right)}_{H}}}{\Gamma_p}\times \frac{SU{{\left( 2 \right)}_{L}}\times U{{\left( 1 \right)}_{X}}}{\Gamma_m}\times \frac{SU{{\left( 3 \right)}_{C}}}{\Gamma_n}$, which is one of the 4 possibilities of case (C).  Suppose $\Gamma={\boldsymbol{Z}_{2H}}\times {\boldsymbol{Z}_{2L}}\times {\boldsymbol{Z}_{3}}$. The Wilson lines are $(q=x_2^e\bmod2;x_2^e=0,1)$, $(h=z_2^e\bmod2;z_2^e=0,1)$ and $(z_3^e=0)$. The corresponding 't Hooft lines are $(2g=x_2^m\bmod2;x_2^m=0,1\bmod2)$, $(2k=z_2^m\bmod2;z_2^m=0,1\bmod2)$ and $(z_3^m=0,1,2\bmod3)$, respectively. 
Taking $G=\frac{U{{\left( 1 \right)}_{A}}\times SU{{\left( 2 \right)}_{L}}\times SU{{\left( 3 \right)}_{C}}}{\Gamma_p}\times \frac{SU{{\left( 2 \right)}_{H}}\times U{{\left( 1 \right)}_{V}}}{\Gamma_m}$ as another example. When $\Gamma={\boldsymbol{Z}_{3}}\times  {\boldsymbol{Z}_{2H}}$, the Wilson lines are $(q_-=z_3^e\bmod3;z_2^e=0,1;z_3^e=0,1,2)$ and $(q_+=x_2^e\bmod2;x_2^e=0,1)$. The corresponding 't Hooft lines are $(g_-=z_3^m\bmod3;z_2^m=0\bmod2;z_3^m=0,1,2\bmod3)$ and $(g_+=x_2^m\bmod2;x_2^m=0,1\bmod2)$. 
When $\Gamma={\boldsymbol{Z}_{6L}} \times {\boldsymbol{Z}_{2H}}$, the Wilson lines are $(q_-=(3z_2^e-2z_3^e)\bmod6;z_2^e=0,1;z_3^e=0,1,2)$ and $(q_+=x_2^e\bmod2;x_2^e=0,1)$. The corresponding 't Hooft lines are $(g_-=(3z_2^m+2z_3^m)\bmod6;z_2^m=0,1\bmod2;z_3^m=0,1,2\bmod3)$ and $(g_+=x_2^m\bmod2;x_2^m=0,1\bmod2)$.

\section{Theta-Angles -- Witten Effect} \label{sec:theta}

In this section, we discuss the spectra of line operators under the variations of the $CP$ violating $\theta $-angles in minimal G2HDM, as illustrated in Fig.~\ref{fig:SU(2)theta2pi} (c) for the case of $SU(2)/\boldsymbol{Z}_2$ and Figs.~\ref{fig:SU(3)theta2&4pi} (c) and (d) for the case of $SU(3)/\boldsymbol{Z}_3$. The $\theta$-angle of $U(1)$ does not change the spectra or correlation functions of the local operators. However, in the presence of monopoles or spatial boundaries, it alters the spectra of line operators through the Witten effect. 
For example, in the case of $(U(1)\times SU(2))/\boldsymbol{Z}_2$, the Abelian lines are $(q, g) = \{(2, 0), (0, 1)\}$, with $\tilde{\theta} \in [0, 8\pi)$. Hence, the Witten effect for the Abelian lines in this case has $4 \times 3$ possible combinations, associated with the different values of $z_2^e=0,1$, $z_2^m=0,1$, and $\tilde{\theta} = 2\pi, 4\pi, 6\pi$. The Witten effect for these twelve Abelian lines of the $\boldsymbol{Z}_2$ case are shown in~Figs.~\ref{app:WittenEffectsZ2a}, \ref{app:WittenEffectsZ2b} and \ref{app:WittenEffectsZ2c} in Appendix~\ref{sec:appendix}.
For the case of $(U(1)\times SU(3))/\boldsymbol{Z}_3$, there will be $8 \times 9=72$ combinations of Abelian lines with Witten effect, and for $(U(1)\times SU(2) \times SU(3))/\boldsymbol{Z}_6$, there will be $35 \times 36 = 1260$ combinations. Therefore, for these $\boldsymbol{Z}_3$ and $\boldsymbol{Z}_6$ cases, we have selected two representative values of the $\tilde \theta$ to show the Witten effect in the Abelian lines in Figs.~\ref{app:WittenEffectsZ3} and~\ref{app:WittenEffectsZ6} respectively in Appendix~\ref{sec:appendix}.
As mentioned in Section~\ref{sec:G2HDM}, we will ignore global anomalies and focus only on the spectra of line operators and the Witten effect. To be more specific, we will study the impacts from the Witten effect to the line operators in G2HDM when the $\theta$-angles vary from the different choices of the quotients.

Before diving into the minimal G2HDM, let's consider the general cases first.
Introduce two canonically normalized gauge fields, $\tilde a$ and $a$ with the corresponding strengths $\tilde f$ and $f$ for $U(1)$ and $SU(N)$ respectively. Without any quotienting, the form of $U(1)\times SU(N)$ $\theta$-term is
\be
{{S}_{\theta }}=\frac{{{\theta }_{N}}}{16{{\pi }^{2}}}\int{d^4 x}\, {\rm tr}\,\left( {}^{\star }ff \right)+\frac{{\tilde{\theta }}}{16{{\pi }^{2}}}\int{d^4 x}\,{}^{\star }\tilde{f}\tilde{f} \; ,
\ee
with both $\tilde{\theta}$ and $\theta_N \in [0, 2 \pi)$.
Using these two gauge fields, one can construct a combined gauge field $a+\tilde{a}{\boldsymbol{1}_{N}}$ with its field strength denoted as $F$. Then the $\theta$-term for the $\frac{U\left( 1 \right)\times SU\left( N \right)}{\boldsymbol{Z}_N}$ theory can be expressed as follows~\cite{Tong:2017oea},
\begin{equation}\label{Stheta1}
{{S}_{\theta }}=\frac{{{\theta }_{N}}}{16{{\pi }^{2}}}\int{d^4 x} \, {\rm tr}\left( {}^{\star }FF \right)+\frac{\tilde{\theta }-N{{\theta }_{N}}}{16{{\pi }^{2}}{{N}^{2}}}\int{{d^4 x} \,
  {}^{\star }\left( {\rm tr}\, F \right)\left( {\rm tr}\, F \right)} \; .
\end{equation}
The periodicity for the $\theta$-angles can then be easily read off as ${{\theta }_{N}}\in \left[ 0,2\pi  \right)$ and $\tilde{\theta }\in \left[ 0,2\pi {{N}^{2}} \right)$, in accord with Table~\ref{tab:periodicitySUN}.

We now generalize to the case of 
$ \frac{U\left( 1 \right)\times SU\left( N \right)\times SU\left( M \right)}{{\boldsymbol{Z}_{N\times M}}}$ that is relevant to us, where $M$ and $N$ are coprime, {\it i.e.} $\mathrm{g.c.d.} (M,N) = 1$, such that $\boldsymbol{Z}_N \times \boldsymbol{Z}_M = \boldsymbol{Z}_{N \times M}$. 
Without any quotienting, the $\theta$-term for the $U\left( 1 \right)\times SU\left( N \right)\times SU\left( M \right)$ theory is
\begin{equation}\label{Stheta2}
{{S}_{\theta }}=\frac{M{{\theta }_{N}}}{16{{\pi }^{2}}}\int{{d^4 x}\, {\rm tr}\left( {}^{\star }{{f}_{N}}{\boldsymbol{1}_{M}}{{f}_{N}}{\boldsymbol{1}_{M}} \right)}+\frac{N{{\theta }_{M}}}{16{{\pi }^{2}}}\int{{d^4 x}\,{\rm tr}\left( {}^{\star }{{f}_{M}}{\boldsymbol{1}_{N}}{{f}_{M}}{\boldsymbol{1}_{N}} \right)}+\frac{{\tilde{\theta }}}{16{{\pi }^{2}}}\int{{d^4 x}\, {}^{\star }\tilde{f}\tilde{f}} \; .
\end{equation}
 Introduce two gauge fields, denoted ${{a}_{N}}+\tilde{a}{\boldsymbol{1}_{N}}$ and ${{a}_{M}}+\tilde{a}{\boldsymbol{1}_{M}}$, with the corresponding field strengths
\be
\begin{aligned}
&{{F}_{N}}={{f}_{N}}+\tilde{f}{\boldsymbol{1}_{N}} \; , \\ 
&{{F}_{M}}={{f}_{M}}+\tilde{f}{\boldsymbol{1}_{M}} \; .\\
\end{aligned}
\ee
The $\theta$-term for the $\frac{U\left( 1 \right)\times SU\left( N \right)\times SU\left( M \right)}{{\boldsymbol{Z}_{N\times M}}}$ theory is then given as 
\begin{equation}\label{Stheta3}
	\begin{aligned}
	{{S}_{\theta }} 
    &=\frac{M{{\theta }_{N}}}{16{{\pi }^{2}}}\int{{d^4 x}\, {\rm tr}\left( {}^{\star }{{F}_{N}}{\boldsymbol{1}_{M}}{{F}_{N}}{\boldsymbol{1}_{M}} \right)}+\frac{N{{\theta }_{M}}}{16{{\pi }^{2}}}\int{d^4 x}\,{\rm tr}\left( {}^{\star }{{F}_{M}}{\boldsymbol{1}_{N}}{{F}_{M}}{\boldsymbol{1}_{N}} \right)\\
	&+\frac{\tilde{\theta }-N{{M}^{2}}{{\theta }_{N}}-M{{N}^{2}}{{\theta }_{M}}}{16{{\pi }^{2}}{{\left( N\times M \right)}^{2}}}\int{d^4 x}\, {}^{\star }{\rm tr}\left( \tilde{f}{\boldsymbol{1}_{N\times M}} \right){\rm tr}\left( \tilde{f}{\boldsymbol{1}_{N\times M}} \right) \; .\\
\end{aligned}
\end{equation}
The last factor ${\rm tr}\left( \tilde{f}{\boldsymbol{1}_{N\times M}} \right)$ in the second line can be simplified to ${\rm tr}\left( {{F}_{N}}{\boldsymbol{1}_{M}} \right)$ or ${\rm tr}\left( {{F}_{M}}{\boldsymbol{1}_{N}} \right)$. The periodicity of the $\theta$-angles can then be read off from Eq.~(\ref{Stheta3}): ${{\theta }_{N}}\in \left[ 0,2\pi  \right)$, ${{\theta }_{M}}\in \left[ 0,2\pi  \right)$, while $\tilde{\theta }\in \left[ 0,2\pi {{N}^{2}}{{M}^{2}} \right)$, in accord with Table~\ref{tab:periodicitySUN}.

We can now consider how the choice of each quotient extend the ranges of $\theta$-angles in minimal G2HDM. Specific values of these $\theta$-angles corresponding to formally $CP$ invariant (or equivalently, time reversal invariant) theories at the Lagrangian level are also shown. We will not consider the possibilities of 
spontaneously symmetry breakings at these specific angles in this work.

According to the above analysis,
for the quotient group $G$ we first studied in Eq.~(\ref{eq:CaseAex1}), namely,
\be
\label{eq:CaseAex1X}
G=\frac{U{{\left( 1 \right)}_{Y}}\times SU{{\left( 2 \right)}_{L}}\times SU{{\left( 3 \right)}_{C}}}{\Gamma_n}\times \frac{ U{{\left( 1 \right)}_{X}} \times SU{{\left( 2 \right)}_{H}}  }{\Gamma_m} \; ,
\ee
we have the following $\theta$-term,
\begin{equation}\label{Stheta4}
\begin{aligned}
	{{S}_{\theta }}& =\frac{3{{\theta }_{2L}}}{16{{\pi }^{2}}}\int{d^4 x}\, {\rm tr}\left( {}^{\star }{{F}_{2L}}{\boldsymbol{1}_{3}}{{F}_{2L}}{\boldsymbol{1}_{3}} \right)+\frac{2{{\theta }_{3}}}{16{{\pi }^{2}}}\int{d^4 x}\, {\rm tr}\left( {}^{\star }{{F}_{3}}{\boldsymbol{1}_{2}}{{F}_{3}}{\boldsymbol{1}_{2}} \right)  \\
	& +\frac{{{{\tilde{\theta }}}_{Y}}-18{{\theta }_{2L}}-12{{\theta }_{3}}}{16{{\pi }^{2}} \cdot 36}\int{d^4 x}\, {}^{\star }{\rm tr}\left( {{{\tilde{f}}}_{Y}}{\boldsymbol{1}_{6}} \right){\rm tr}\left( {{{\tilde{f}}}_{Y}}{\boldsymbol{1}_{6}} \right) \\ 
	& +\frac{{{\theta }_{2H}}}{16{{\pi }^{2}}}\int{d^4 x}\, {\rm tr}\left( {}^{\star }{{F}_{2H}}{{F}_{2H}} \right)+\frac{{{{\tilde{\theta }}}_{X}}-2{{\theta }_{2H}}}{16{{\pi }^{2}} \cdot 4}\int{d^4 x}\, {}^{\star }{\rm tr}\left( {{{\tilde{f}}}_{X}}{\boldsymbol{1}_{2}} \right){\rm tr}\left( {{{\tilde{f}}}_{X}}{\boldsymbol{1}_{2}} \right) \; . \\ 
\end{aligned}
\end{equation}

\begin{table}[htbp!]
	\begin{center}
		\caption{Ranges of the $CP$ violating $\theta$-angles and their specific values in $CP$ invariant theories, corresponding to different choices of $\Gamma$ for the quotient group in Eq.~(\ref{eq:CaseAex1X}).
        }
        \label{tab:thetas4ex1}
        \resizebox{\textwidth}{!}{
		\begin{tabular}{|c|c|c|c|}
			\hline
			$\Gamma$ & \text{Ranges of} $\theta_3, \theta_{2L} , \theta_{2H}$ & \text{Ranges  of} ${\tilde \theta}_Y , {\tilde \theta}_X$ & $CP$ \text{invariant theory} \\ 
			\hline\hline
            $\boldsymbol{1} \times \boldsymbol{1}$ & $[0, 2 \pi) $ & $[0, 2 \pi) $ & \text{All angles} $0$ \text{or} $\pi$  \\ 
            \hline
            $\boldsymbol{Z}_{2L} \times \boldsymbol{1}$ & $[0, 2 \pi) $ & $ \begin{aligned}&\tilde{\theta}_Y \in [0,8\pi), \\
            &\tilde{\theta}_X \in [0,2\pi) . \end{aligned}$ & $\begin{aligned}&\theta_{2H},\theta_3,\tilde{\theta}_X = 0, \pi ;\\ & \theta_{2L}=0 \Rightarrow \tilde{\theta}_Y = 0, 4 \pi ; \, \theta_{2L}=\pi \Rightarrow \tilde{\theta}_Y = 2\pi, 6 \pi .\end{aligned}$ \\
            \hline
             $\boldsymbol{1} \times {\boldsymbol{Z}_{2H}}$ & $[0, 2 \pi) $ 
             & 
             $ \begin{aligned} 
            &\tilde{\theta}_Y \in [0,2\pi), \\
            &\tilde{\theta}_X \in [0,8\pi) .     
            \end{aligned}$ & 
            $\begin{aligned}
             & \theta_{2L},\theta_3,\tilde{\theta}_Y = 0, \pi ; \\
             & \theta_{2H} = 0 \Rightarrow \tilde{\theta}_X = 0, 4 \pi ; \, \theta_{2H} = \pi \Rightarrow \tilde{\theta}_X = 2\pi, 6 \pi.\\
            \end{aligned}$ \\ \hline
            ${\boldsymbol{Z}_{3}} \times \boldsymbol{1}$ & $[0, 2 \pi) $ & $ \begin{aligned} &\tilde{\theta}_Y \in [0,18\pi), \\
            & \tilde{\theta}_X \in [0,2\pi) . \end{aligned}$ & 
            $
            \begin{aligned}
            & \theta_{2L}, \theta_{2H},\tilde{\theta}_X = 0, \pi ; \\
            & \theta_3 = 0 \Rightarrow \tilde{\theta}_Y = 0, 9 \pi ; \, \theta_3 = \pi \Rightarrow \tilde{\theta}_Y = 3\pi, 12 \pi .
            \end{aligned}
            $
            \\ \hline
            $\boldsymbol{Z}_{2L} \times {\boldsymbol{Z}_{2H}}$ & $[0, 2 \pi) $ & $ \tilde{\theta}_Y , \tilde{\theta}_X \in [0,8\pi), $ & 
            $
            \begin{aligned}
                & \theta_3 = 0, \pi; \\
                & \theta_{2L}=0 \Rightarrow \tilde{\theta}_Y=0,4\pi; \, \theta_{2L}= \pi  \Rightarrow \tilde{\theta}_Y=2\pi,6 \pi; \\
                & \theta_{2H}=0 \Rightarrow \tilde{\theta}_X=0,4\pi; \, \theta_{2H}= \pi  \Rightarrow \tilde{\theta}_X=2\pi,6 \pi .
            \end{aligned}
            $
            \\ \hline
            $\boldsymbol{Z}_{6L} \times \boldsymbol{1}$ & $[0, 2 \pi) $ & $ \begin{aligned}&\tilde{\theta}_Y \in [0,72\pi), \\
            &\tilde{\theta}_X \in [0,2\pi) . \end{aligned}$ & 
            $\begin{aligned}
	        & \theta_{2H}, \tilde \theta_X = 0, \pi;\\
            &{{\theta }_{2L}}=0, {{\theta }_{3}}=0 \Rightarrow \tilde{\theta }_Y=0,36\pi  ; \\
	           &{{\theta }_{2L}}=0, {{\theta }_{3}}=\pi \Rightarrow \tilde{\theta }_Y=12\pi ,48\pi  ; \\
	           &{{\theta }_{2L}}=\pi, {{\theta }_{3}}=0 \Rightarrow \tilde{\theta }_Y=18\pi ,54\pi  ;\\
	           &{{\theta }_{2L}}=\pi, {{\theta }_{3}}=\pi \Rightarrow \tilde{\theta }_Y=30\pi ,66\pi    .
            \end{aligned}$ \\ \hline
            $\boldsymbol{Z}_{3} \times {\boldsymbol{Z}_{2H}}$ & $[0, 2 \pi) $ & $ \begin{aligned}&\tilde{\theta}_Y \in [0,18\pi), \\
            &\tilde{\theta}_X \in [0,8\pi) . \end{aligned}$ & $\begin{aligned}
            & \theta_{2L}=0,\pi ;\\
            & {{\theta }_{3}}=0 \Rightarrow  \tilde{\theta }_Y=0,9\pi  , \;   {{\theta }_{2H}}=0 \Rightarrow \tilde{\theta }_X=0,4\pi  ; \\
            & {{\theta }_{3}}=\pi \Rightarrow \tilde{\theta }_Y=3\pi ,12\pi , \;  {{\theta }_{2H}}=\pi \Rightarrow \tilde{\theta }_X=2\pi , 6\pi .
            \end{aligned}$ \\ \hline
            $\boldsymbol{Z}_{6L} \times {\boldsymbol{Z}_{2H}}$ & $[0, 2 \pi) $ & $ \begin{aligned}&\tilde{\theta}_Y \in [0,72\pi), \\
            &\tilde{\theta}_X \in [0,8\pi) . \end{aligned}$ & 
           $\begin{aligned} & \theta_{2H} = 0 \Rightarrow  \tilde{\theta}_X = 0, 4 \pi ; \\& \theta_{2H} = \pi \Rightarrow  \tilde{\theta}_X = 2 \pi , 6 \pi ;\\
	       &{{\theta }_{2L}}=0, {{\theta }_{3}}=0 \Rightarrow \tilde{\theta }_Y=0,36\pi  ; \\
	    &{{\theta }_{2L}}=0, {{\theta }_{3}}=\pi \Rightarrow \tilde{\theta }_Y=12\pi ,48\pi  ; \\
	       &{{\theta }_{2L}}=\pi, {{\theta }_{3}}=0 \Rightarrow \tilde{\theta }_Y=18\pi ,54\pi  ;\\
	       &{{\theta }_{2L}}=\pi, {{\theta }_{3}}=\pi \Rightarrow \tilde{\theta }_Y=30\pi ,66\pi  .
            \end{aligned}$ \\
			\hline   
		\end{tabular}
        }
	\end{center}
\end{table}

$\Gamma = \boldsymbol{1} \times \boldsymbol{1}$: All $\theta$-angles have periodicity $2\pi $. When all of them take $0$ or $\pi $, the resulting theories are invariant under the $CP$ transformation.

$\Gamma ={\boldsymbol{Z}_{2L}} \times \boldsymbol{1} $: The periodicity of $\tilde{\theta}_Y$ extends, becoming $8\pi$, while all other angles have the same range $[0, 2\pi)$. The CP invariant theories are specified by: ${{\theta }_{3}},\tilde{\theta}_X,{{\theta }_{2H}}=0,\pi $; from Eq.~\eqref{Stheta4},  $\tilde{\theta }_Y=0,4\pi $ when ${{\theta }_{2L}}=0$, and $\tilde{\theta }_Y=2\pi ,6\pi $ when ${{\theta }_{2L}}=\pi $.

$\Gamma = \boldsymbol{1} \times  {\boldsymbol{Z}_{2H}}$: The periodicity of $\tilde{\theta }_{X}$ is $8\pi$, while all other angles have the same range $[0, 2\pi)$. 
The CP invariant theories are specified by: ${{\theta }_{3}},\tilde{\theta}_Y,{{\theta }_{2L}}=0,\pi $; from Eq.~\eqref{Stheta4}, $\tilde{\theta }_X=0,4\pi $ when ${{\theta }_{2H}}=0$, and $\tilde{\theta }_X=2\pi ,6\pi $ when ${{\theta }_{2H}}=\pi$.

$\Gamma ={\boldsymbol{Z}_{3}} \times \boldsymbol{1}$: The periodicity of $\tilde{\theta }_Y$ is $18\pi$, while all other angles have the same range $[0, 2\pi)$. 
The CP invariant theories are specified by:
${{\theta }_{2L}},\tilde{\theta }_X,{{\theta }_{2H}}=0,\pi $; 
when ${{\theta }_{3}}=0$, one has $\tilde{\theta }_Y=0,9\pi $; and 
when ${{\theta }_{3}}=\pi $, one has $\tilde{\theta }_Y= 3\pi,12\pi $.

$\Gamma ={\boldsymbol{Z}_{2L}}\times {\boldsymbol{Z}_{2H}}$: Both $\tilde{\theta }_Y$ and $\tilde{\theta }_X$ have periodicity $8\pi$, while all other angles have the same range $[0,2\pi)$. The CP invariant theories are specified by: 
${{\theta }_{3}}=0,\pi $; 
when ${{\theta }_{2L}}=0$, one has $\tilde{\theta }_Y=0,4\pi $; 
when ${{\theta }_{2L}}=\pi $, one has $\tilde{\theta }_Y=2\pi ,6\pi $; 
when ${{\theta }_{2H}}=0$, one has $\tilde{\theta }_X=0,4\pi $; and 
when ${{\theta }_{2H}}=\pi $, one has $\tilde{\theta }_X=2\pi ,6\pi $ .

$\Gamma ={\boldsymbol{Z}_{6L}} \times \boldsymbol{1} $: The periodicity of $\tilde{\theta }_Y$ is $72\pi$, while all other angles have the same range $[0, 2\pi)$. The $\boldsymbol{Z}_{6L}$ part is exactly the same as in SM~\cite{Tong:2017oea}, so for the $CP$ invariant theories, we have
\be
\begin{aligned}
	&{{\theta }_{2L}}=0, {{\theta }_{3}}=0 \Rightarrow \tilde{\theta }_Y=0,36\pi  \; ; \\
	&{{\theta }_{2L}}=0, {{\theta }_{3}}=\pi \Rightarrow \tilde{\theta }_Y=12\pi ,48\pi  \; ; \\
	&{{\theta }_{2L}}=\pi, {{\theta }_{3}}=0 \Rightarrow \tilde{\theta }_Y=18\pi ,54\pi  \; ;\\
	&{{\theta }_{2L}}=\pi, {{\theta }_{3}}=\pi \Rightarrow \tilde{\theta }_Y=30\pi ,66\pi   \; .
\end{aligned}
\ee
Meanwhile, we can have $\tilde{\theta }_X$ and $\theta_{2H} $ taking $0$ or $\pi $, which are both $CP$ invariant.

$\Gamma = {\boldsymbol{Z}_{3}} \times {\boldsymbol{Z}_{2H}}$: 
The periodicity of $\tilde \theta_X$ and $\tilde \theta_Y$ are $8 \pi$ and $18 \pi$ respectively, while all other angles have the same range $[0,2\pi)$.
The $CP$ invariant theories are just as in the theories of $\Gamma ={\boldsymbol{Z}_{3}} \times \boldsymbol{1}$ and $\Gamma = \boldsymbol{1} \times {\boldsymbol{Z}_{2H}}$. Among these, $\tilde{\theta }_Y$, ${{\theta }_{2L}}$ and ${{\theta }_{3}}$ are similar to $\Gamma ={\boldsymbol{Z}_{3}} \times \boldsymbol{1}$, and $\tilde{\theta }_X$, ${{\theta }_{2H}}$ are similar to $\Gamma =\boldsymbol{1} \times  {\boldsymbol{Z}_{2H}}$. Thus ${{\theta }_{2L}}=0,\pi$; and
\be
\begin{aligned}
{{\theta }_{3}}=0 \Rightarrow  \tilde{\theta }_Y=0,9\pi \; , & \text{\quad\quad\quad}  {{\theta }_{2H}}=0 \Rightarrow \tilde{\theta }_X=0,4\pi \; , \\
{{\theta }_{3}}=\pi \Rightarrow \tilde{\theta }_Y=3\pi ,12\pi  \; ,  &  \text{\quad\quad\quad}  {{\theta }_{2H}}=\pi \Rightarrow \tilde{\theta }_X=2\pi , 6\pi \; .
\end{aligned}
\ee

$\Gamma = {\boldsymbol{Z}_{6L}} \times {\boldsymbol{Z}_{2H}}$: 
$\tilde{\theta }_Y$, ${{\theta }_{2L}}$ and ${{\theta }_{3}}$ are just like the case in $\Gamma ={\boldsymbol{Z}_{6L}} \times \boldsymbol{1}$, and $\tilde{\theta }_X$, ${{\theta }_{2H}}$ are just like the case in $\Gamma = \boldsymbol{1} \times {\boldsymbol{Z}_{2H}}$.

\begin{table}
	\begin{center}
		\caption{
        Same as Table~\ref{tab:thetas4ex1} for the quotient group in Eq.~(\ref{eq:CaseCex1}).
        }
        \label{tab:thetas4ex2}
        \resizebox{\textwidth}{!}{
		\begin{tabular}{|c|c|c|c|}
			\hline
			$\Gamma$ & \text{Ranges of} $\theta_3, \theta_{2L} , \theta_{2H}$ & \text{Ranges  of} ${\tilde \theta}_V , {\tilde \theta}_A$ & $CP$ \text{invariant theory} \\ 
			\hline
            
            \hline
            $\boldsymbol{1} \times \boldsymbol{1} \times \boldsymbol{1} $ & $[0, 2 \pi) $ & $[0, 2 \pi) $ & \text{All angles 0 or $\pi$.} \\
            
            \hline
            $\boldsymbol{1} \times \boldsymbol{Z}_{2L} \times \boldsymbol{1}$ & $[0, 2 \pi) $ & 
            
            $\begin{aligned}
            &{\tilde \theta}_V \in [0, 2 \pi);\\ 
            & {\tilde \theta}_A \in [0, 8\pi) 
            \end{aligned}$ & 
            
            $\begin{aligned}&\theta_3, \theta_{2H} = 0, \pi; \;\tilde{\theta}_{V} = 0,\pi;\\
	&{{\theta }_{2L}}=0 \Rightarrow\tilde{\theta }_A=0,4\pi ;  \\
	&{{\theta }_{2L}}=\pi \Rightarrow \tilde{\theta }_A=2\pi ,6\pi .
\end{aligned}$ \\

\hline
    $ {\boldsymbol{Z}_{2H}} \times \boldsymbol{1} \times \boldsymbol{1} $ & $[0, 2 \pi) $ &  $\begin{aligned}
            &{\tilde \theta}_V \in [0, 8 \pi);\\ 
            & {\tilde \theta}_A \in [0, 2\pi) 
            \end{aligned}$ & $\begin{aligned}
    &\theta_3, \theta_{2L} = 0, \pi; \; \tilde{\theta}_{A} = 0,\pi; \\
	&{{\theta }_{2H}}=0 \Rightarrow\tilde{\theta }_V=0,4\pi  ,   \\
	&{{\theta }_{2H}}=\pi \Rightarrow \tilde{\theta }_V=2\pi ,6\pi .
\end{aligned}$ \\

\hline   
             $\boldsymbol{1} \times \boldsymbol{1} \times {\boldsymbol{Z}_{3}} $ & 
             $\begin{aligned}
             &\theta_{2L},\theta_{2H} \in [0, 2 \pi);\\
             &\theta_3 \in [0, 6 \pi)
             \end{aligned} $ & 
             ${\tilde \theta}_V, {\tilde \theta}_A\in [0, 2\pi)$ & $\theta_3=0, 3 \pi; \; \text{others} =0, \pi$. \\
            
\hline
            $\boldsymbol{Z}_{2L} \times {\boldsymbol{Z}_{2H}} \times \boldsymbol{1} $ & $[0, 2 \pi) $ & ${\tilde \theta}_V, {\tilde \theta}_A\in [0, 8\pi)$ & 
    $\theta_3 = 0, \pi;$ \,
    $\begin{aligned} &{{\theta }_{2L}}=0 \Rightarrow\tilde{\theta }_A=0,4\pi  ;   \\
	&{{\theta }_{2L}}=\pi \Rightarrow \tilde{\theta }_A=2\pi ,6\pi  ;\\
    &{{\theta }_{2H}}=0 \Rightarrow\tilde{\theta }_V=0,4\pi  ;   \\
    &{{\theta }_{2H}}=\pi \Rightarrow \tilde{\theta }_V=2\pi ,6\pi .\\
\end{aligned}$  \\ 

\hline
            $\boldsymbol{1} \times \boldsymbol{Z}_{2L} \times \boldsymbol{Z}_3$ &  
            $\begin{aligned}
             &\theta_{2L},\theta_{2H} \in [0, 2 \pi);\\
             &\theta_3 \in [0, 6 \pi)
             \end{aligned} $ & 
             $\begin{aligned}
            &{\tilde \theta}_V \in [0, 2 \pi);\\ 
            & {\tilde \theta}_A \in [0, 8\pi) 
            \end{aligned}$ &
            
            $\begin{aligned} 
            &\theta_3= 0, 3\pi; \; \theta_{2H} = 0, \pi; \; \tilde{\theta}_{V} = 0,\pi; \\
	&{{\theta }_{2L}}=0 \Rightarrow\tilde{\theta }_A=0,4\pi ;  \\
	&{{\theta }_{2L}}=\pi \Rightarrow \tilde{\theta }_A=2\pi ,6\pi .
\end{aligned}$ \\

\hline
        $\boldsymbol{Z}_{2H} \times \boldsymbol{1} \times \boldsymbol{Z}_3$  &  $\begin{aligned}
             &\theta_{2L},\theta_{2H} \in [0, 2 \pi);\\
             &\theta_3 \in [0, 6 \pi)
             \end{aligned} $
    &  $\begin{aligned}
            &{\tilde \theta}_V \in [0, 8 \pi);\\ 
            & {\tilde \theta}_A \in [0, 2\pi) 
            \end{aligned}$ & 
        $\begin{aligned}
    & \theta_3 = 0, 3\pi; \; \theta_{2L} = 0, \pi; \; \tilde{\theta}_{A} = 0,\pi;\\
    &{{\theta }_{2H}}=0 \Rightarrow\tilde{\theta }_V=0,4\pi  ;   \\
	&{{\theta }_{2H}}=\pi \Rightarrow \tilde{\theta }_V=2\pi ,6\pi  .\\
\end{aligned}$  \\ 

\hline
     $\boldsymbol{Z}_{2L} \times {\boldsymbol{Z}_{2H}} \times \boldsymbol{Z}_3 $  & 
     $\begin{aligned}
             &\theta_{2L},\theta_{2H} \in [0, 2 \pi);\\
             &\theta_3 \in [0, 6 \pi)
             \end{aligned} $
    & ${\tilde \theta}_V, {\tilde \theta}_A\in [0, 8\pi)$ & 
    $\theta_3 = 0, 3\pi$; \,
    $\begin{aligned} &{{\theta }_{2L}}=0 \Rightarrow\tilde{\theta }_A=0,4\pi  ;   \\
	&{{\theta }_{2L}}=\pi \Rightarrow \tilde{\theta }_A=2\pi ,6\pi  ;\\
    &{{\theta }_{2H}}=0 \Rightarrow\tilde{\theta }_V=0,4\pi  ;   \\
    &{{\theta }_{2H}}=\pi \Rightarrow \tilde{\theta }_V=2\pi ,6\pi .\\
\end{aligned}$ \\ 

\hline
		\end{tabular} }
	\end{center}
    \end{table}

The results of the above analysis for this example of $G$ in Eq.~(\ref{eq:CaseAex1X}) are summarized at Table~\ref{tab:thetas4ex1}. For the Witten effect in the Abelian lines for this quotient group, 
one can refer to Figs.~\ref{fig:SU(2)theta2pi}, \ref{fig:SU(3)theta2&4pi}, \ref{app:WittenEffectsZ2a}, \ref{app:WittenEffectsZ2b}, 
\ref{app:WittenEffectsZ2c}, \ref{app:WittenEffectsZ3},  
and \ref{app:WittenEffectsZ6}.

Next, let us discuss an alternative quotient
\be
\label{eq:CaseCex1}
G=\frac{U{{\left( 1 \right)}_{V}}\times SU{{\left( 2 \right)}_{H}}}{\Gamma_p}\times \frac{ U{{\left( 1 \right)}_{A}} \times SU{{\left( 2 \right)}_{L}}}{\Gamma_m}\times \frac{SU{{\left( 3 \right)}_{C}}}{\Gamma_n} \; ,
\ee
which is one of the 4 possibilities of case (C).
The $\theta$-angles $\tilde{\theta}_X$ and $\tilde{\theta}_Y$ turn into $\tilde{\theta}_V$ and $\tilde{\theta}_A$. Specifically, $SU(3)_C/\boldsymbol{Z}_3$ is now standalone and not connected with any $U(1)$, so now the period of $\theta_3$ is $[0,2\pi\cdot 3)$ according to Table~\ref{tab:periodicitySUN}.
The analysis is identical to that of the previous example, so we will not present the details here. 
The results for this example of $G$ in Eq.~(\ref{eq:CaseCex1}) are summarized at Table~\ref{tab:thetas4ex2}. Similarly, for the Witten effect in the Abelian lines for this quotient group, 
one can refer to Figs.~\ref{fig:SU(2)theta2pi}, \ref{fig:SU(3)theta2&4pi}, \ref{app:WittenEffectsZ2a}, \ref{app:WittenEffectsZ2b}, 
\ref{app:WittenEffectsZ2c}, \ref{app:WittenEffectsZ3},  
and \ref{app:WittenEffectsZ6}.

The $\theta$-angles in the two quotient groups $G$ discussed above are merely representative examples. For the remaining possible center quotienting for $G$, the corresponding periodicity of the $\theta$-angles and their specific values for $CP$ invariant theories are similar to these two examples and will not be elaborated further here.

\section{ Symmetry Breaking }\label{sec:symmetrybreaking}

Vacuum expectation values of line operators provide powerful gauge invariant ways to probe the phase structures of gauge theories, provided that the lines are infinitely long or closed loop. Distinctive features are observed with or without the spontaneous symmetry breaking by the Higgs mechanism in the theories. 

The spontaneous breaking of higher-form symmetry impacts the non-local object that it acts upon. For the breaking of 1-forms, it acts on Wilson and 't Hooft lines, even the more general dyonic lines. The vacuum expectation value of a Wilson line $W_R [\mathcal{C}]$ 
(or 't Hooft line $T_R[\mathcal{C}]$ in the magnetic sector or dyonic line in the mixed sector) follows an area law without Higgs field condensates, or a perimeter law with Higgs condensates. For large loops, the area law decays much faster than the perimeter law. Thus area law indicates a confinement phase corresponding to a symmetry-preserving scenario (like the quantum chromodynamics in SM), whereas perimeter law suggests a deconfined Higgs phase with symmetry breaking, leading to short range screening Yukawa force (like the electroweak sector in SM). The Coulomb phase is another possible deconfined phase independent of the above two extremes, having a long-range force (like the electromagnetism in SM). The area and perimeter laws are scale dependent, while the Coulomb law is scale independent. Therefore, the 1-form symmetry breaking exhibits behaviors very distinctive from those of ordinary 0-form symmetry breaking, where the key difference lies in the order parameters, which are the vacuum expectations of the non-local operators rather than the local ones. As in the case of 0-form symmetry,
spontaneous symmetry breaking of 1-form symmetry is accompanied by a Goldstone excitation, which is exactly the massless spin 1 gauge boson for the $U(1)$ gauge theory. 
In minimal G2HDM, the photon and its hidden partner dark photon, can be interpreted as these Goldstone excitations due to spontaneous symmetry breakings of the associated 1-form symmetries. For a recent introductory review on higher-form symmetries, see for example~\cite{Gomes:2023ahz}.

In the following, we will explore the behaviors of the line operators of the electroweak SM $SU(2)_L \times U(1)_Y$ and its hidden replica $SU(2)_H \times U(1)_X$ under the influence of spontaneous symmetry breaking by the Higgs mechanism in minimal G2HDM.
In the minimal G2HDM framework, two Higgs fields contribute to the breaking of the original gauge symmetries. 
The first step involves the hidden Higgs field,  $H{(\boldsymbol{1}, \boldsymbol{2}, \boldsymbol{1})}_{0,1}$ with the SM and dark hypercharges $q=0$ and $h=1$ respectively to break the  $SU(2)_H \times U(1)_X$ symmetry at a high scale down to the so-called ``dark electromagnetism''  $U(1)_D$.
The second step involves the bi-doublet Higgs field,  $H{(\boldsymbol{2}, \boldsymbol{2}, \boldsymbol{1})}_{3,-1}$ which condenses to break the  $SU(2)_L \times U(1)_Y$ symmetries down to the ordinary electromagnetism  $U(1)_{\text{em}}$. This latter Higgs field belongs to a two-dimensional representation of both  $SU(2)_L$ and $SU(2)_H$ with $q = 3$ and  $h=-1$, respectively.

According to the SM Higgs mechanism, the allowed electromagnetic electric charge  $Q_{\text{em}}$ satisfies the Gell-Mann-Nishijima formula:
\be
\label{eq:Qem}
Q_{\text{em}} = \frac{q}{6} + \frac{\lambda_2^e}{2} \; ,
\ee
where  $q/6$ represents the  $U(1)_Y$ hypercharge, and  
$\lambda_2^e \in \boldsymbol{Z}$ is two times the Cartan generator of $SU(2)_L$ with $\lambda_2^e = 0$ for the one-dimensional representation (singlet), $\lambda_2^e = \pm1$ for the fundamental representation, $\lambda_2^e = \pm 2,0$ for the adjoint representation, and so on.
Note that $z_2^e$, entered in the Wilson lines and GDQC, satisfies $z_2^e = \lambda_2^e  \bmod2$, while its magnetic counterpart is $z_2^m = \lambda_2^m \bmod2$ with 
$\lambda_2^m \in \boldsymbol{Z}$.
Similarly, the dark electric charge  $Q_D$ in the dark sector is given by~\footnote{When the 
bi-doublet $H{(\boldsymbol{2}, \boldsymbol{2}, \boldsymbol{1})}_{3,-1}$ condenses and breaks the SM down to $U(1)_{\rm em}$ in the second step, it will mix the visible SM $SU(2)_L \times U(1)_Y$ sector and the hidden $SU(2)_H \times U(1)_X$ sector. For simplicity, we will ignore such small mixings here, since these effects are suppressed by the ratio of the SM and hidden vacuum expectation values, which is assumed to be small.}
\be
\label{eq:Qdem}
Q_D = \frac{h}{2} + \frac{\rho_2^e}{2} \; ,
\ee
where  $h/2$ is the dark hypercharge of  $U(1)_X$, 
$\rho_2^{e,m } \in \boldsymbol{Z}$, $x_2^{e,m} = \rho_2^{e,m} \bmod 2$ for the hidden $SU(2)_H$.

The higher-form symmetry breaking follows the Higgs condensation. After symmetry breaking, while the electric charges of the Wilson lines are determined by 
Eqs.~(\ref{eq:Qem}) and (\ref{eq:Qdem}), most expectation values of the 't Hooft and dyonic lines associated with the $U(1)_{\widetilde Y}$ and $SU(2)_L$ (as well as their hidden partners $U(1)_{\widetilde X}$ and $SU(2)_{H}$) exhibit area laws and remain confined in the process. Only those lines in the Higgs phase become deconfined, with the magnetic charges determined by GDQC as 
\be
\label{eq:deconfinedcharges}
\begin{aligned}
& 6g = \lambda_2^m \quad \Rightarrow  \quad 6g = z_2^m \pmod 2 \quad \Rightarrow \quad G_{\rm em} = 6g \;, \\
& 2k = \rho_2^m \quad \Rightarrow  \quad 2k = x_2^m \pmod 2 \quad \Rightarrow \quad G_D = 2k \;. \\
\end{aligned}
\ee
The magnetic charges $G_{\rm em}$ and $G_D$ are representing the magnetic counterparts of the $U(1)_{\rm em}$ and $U(1)_D$ symmetries. One can easily check that the two sets of formulas in Eq.~(\ref{eq:deconfinedcharges}) can be obtained by applying the GDQC of Eq.~(\ref{eq:DiracCondition}) to SM Higgs doublet $H_1$ and hidden Higgs doublet $\Phi_H$ respectively.
In general, by examining different quotients, we can determine the minimal electric and magnetic charges of the line operators under symmetry breaking.

Consider first the following quotient that we have analyzed its spectrum in Section~\ref{sec:LO-G2HDM}:
\be
\label{eq:SBex1}
G=\frac{U{{\left( 1 \right)}_{Y}}\times SU{{\left( 2 \right)}_{L}}\times SU{{\left( 3 \right)}_{C}}}{\Gamma_p}\times \frac{ U{{\left( 1 \right)}_{X}} \times SU{{\left( 2 \right)}_{H}} }{\Gamma_m}
\; .
\ee
The different choices of $\Gamma$ may lead to distinct symmetry-breaking patterns, which in turn dictate the electric and magnetic charge spectra in both the visible and dark sectors.

Here and henceforth, we will adopt the convention similar to that of the SM~\cite{Tong:2017oea}, representing the Wilson line before symmetry breaking by $W(R_{SU(2)_L},R_{SU(2)_H},R_{SU(3)_C})_{q,h}$, just as representing the particle content in the model discussed in Section~\ref{sec:G2HDM}. After symmetry breaking of the dark sector in the first step, we represent the Wilson line by $W(R_{SU(2)_L},R_{SU(3)_C})_q$.

%Table V
\begin{table}
	\begin{center}
		\caption{
       The minimal (dark) electric charge and minimal (dark) magnetic charge resulting from the two-step symmetry breaking of the gauge group in Eq.~(\ref{eq:SBex1}), along with the corresponding Wilson and 't Hooft lines.
        }
        \label{tab:minimaicharge1}
        \resizebox{\textwidth}{!}{
		\begin{tabular} {|c|c|c|c|c|}
			
            \hline
			$\Gamma$ & \text{\makecell[c] {Wilson lines and the\\corresponding minimal \\dark electric charge \\in the first step of \\ symmetry breaking} }
            & \text{\makecell [c] {'t Hooft lines and the \\ corresponding minimal \\dark magnetic charge \\in the first step of \\ symmetry breaking}}
            & \text{\makecell [c] {Wilson lines and the \\ corresponding minimal \\electric charge \\in the second step of \\ symmetry breaking}}
            & \text{\makecell [c] {'t Hooft lines and the \\corresponding minimal \\magnetic charge \\in the second step of \\ symmetry breaking}} \\
			\hline
            $\boldsymbol{1} \times \boldsymbol{1}$ & $W(\boldsymbol{1},\boldsymbol{1},\boldsymbol{1})_{0,1}\Rightarrow Q_D=\frac{1}{2}$ & $\rho_2^m=2, k=1\Rightarrow G_D=2$ & $W(\boldsymbol{1},\boldsymbol{1})_1\Rightarrow Q_{\rm em}=\frac{1}{6}$ & $\lambda_2^m=6, g=1\Rightarrow G_{\rm em}=6$   \\
            \hline
            $\boldsymbol{Z}_{2L}\times \boldsymbol{1}$ & $W(\boldsymbol{1},\boldsymbol{1},\boldsymbol{1})_{0,1}\Rightarrow Q_D=\frac{1}{2}$ & $\rho_2^m=2, k=1\Rightarrow G_D=2$ & $W(\boldsymbol{1},\boldsymbol{1})_2\Rightarrow Q_{\rm em}=\frac{1}{3}$ & $\lambda_2^m=3, g=\frac{1}{2}\Rightarrow G_{\rm em}=3$   \\
            \hline
            
            $\boldsymbol{1} \times \boldsymbol{Z}_{2H}$ & $W(\boldsymbol{1},\boldsymbol{1},\boldsymbol{1})_{0,2}\Rightarrow Q_D=1$ & $\rho_2^m=1, k=\frac{1}{2}\Rightarrow G_D=1$ & $W(\boldsymbol{1},\boldsymbol{1})_1\Rightarrow Q_{\rm em}=\frac{1}{6}$ & $\lambda_2^m=6, g=1\Rightarrow G_{\rm em}=6$   \\
            \hline
            $\boldsymbol{Z}_{3}\times \boldsymbol{1}$ & $W(\boldsymbol{1},\boldsymbol{1},\boldsymbol{1})_{0,1}\Rightarrow Q_D=\frac{1}{2}$ & $\rho_2^m=2, k=1\Rightarrow G_D=2$ & $W(\boldsymbol{1},\boldsymbol{3})_1\Rightarrow Q_{\rm em}=\frac{1}{6}$ & $\lambda_2^m=2, g=\frac{1}{3}\Rightarrow G_{\rm em}=2$   \\
            \hline
            $\boldsymbol{Z}_{2L}\times \boldsymbol{Z}_{2H}$ & $W(\boldsymbol{1},\boldsymbol{1},\boldsymbol{1})_{0,2}\Rightarrow Q_D=1$ & $\rho_2^m=1, k=\frac{1}{2}\Rightarrow G_D=1$ & $W(\boldsymbol{1},\boldsymbol{1})_2\Rightarrow Q_{\rm em}=\frac{1}{3}$ & $\lambda_2^m=3, g=\frac{1}{2}\Rightarrow G_{\rm em}=3$   \\
            \hline
            $\boldsymbol{Z}_{6L}\times \boldsymbol{1}$ & $W(\boldsymbol{1},\boldsymbol{1},\boldsymbol{1})_{0,1}\Rightarrow Q_D=\frac{1}{2}$ & $\rho_2^m=2, k=1\Rightarrow G_D=2$ & $W(\boldsymbol{2},\boldsymbol{3})_{-1}\Rightarrow Q_{\rm em}=\frac{1}{3}$ & $\lambda_2^m=1, g=\frac{1}{6}\Rightarrow G_{\rm em}=1$   \\
            \hline
            $\boldsymbol{Z}_{3}\times \boldsymbol{Z}_{2H}$ & $W(\boldsymbol{1},\boldsymbol{1},\boldsymbol{1})_{0,2}\Rightarrow Q_D=1$ & $\rho_2^m=1, k=\frac{1}{2}\Rightarrow G_D=1$ & $W(\boldsymbol{1},\boldsymbol{3})_1\Rightarrow Q_{\rm em}=\frac{1}{6}$ & $\lambda_2^m=2, g=\frac{1}{3}\Rightarrow G_{\rm em}=2$   \\
            \hline
            $\boldsymbol{Z}_{6L}\times \boldsymbol{Z}_{2H}$ & $W(\boldsymbol{1},\boldsymbol{1},\boldsymbol{1})_{0,2}\Rightarrow Q_D=1$ & $\rho_2^m=1, k=\frac{1}{2}\Rightarrow G_D=1$ & $W(\boldsymbol{2},\boldsymbol{3})_{-1}\Rightarrow Q_{\rm em}=\frac{1}{3}$ & $\lambda_2^m=2, g=\frac{1}{6}\Rightarrow G_{\rm em}=1$   \\
            \hline
		\end{tabular} }
	\end{center}
\end{table}

$\Gamma =\boldsymbol{1} \times \boldsymbol{1}$: In the first step of symmetry breaking, the minimum dark electric charge ${{Q}_{D}}=\frac{1}{2}$ arises from the Wilson line in the representation ${{\left( \boldsymbol{1},\boldsymbol{1},\boldsymbol{1} \right)}_{0,1}}$.
In the second step of symmetry breaking, the minimal electric charge $Q_{\rm em}=\frac{1}{6}$ is derived from the Wilson line in the representation 
${{\left( \boldsymbol{1},\boldsymbol{1} \right)}_{1}}$.
The corresponding dark magnetic and magnetic charges are ${{G}_{D}}=2$ and $G_{\rm em}=6$, emerging from the ’t Hooft lines with $\rho _{2}^{m}=2$, $k=1$ and $\lambda _{2}^{m}=6$, $g=1$ respectively. 

$\Gamma ={\boldsymbol{Z}_{2L}}\times \boldsymbol{1}$: The minimum dark electric charge remains ${{Q}_{D}}=\frac{1}{2}$ from the Wilson line ${{\left( \boldsymbol{1},\boldsymbol{1},\boldsymbol{1} \right)}_{0,1}}$, and the minimal electric charge increases to $Q_{\rm em}=\frac{1}{3}$, derived from the Wilson line ${{\left( \boldsymbol{1},\boldsymbol{1} \right)}_{2}}$. The corresponding dark magnetic and magnetic charges are ${{G}_{D}}=2$ and $G_{\rm em}=3$, from the ’t Hooft lines with $\rho _{2}^{m}=2$, $k=1$ and $\lambda _{2}^{m}=3$, $g=\frac{1}{2}$ respectively.

$\Gamma =\boldsymbol{1} \times {\boldsymbol{Z}_{2H}}$: The minimum dark electric charge increases to ${{Q}_{D}}=1$, originating from the Wilson line ${{\left( \boldsymbol{1},\boldsymbol{1},\boldsymbol{1} \right)}_{0,2}}$, while the minimal electric charge returns to $Q_{\rm em}=\frac{1}{6}$, from the Wilson line ${{\left( \boldsymbol{1},\boldsymbol{1} \right)}_{1}}$. The corresponding dark magnetic and magnetic charges are ${{G}_{D}}=1$ and $G_{\rm em}=6$,  with the parameters of the 't Hooft lines $\rho _{2}^{m}=1$, $k=\frac{1}{2}$ and $\lambda _{2}^{m}=6$, $g=1$ respectively.

$\Gamma ={\boldsymbol{Z}_{3}} \times \boldsymbol{1}$: The minimum dark electric charge is ${{Q}_{D}}=\frac{1}{2}$ from the Wilson line ${{\left( \boldsymbol{1},\boldsymbol{1},\boldsymbol{1} \right)}_{0,1}}$, and the minimal electric charge $Q_{\rm em}=\frac{1}{6}$ comes from the Wilson line ${{\left( \boldsymbol{1},\boldsymbol{3} \right)}_{1}}$. The corresponding dark magnetic and magnetic charges are ${{G}_{D}}=2$ and $G_{\rm em}=2$, respectively. These dark magnetic and magnetic charges arise from the ’t Hooft line with $\rho _{2}^{m}=2$, $k=1$ and $\lambda _{2}^{m}=2$, $g=\frac{1}{3}$, respectively. Note that $Q_{\rm em} G_{\rm em} = \frac{1}{3}$, while $Q_D G_D = 1$.

$\Gamma ={\boldsymbol{Z}_{2L}}\times {\boldsymbol{Z}_{2H}}$: The minimum dark electric charge is ${{Q}_{D}}=1$, derived from the Wilson line ${{\left( \boldsymbol{1},\boldsymbol{1},\boldsymbol{1} \right)}_{0,2}}$. And the minimal electric charge $Q_{\rm em}=\frac{1}{3}$ is derived from the Wilson line ${{\left( \boldsymbol{1},\boldsymbol{1} \right)}_{2}}$. The corresponding dark magnetic and magnetic charges are ${{G}_{D}}=1$ and $G_{\rm em}=3$, from the ’t Hooft lines with $\rho _{2}^{m}=1$, $k=\frac{1}{2}$ and $\lambda _{2}^{m}=3$, $g=\frac{1}{2}$ respectively.

$\Gamma ={\boldsymbol{Z}_{6L}} \times \boldsymbol{1}$: The minimum dark electric charge is ${{Q}_{D}}=\frac{1}{2}$ from the Wilson line ${{\left( \boldsymbol{1},\boldsymbol{1},\boldsymbol{1} \right)}_{0,1}}$ and the minimal electric charge $Q_{\rm em}=\frac{1}{3}$ from the Wilson line ${{\left( \boldsymbol{2},\boldsymbol{3} \right)}_{-1}}$. The corresponding dark magnetic and magnetic charges are ${{G}_{D}}=2$ and $G_{\rm em}=1$, respectively. These arise from the ’t Hooft lines with $\rho _{2}^{m}=2$, $k=1$ and $\lambda _{2}^{m}=1$, $g=\frac{1}{6}$ respectively. Note that $Q_{\rm em} G_{\rm em} = \frac{1}{3}$, while $Q_D G_D = 1$.

$\Gamma ={\boldsymbol{Z}_{3}}\times {\boldsymbol{Z}_{2H}}$: The minimum dark electric charge is ${{Q}_{D}}=1$ from the Wilson line ${{\left( \boldsymbol{1},\boldsymbol{1},\boldsymbol{1} \right)}_{0,2}}$. And the minimal electric charge is $Q_{\rm em}=\frac{1}{6}$ from the Wilson line ${{\left( \boldsymbol{1},\boldsymbol{3} \right)}_{1}}$. The corresponding dark magnetic and magnetic charges are ${{G}_{D}}=1$ and $G_{\rm em}=2$, arise from the ’t Hooft lines with $\rho _{2}^{m}=1$, $k=\frac{1}{2}$ and $\lambda _{2}^{m}=2$, $g=\frac{1}{3}$ respectively. Note that $Q_{\rm em} G_{\rm em} = \frac{1}{3}$, while $Q_D G_D = 1$.

$\Gamma ={\boldsymbol{Z}_{6L}}\times {\boldsymbol{Z}_{2H}}$: The minimum dark electric charge is ${{Q}_{D}}=1$, derived from the Wilson line ${{\left( \boldsymbol{1},\boldsymbol{1},\boldsymbol{1} \right)}_{0,2}}$ and the minimal electric charge $Q_{\rm em}=\frac{1}{6}$ from the Wilson line ${{\left( \boldsymbol{2},\boldsymbol{3} \right)}_{-1}}$. The corresponding dark magnetic and magnetic charges are ${{G}_{D}}=1$ and $G_{\rm em}=1$, respectively. These arise from the ’t Hooft lines with $\rho _{2}^{m}=1$, $k=\frac{1}{2}$ and $\lambda _{2}^{m}=2$, $g=\frac{1}{6}$ respectively. Note that $Q_{\rm em} G_{\rm em} = \frac{1}{6}$, while $Q_D G_D = 1$.

We note that when $U(1)_{Y}$ and $SU(3)_C$ are combined and quotiented by a $\boldsymbol{Z}_{3}$, and also when $U(1)_{Y}$, $SU(2)_L$ and $SU(3)_C$ are combined and then quotiented by a $\boldsymbol{Z}_{6L}$, the resulting spectrum is no longer consistent with the electromagnetic Dirac quantization condition (DQC) $Q_{\rm em} G_{\rm em} \in \boldsymbol{Z}$. This inconsistency suggests that the minimal Dirac magnetic monopole does not consistent with the fractional charge of the quarks. Consequently, it implies that magnetic monopoles must also carry a color magnetic charge, reflecting by an additional term in the GDQC and thereby rendering the spectrum more consistent. This was pointed out some time ago in~\cite{Corrigan:1976wk} and demonstrated by Tong~\cite{Tong:2017oea} from the perspective of SM line operators. The monopole given by the GDQC is now aligned with the fractional charge of the quark. Indeed, as seen in Figs.~\ref{fig:LOex1case4}, \ref{fig:LOex1case6}, \ref{fig:LOex1case7} and \ref{fig:LOex1case8}, the associated 't Hooft lines do carry $SU(3)_C$ magnetic charges. This implies that in these cases, the gauge group of physical theory at lower energy is actually $U(3)_C$ rather than $U(1)_{\rm em} \times SU(3)_C$.

The results of the above minimal charges for the gauge quotient group in Eq.~(\ref{eq:SBex1}) are summarized in Table~\ref{tab:minimaicharge1}.

We next turn our attention to the quotient:
\be
\label{eq:SBex2}
G=\frac{U{{\left( 1 \right)}_{Y}}\times SU{{\left( 2 \right)}_{H}}\times SU{{\left( 3 \right)}_{C}}}{\Gamma_n}\times \frac{ U{{\left( 1 \right)}_{X}} \times SU{{\left( 2 \right)}_{L}} }{\Gamma_m} \; .
\ee
Although this quotient group differs slightly from the previous one with the two $SU(2)$ factors interchanged, this leads to subtle differences in their spectra.

When searching for the minimal (dark) electric charge and (dark) magnetic charge under different elements of $\Gamma$, they differ from the previous case. The main difference in this case compared to the previous one is that the two non-Abelian  $SU(2)_L$ and $SU(2)_H$ are separated from their respective $U(1)_Y$ and $U(1)_X$ first and reorganized later, since the symmetry breaking still follows the patterns of $SU(2)_L \times U(1)_Y \to U(1)_{\rm em}$ and $SU(2)_H \times U(1)_X \to U(1)_D$, with the constraints on charges imposed by the invariance under any group element of the center. In the calculation of the minimal (dark) electric charge, both the constraints from the center as well as  the Gell-Mann-Nishijima formulae Eq.~(\ref{eq:Qem}) and its hidden partner Eq.~(\ref{eq:Qdem}) must be taken into account. For the calculation of the minimal (dark) magnetic charge, the main focus is on the constraints from the center invariance and the deconfined magnetic charge Eq.~(\ref{eq:deconfinedcharges}). Nevertheless, the analysis for this quotient group in Eq.~(\ref{eq:SBex2}) is similar to previous case. Details are given as follows.

%Table VI
\begin{table}
	\begin{center}
		\caption{
        The minimal (dark) electric charge and minimal (dark) magnetic charge resulting from the two-step symmetry breaking of the gauge group in Eq.~(\ref{eq:SBex2}), along with the corresponding Wilson and 't Hooft lines.
        }
        \label{tab:minimaicharge2}
        \resizebox{\textwidth}{!}{
		\begin{tabular} {|c|c|c|c|c|}
			
            \hline
			$\Gamma$ & \text{\makecell[c] {Wilson lines and the\\corresponding minimal \\dark electric charge \\in the first step of \\ symmetry breaking} }
            & \text{\makecell [c] {'t Hooft lines and the \\ corresponding minimal \\dark magnetic charge \\in the first step of \\ symmetry breaking}}
            & \text{\makecell [c] {Wilson lines and the \\ corresponding minimal \\electric charge \\in the second step of \\ symmetry breaking}}
            & \text{\makecell [c] {'t Hooft lines and the \\corresponding minimal \\magnetic charge \\in the second step of \\ symmetry breaking}} \\
			\hline
            $\boldsymbol{1} \times \boldsymbol{1}$ & $W(\boldsymbol{1},\boldsymbol{1},\boldsymbol{1})_{0,1}\Rightarrow Q_D=\frac{1}{2}$ & $\rho_2^m=2, k=1\Rightarrow G_D=2$ & $W(\boldsymbol{1},\boldsymbol{1})_1\Rightarrow Q_{\rm em}=\frac{1}{6}$ & $\lambda_2^m=6, g=1\Rightarrow G_{\rm em}=6$   \\
            \hline
            $\boldsymbol{1} \times \boldsymbol{Z}_{2L}$ & $W(\boldsymbol{2},\boldsymbol{1},\boldsymbol{1})_{0,1}\Rightarrow Q_D=\frac{1}{2}$ & $\rho_2^m=2, k=1\Rightarrow G_D=2$ & $W(\boldsymbol{1},\boldsymbol{1})_1\Rightarrow Q_{\rm em}=\frac{1}{6}$ & $\lambda_2^m=6, g=1\Rightarrow G_{\rm em}=6$   \\
            \hline
            $\boldsymbol{Z}_{2H} \times \boldsymbol{1}$ & $W(\boldsymbol{1},\boldsymbol{1},\boldsymbol{1})_{0,1}\Rightarrow Q_D=\frac{1}{2}$ & $\rho_2^m=2, k=1\Rightarrow G_D=2$ & $W(\boldsymbol{1},\boldsymbol{1})_1\Rightarrow Q_{\rm em}=\frac{1}{6}$ & $\lambda_2^m=6, g=1\Rightarrow G_{\rm em}=6$   \\
            \hline
            $\boldsymbol{Z}_{3} \times \boldsymbol{1}$ & $W(\boldsymbol{1},\boldsymbol{1},\boldsymbol{1})_{0,1}\Rightarrow Q_D=1$ & $\rho_2^m=2, k=1\Rightarrow G_D=2$ & $W(\boldsymbol{1},\boldsymbol{3})_1\Rightarrow Q_{\rm em}=\frac{1}{6}$ & $\lambda_2^m=2, g=\frac{1}{3}\Rightarrow G_{\rm em}=2$   \\
            \hline
            $\boldsymbol{Z}_{2H} \times \boldsymbol{Z}_{2L}$ & $W(\boldsymbol{2},\boldsymbol{1},\boldsymbol{1})_{0,1}\Rightarrow Q_D=\frac{1}{2}$ & $\rho_2^m=2, k=1\Rightarrow G_D=2$ & $W(\boldsymbol{1},\boldsymbol{1})_1\Rightarrow Q_{\rm em}=\frac{1}{6}$ & $\lambda_2^m=6, g=1\Rightarrow G_{\rm em}=6$   \\
            \hline
            $\boldsymbol{Z}_{6H} \times \boldsymbol{1}$ & $W(\boldsymbol{1},\boldsymbol{1},\boldsymbol{1})_{0,1}\Rightarrow Q_D=\frac{1}{2}$ & $\rho_2^m=2, k=1\Rightarrow G_D=2$ & $W(\boldsymbol{1},\boldsymbol{3})_1\Rightarrow Q_{\rm em}=\frac{1}{6}$ & $\lambda_2^m=2, g=\frac{1}{3}\Rightarrow G_{\rm em}=2$   \\
            \hline
            $\boldsymbol{Z}_{3} \times \boldsymbol{Z}_{2L}$ & $W(\boldsymbol{2},\boldsymbol{1},\boldsymbol{1})_{0,1}\Rightarrow Q_D=\frac{1}{2}$ & $\rho_2^m=2, k=1\Rightarrow G_D=2$ & $W(\boldsymbol{1},\boldsymbol{3})_1\Rightarrow Q_{\rm em}=\frac{1}{6}$ & $\lambda_2^m=2, g=\frac{1}{3}\Rightarrow G_{\rm em}=2$   \\
            \hline
            $\boldsymbol{Z}_{6H} \times \boldsymbol{Z}_{2L}$ & $W(\boldsymbol{2},\boldsymbol{1},\boldsymbol{1})_{0,1}\Rightarrow Q_D=\frac{1}{2}$ & $\rho_2^m=2, k=1\Rightarrow G_D=2$ & $W(\boldsymbol{1},\boldsymbol{3})_1\Rightarrow Q_{\rm em}=\frac{1}{6}$ & $\lambda_2^m=2, g=\frac{1}{3}\Rightarrow G_{\rm em}=2$   \\
            \hline
		\end{tabular} }
	\end{center}
\end{table}

$\Gamma =\boldsymbol{1} \times \boldsymbol{1}$: The minimum dark electric charge ${{Q}_{D}}=\frac{1}{2}$ is derived from the Wilson line ${{\left( \boldsymbol{1},\boldsymbol{1},\boldsymbol{1} \right)}_{0,1}}$. The minimal electric charge $Q_{\rm em}=\frac{1}{6}$ originates from the Wilson line ${{\left( \boldsymbol{1},\boldsymbol{1} \right)}_{1}}$. The corresponding dark magnetic and magnetic charges are ${{G}_{D}}=2$ and $G_{\rm em}=6$, arising from the ’t Hooft lines with $\rho _{2}^{m}=2$, $k=1$ and $\lambda _{2}^{m}=6$, $g=1$ respectively, identical to those discussed above.

$\Gamma =\boldsymbol{1} \times {\boldsymbol{Z}_{2L}}$: The minimum dark electric charge ${{Q}_{D}}=\frac{1}{2}$ is derived from the Wilson line ${{\left( \boldsymbol{2},\boldsymbol{1},\boldsymbol{1} \right)}_{0,1}}$, and the minimal electric charge $Q_{\rm em}=\frac{1}{6}$ is obtained from the Wilson line ${{\left( \boldsymbol{1},\boldsymbol{1} \right)}_{1}}$. Since $h=z_{2}^{e}\bmod2$ now and $Q_{D}$ is only related to $x_{2}^{e}$ but not $z_{2}^{e}$, we can conveniently set $x_{2}^{e}=0$. The corresponding dark magnetic and magnetic charges are ${{G}_{D}}=2$ and $G_{\rm em}=6$, respectively. Although one may initially think $G_{\rm em}=3$ due to $2k={z}_{2}^{m}$ and ${z}_{2}^{m}=0,1\bmod{2}$, considering the conditions of Eq.~\eqref{eq:deconfinedcharges} and that ${x}_{2}^{m}$ can only take values $0\bmod2$, the magnetic charges must be even, hence the minimal $G_{\rm em}=6$. These charges arise from the ’t Hooft lines with $\rho _{2}^{m}=2$, $k=1$ and $\lambda _{2}^{m}=6$, $g=1$ respectively.

$\Gamma ={\boldsymbol{Z}_{2H}} \times \boldsymbol{1} $: The minimum dark electric charge ${{Q}_{D}}=\frac{1}{2}$ is derived from the Wilson line ${{\left( \boldsymbol{1},\boldsymbol{1},\boldsymbol{1} \right)}_{0,1}}$ and the minimal electric charge $Q_{\rm em}=\frac{1}{6}$ comes from the Wilson line ${{\left( \boldsymbol{1},\boldsymbol{1} \right)}_{1}}$. Since $q=x_{2}^{e}\bmod2$ and $Q_{\rm em}$ are only related to $z_{2}^{e}$, we can set $z_{2}^{e}=0$. The corresponding dark magnetic and magnetic charges are ${{G}_{D}}=2$ and $G_{\rm em}=6$, respectively, because the magnetic charges are required to be even by Eq.~\eqref{eq:deconfinedcharges}. These charges arise from the ’t Hooft lines with $\rho _{2}^{m}=2$, $k=1$ and $\lambda _{2}^{m}=6$, $g=1$ respectively.

$\Gamma ={\boldsymbol{Z}_{3}} \times \boldsymbol{1} $: The minimum dark electric charge ${{Q}_{D}}=1$ is derived from the Wilson line ${{\left( \boldsymbol{1},\boldsymbol{1},\boldsymbol{1} \right)}_{0,1}}$ and the minimal electric charge $Q_{\rm em}=\frac{1}{6}$ comes from the Wilson line ${{\left( \boldsymbol{1},\boldsymbol{3} \right)}_{1}}$. The corresponding dark magnetic and magnetic charges are ${{G}_{D}}=2$ and $G_{\rm em}=2$, respectively. These arise from the ’t Hooft lines with $\rho _{2}^{m}=2$, $k=1$ and $\lambda _{2}^{m}=2$, $g=\frac{1}{3}$, respectively. Note that $Q_{\rm em} G_{\rm em} = \frac{1}{3}$, while $Q_D G_D = 2$.

$\Gamma ={\boldsymbol{Z}_{2H}}\times {\boldsymbol{Z}_{2L}}$: The minimum dark electric charge ${{Q}_{D}}=\frac{1}{2}$ is originated from the Wilson line ${{\left( \boldsymbol{2},\boldsymbol{1},\boldsymbol{1} \right)}_{0,1}}$. And the minimal electric charge $Q_{\rm em}=\frac{1}{6}$ is derived from the Wilson line ${{\left( \boldsymbol{1},\boldsymbol{1} \right)}_{1}}$. The corresponding dark magnetic and magnetic charges are ${{G}_{D}}=2$ and $G_{\rm em}=6$, respectively. These charges arise from the ’t Hooft lines with $\rho _{2}^{m}=2$, $k=1$ and $\lambda _{2}^{m}=6$, $g=1$ respectively.

$\Gamma ={\boldsymbol{Z}_{6H}} \times \boldsymbol{1}$: The minimum dark electric charge ${{Q}_{D}}=\frac{1}{2}$ arises from the Wilson line ${{\left( \boldsymbol{1},\boldsymbol{1},\boldsymbol{1} \right)}_{0,1}}$. And the minimal electric charge $Q_{\rm em}=\frac{1}{6}$ is derived from the Wilson line ${{\left( \boldsymbol{1},\boldsymbol{3} \right)}_{1}}$. This results from $q=3x_{2}^{e}-2z_{3}^{e}\bmod6$, where $Q_{\rm em}$ is only dependent on $z_{2}^{e}$ but not $x_{2}^{e}$, allowing us to set $z_{2}^{e}=0$. The corresponding dark magnetic and magnetic charges are ${{G}_{D}}=2$ and $G_{\rm em}=2$, respectively, since $G_{\rm em}$ is required to be even according to Eq.~\eqref{eq:deconfinedcharges}. These charges arise from the ’t Hooft lines with $\rho _{2}^{m}=2$, $k=1$ and $\lambda _{2}^{m}=2$, $g=\frac{1}{3}$, respectively. Note that $Q_{\rm em} G_{\rm em} = \frac{1}{3}$, while $Q_D G_D = 1$.

$\Gamma ={\boldsymbol{Z}_{3}}\times {\boldsymbol{Z}_{2L}}$: The minimum dark electric charge ${{Q}_{D}}=\frac{1}{2}$ is originated from the Wilson line ${{\left( \boldsymbol{2},\boldsymbol{1},\boldsymbol{1} \right)}_{0,1}}$. And the minimal electric charge $Q_{\rm em}=\frac{1}{6}$ is derived from the Wilson line ${{\left( \boldsymbol{1},\boldsymbol{3} \right)}_{1}}$. The corresponding dark magnetic and magnetic charges are ${{G}_{D}}=2$ and $G_{\rm em}=2$, respectively. These charges arise from the ’t Hooft lines with $\rho _{2}^{m}=2$, $k=1$ and $\lambda _{2}^{m}=2$, $g=\frac{1}{3}$ respectively. Note that $Q_{\rm em} G_{\rm em} = \frac{1}{3}$, while $Q_D G_D = 1$.

$\Gamma ={\boldsymbol{Z}_{6H}}\times {\boldsymbol{Z}_{2L}}$: The minimum dark electric charge ${{Q}_{D}}=\frac{1}{2}$ arises from the Wilson line ${{\left( \boldsymbol{2},\boldsymbol{1},\boldsymbol{1} \right)}_{0,1}}$, and the minimal electric charge $Q_{\rm em}=\frac{1}{6}$ is derived from the Wilson line ${{\left( \boldsymbol{1},\boldsymbol{3} \right)}_{1}}$. The corresponding dark magnetic and magnetic charges are ${{G}_{D}}=2$ and $G_{\rm em}=2$, respectively, arising from the ’t Hooft lines with $\rho _{2}^{m}=2$, $k=1$ and $\lambda _{2}^{m}=2$, $g=\frac{1}{3}$ respectively. Note that $Q_{\rm em} G_{\rm em} = \frac{1}{3}$, while $Q_D G_D = 1$.

Again, in this case, we find that if $\Gamma$ contains the center of $SU(3)_C$, the electromagnetic DQC is violated. The above results of the minimal charges for the quotient in Eq.~(\ref{eq:SBex2}) are summarized in Table~\ref{tab:minimaicharge2}.

Now, let us continue with the exotic quotient whose line operators have been analyzed in Section~\ref{sec:lineoperators}:
\be
\label{eq:SBex3}
G = \frac{U(1)_V \times SU(2)_L \times SU(3)_C}{\Gamma_n} \times \frac{ U(1)_A \times SU(2)_H }{\Gamma_m} \; .
\ee

While Nature may not be kind enough to select this configuration at her unbroken phase, we have explored it for completeness. The GDQC to be used for this quotient group has been given previously in Eq.~(\ref{eq:DiracConditionAlt}). 
We note that 
$g_+ = \frac{1}{\sqrt 2} ( k\cos\alpha  + g \sin\alpha )$ 
and $g_-=  \frac{1}{\sqrt 2}( -k\sin \alpha  + g \cos\alpha )$ with an arbitrary angle $\alpha$ are possible as well, should one use $q_+=\sqrt{2}(h\cos\alpha+q\sin\alpha)$ and $q_-=\sqrt{2}(-h\sin\alpha+q\cos\alpha)$ to begin with. Requiring \( q_\pm \) to be quantized just like $q$ and $h$ are, 
one must have \( \alpha = \frac{\pi}{4} \), leading to \( g_\pm = \frac{1}{2}(g \pm k) \).

%%%

\begin{table}
	\begin{center}
		\caption{
        The minimal (dark) electric charge resulting from the two-step symmetry breaking of the gauge group in Eq.~(\ref{eq:SBex3}), along with the corresponding Wilson and 't Hooft lines.
        }
        \label{tab:minimaicharge3}
        \resizebox{\textwidth}{!}{
		\begin{tabular} {|c|c|c|c|c|}
			
            \hline
			$\Gamma$ & \text{\makecell[c] {Wilson lines and the\\corresponding minimal \\dark electric charge \\in the first step of \\ symmetry breaking} }
            & \text{\makecell [c] {'t Hooft lines and the \\ corresponding minimal \\dark magnetic charge \\in the first step of \\ symmetry breaking}}
            & \text{\makecell [c] {Wilson lines and the \\ corresponding minimal \\electric charge \\in the second step of \\ symmetry breaking}}
            & \text{\makecell [c] {'t Hooft lines and the \\corresponding minimal \\magnetic charge \\in the second step of \\ symmetry breaking}} \\
			\hline
            
            $\boldsymbol{1} \times \boldsymbol{1}$ & $W(\boldsymbol{1},\boldsymbol{1},\boldsymbol{1})_{0,1}\Rightarrow Q_D=\frac{1}{2}$  & 
            $\rho_2^m=2,k=1\Rightarrow G_D=2$ &
            $W(\boldsymbol{1},\boldsymbol{1})_1\Rightarrow Q_{\rm em}=\frac{1}{6}$ &
            $\lambda_2^m=6,g=1\Rightarrow G_{\rm em}=6$ \\
            \hline
            $\boldsymbol{Z}_{2L} \times \boldsymbol{1}$ & $W(\boldsymbol{1},\boldsymbol{1},\boldsymbol{1})_{0,1}\Rightarrow Q_D=\frac{1}{2}$  & 
            $\rho_2^m=2,k=1\Rightarrow G_D=2$ &
            $W(\boldsymbol{1},\boldsymbol{1})_1\Rightarrow Q_{\rm em}=\frac{1}{6}$   &
            $\lambda_2^m=6,g=1\Rightarrow G_{\rm em}=6$\\
            \hline
            
            $\boldsymbol{1} \times \boldsymbol{Z}_{2H}$ & $W(\boldsymbol{1},\boldsymbol{1},\boldsymbol{1})_{0,1}\Rightarrow Q_D=\frac{1}{2}$ & 
            $\rho_2^m=2,k=1\Rightarrow G_D=2$ &
            $W(\boldsymbol{1},\boldsymbol{1})_1\Rightarrow Q_{\rm em}=\frac{1}{6}$   &
            $\lambda_2^m=6,g=1\Rightarrow G_{\rm em}=6$\\
            \hline
            
            $\boldsymbol{Z}_{3} \times \boldsymbol{1}$ & $W(\boldsymbol{1},\boldsymbol{1},\boldsymbol{3})_{0,1}\Rightarrow Q_D=\frac{1}{2}$ & 
            $\rho_2^m=2,k=1\Rightarrow G_D=2$ &
            $W(\boldsymbol{1},\boldsymbol{3})_1\Rightarrow Q_{\rm em}=\frac{1}{6}$   &
            $\lambda_2^m=6,g=1\Rightarrow G_{\rm em}=6$\\
            \hline
            $\boldsymbol{Z}_{2L} \times \boldsymbol{Z}_{2H}$ & $W(\boldsymbol{1},\boldsymbol{1},\boldsymbol{1})_{0,1}\Rightarrow Q_D=\frac{1}{2}$ & 
            $\rho_2^m=1,k=\frac{1}{2}\Rightarrow G_D=1$ &
            $W(\boldsymbol{1},\boldsymbol{1})_1\Rightarrow Q_{\rm em}=\frac{1}{6}$   &
            $\lambda_2^m=3,g=\frac{1}{2}\Rightarrow G_{\rm em}=3$\\
            \hline
            $\boldsymbol{Z}_{6L} \times \boldsymbol{1}$ & $W(\boldsymbol{1},\boldsymbol{2},\boldsymbol{3})_{0,-1}\Rightarrow Q_D=\frac{1}{2}$ & 
            $\rho_2^m=2,k=1\Rightarrow G_D=2$ &
            $W(\boldsymbol{2},\boldsymbol{3})_{-3}\Rightarrow Q_{\rm em}=\frac{1}{2}$   &
            $\lambda_2^m=6,g=1\Rightarrow G_{\rm em}=6$\\
            \hline
            
            $\boldsymbol{Z}_{3} \times \boldsymbol{Z}_{2H}$ & $W(\boldsymbol{1},\boldsymbol{1},\boldsymbol{3})_{0,1}\Rightarrow Q_D=\frac{1}{2}$ & 
            $\rho_2^m=2,k=1\Rightarrow G_D=2$ &
            $W(\boldsymbol{1},\boldsymbol{3})_1\Rightarrow Q_{\rm em}=\frac{1}{6}$   &
            $\lambda_2^m=6,g=1\Rightarrow G_{\rm em}=6$\\
            \hline
            $\boldsymbol{Z}_{6L} \times \boldsymbol{Z}_{2H}$ & $W(\boldsymbol{1},\boldsymbol{2},\boldsymbol{3})_{0,-1}\Rightarrow Q_D=\frac{1}{2}$ & 
            $\rho_2^m=1,k=\frac{1}{2}\Rightarrow G_D=1$ &
            $W(\boldsymbol{2},\boldsymbol{3})_{-3}\Rightarrow Q_{\rm em}=\frac{1}{2}$   &
            $\lambda_2^m=3,g=\frac{1}{2}\Rightarrow G_{\rm em}=3$\\
            \hline
		\end{tabular} }
	\end{center}
\end{table}

$\Gamma =\boldsymbol{1} \times \boldsymbol{1}$: The minimum dark electric charge ${{Q}_{D}}=\frac{1}{2}$ originates from the Wilson line ${{\left( \boldsymbol{1},\boldsymbol{1},\boldsymbol{1} \right)}_{0,1}}$. And the minimal electric charge $Q_{\rm em}=\frac{1}{6}$ is derived from the Wilson line ${{\left( \boldsymbol{1},\boldsymbol{1} \right)}_{1}}$. The corresponding dark magnetic and magnetic charges are ${{G}_{D}}=2$ and $G_{\rm em}=6$, respectively, arising from the ’t Hooft lines with $\rho _{2}^{m}=2$, $k=1$ and $\lambda _{2}^{m}=6$, $g=1$ respectively.

$\Gamma ={\boldsymbol{Z}_{2L}} \times \boldsymbol{1}$: The hypercharges are constrained by $q+h=z_2^e\bmod2$. The minimum dark electric charge ${{Q}_{D}}=\frac{1}{2}$ arises from the Wilson line ${{\left( \boldsymbol{1},\boldsymbol{1},\boldsymbol{1} \right)}_{0,1}}$. The minimal electric charge $Q_{\rm em}=\frac{1}{6}$ is derived from the Wilson line ${{\left( \boldsymbol{1},\boldsymbol{1} \right)}_{1}}$. This derived from $q+h=1+1=0\bmod2$.  The corresponding dark magnetic and magnetic charges are ${{G}_{D}}=2$ and $G_{\rm em}=6$, respectively, arising from the ’t Hooft lines with $\rho _{2}^{m}=2$, $k=1$ and $\lambda _{2}^{m}=6$, $g=1$ respectively.

$\Gamma =\boldsymbol{1} \times {\boldsymbol{Z}_{2H}}$: The hypercharges are constrained by $q-h=x_2^e\bmod2$. The minimum dark electric charge ${{Q}_{D}}=\frac{1}{2}+0=\frac{1}{2}$ is derived from the Wilson line ${{\left( \boldsymbol{1},\boldsymbol{1},\boldsymbol{1} \right)}_{0,1}}$. And the minimal electric charge $Q_{\rm em}=\frac{1}{6}+0=\frac{1}{6}$ comes from the Wilson line ${{\left( \boldsymbol{1},\boldsymbol{1} \right)}_{1}}$. The hypercharges follow from $q-h=1-1=0\bmod2$. The corresponding dark magnetic and magnetic charges are ${{G}_{D}}=2$ and $G_{\rm em}=6$, respectively, arising from the ’t Hooft lines with $\rho _{2}^{m}=2$, $k=1$ and $\lambda _{2}^{m}=6$, $g=1$ respectively.

$\Gamma ={\boldsymbol{Z}_{3}} \times \boldsymbol{1}$: The hypercharges are constrained by $q+h=z_3^e\bmod3$. The minimum dark electric charge ${{Q}_{D}}=\frac{1}{2}+0=\frac{1}{2}$ originates from the Wilson line ${{\left( \boldsymbol{1},\boldsymbol{1},\boldsymbol{3} \right)}_{0,1}}$. The minimal electric charge $Q_{\rm em}=\frac{1}{6}+0=\frac{1}{6}$ is derived from the Wilson line ${{\left( \boldsymbol{1},\boldsymbol{3} \right)}_{1}}$. The hypercharges follow from $q+h=1+1=0\bmod3$. The corresponding dark magnetic and magnetic charges are ${{G}_{D}}=2$ and $G_{\rm em}=6$, respectively, arising from the ’t Hooft lines with $\rho _{2}^{m}=2$, $k=1$ and $\lambda _{2}^{m}=6$, $g=1$ respectively. Note that despite the presence of $\boldsymbol{Z}_3$, both $Q_{\rm em} G_{\rm em} = 1$ and $Q_D G_D = 1$ are integers.

$\Gamma ={\boldsymbol{Z}_{2L}}\times {\boldsymbol{Z}_{2H}}$: The hypercharges are constrained by $q+h=z_2^e\bmod2$ and $q-h=x_2^e\bmod2$. The minimum dark electric charge ${{Q}_{D}}=\frac{1}{2}$ arises from the Wilson line ${{\left( \boldsymbol{1},\boldsymbol{1},\boldsymbol{1} \right)}_{0,1}}$. The minimal electric charge $Q_{\rm em}=\frac{1}{6}$ is derived from the Wilson line ${{\left( \boldsymbol{1},\boldsymbol{1} \right)}_{1}}$. The corresponding dark magnetic and magnetic charges are ${{G}_{D}}=1$ and $G_{\rm em}=3$, respectively, arising from the ’t Hooft lines with $\rho _{2}^{m}=1$, $k=\frac{1}{2}$ and $\lambda _{2}^{m}=3$, $g=\frac{1}{2}$ respectively. Note that $Q_{\rm em} G_{\rm em} = \frac{1}{2}$, while $Q_D G_D = \frac{1}{2}$.

$\Gamma ={\boldsymbol{Z}_{6L}} \times \boldsymbol{1}$: The minimum dark electric charge ${{Q}_{D}}=\frac{1}{2}$ is derived from the Wilson line ${{\left( \boldsymbol{1},\boldsymbol{2},\boldsymbol{3} \right)}_{0,-1}}$. The minimal electric charge $Q_{\rm em}=-\frac{3}{6}+1=\frac{1}{2}$ is derived from the Wilson line ${{\left( \boldsymbol{2},\boldsymbol{3} \right)}_{-3}}$. This is because $q+h=3z_{2}^{e}-2z_{3}^{e}\bmod6$, which allows us to set $q+h=-3-1=-4\bmod6$. The corresponding dark magnetic and magnetic charges are ${{G}_{D}}=2$ and $G_{\rm em}=6$, respectively, arising from the ’t Hooft lines with $\rho _{2}^{m}=2$, $k=1$ and $\lambda _{2}^{m}=6$, $g=1$ respectively. Note that despite the presence of $\boldsymbol{Z}_6$, both $Q_{\rm em} G_{\rm em} = 3$ and $Q_D G_D = 1$ are integers.

$\Gamma ={\boldsymbol{Z}_{3}}\times {\boldsymbol{Z}_{2H}}$: The minimum dark electric charge ${{Q}_{D}}=\frac{1}{2}$ arises from the Wilson line ${{\left( \boldsymbol{1},\boldsymbol{1},\boldsymbol{3} \right)}_{0,1}}$. The minimal electric charge $Q_{\rm em}=\frac{1}{6}$ is derived from the Wilson line ${{\left( \boldsymbol{1},\boldsymbol{3} \right)}_{1}}$. This is because $q+h=z_{3}^{e}\bmod3$ and $q-h=x_2^e\bmod2$, allowing us to set $q+h=1+1=0\bmod3$ and $q-h=1-1=0\bmod2$. The corresponding dark magnetic and magnetic charges are ${{G}_{D}}=2$ and $G_{\rm em}=6$, respectively, arising from the ’t Hooft lines with $\rho _{2}^{m}=2$, $k=1$ and $\lambda _{2}^{m}=6$, $g=1$ respectively. Note that despite the presence of $\boldsymbol{Z}_3$, both $Q_{\rm em} G_{\rm em} = 1$ and $Q_D G_D = 1$ are integers.

$\Gamma ={\boldsymbol{Z}_{6L}}\times {\boldsymbol{Z}_{2H}}$: The hypercharges are now restricted by $q+h=3z_2^e-2z_{3}^{e}\bmod6$ and $q-h=x_2^e\bmod2$, allowing us to set $q+h=-3-1=-4\bmod6$ and $q-h=-3+1=0\bmod2$. The minimum dark electric charge ${{Q}_{D}}=\frac{1}{2}$ is derived from the Wilson line ${{\left( \boldsymbol{1},\boldsymbol{2},\boldsymbol{3} \right)}_{0,-1}}$. The minimal electric charge $Q_{\rm em}=\frac{1}{2}$ is derived from the Wilson line ${{\left( \boldsymbol{2},\boldsymbol{3} \right)}_{-3}}$. The corresponding dark magnetic and magnetic charges are ${{G}_{D}}=1$ and $G_{\rm em}=3$, respectively, arising from the ’t Hooft lines with $\rho _{2}^{m}=1$, $k=\frac{1}{2}$ and $\lambda _{2}^{m}=3$, $g=\frac{1}{2}$ respectively. Note that $Q_{\rm em} G_{\rm em} = \frac{3}{2}$, while $Q_D G_D = \frac{1}{2}$.

We note that in the cases of $\boldsymbol{Z}_{2L} \times \boldsymbol{Z}_{2H}$ and $\boldsymbol{Z}_{6L} \times \boldsymbol{Z}_{2H}$, both the electromagnetic DQC and its dark version are violated.
The results of the above minimal charges for the gauge quotient group in Eq.~(\ref{eq:SBex3}) are summarized in Table~\ref{tab:minimaicharge3}.

Besides the above three cases of quotients $G$ studied in this section, others can be analyzed in a similar manner and we will not explore further here.

We now turn our attention to the $\theta$-terms before and after the electroweak symmetry breaking in minimal G2HDM.

We denote ${B}_{\mu \nu}$, ${F}^{a}_{L\mu \nu}$, ${G}^{a}_{\mu \nu}$ as the field strength for $U(1)_{\tilde{Y}}$, $SU(2)_{L}$ and $SU(3)_C$, respectively, and ${C}_{\mu \nu}$ and ${F}^{a}_{H\mu \nu}$ as the field strength for $U(1)_{\tilde{X}}$ and $SU(2)_{H}$. The $\theta$-term of these fields before symmetries breaking is given by
\begin{equation}
\label{Stheta}
	\begin{aligned}
		{{S}_{\theta }} 
        & =\frac{{\tilde{\theta }_Y}}{16{{\pi }^{2}}}\frac{g_{1}^{2}}{36}\int{{}^{\star }{{B}^{\mu \nu }}{{B}_{\mu \nu }}}+\frac{{{\theta }_{2L}}}{16{{\pi }^{2}}}\frac{g_{2}^{2}}{2}\int{{}^{\star }{{F}^{a\mu \nu }_L}F_{L\mu \nu }^{a}}+\frac{{{\theta }_{3}}}{16{{\pi }^{2}}}\frac{g_{3}^{2}}{2}\int{{}^{\star }{{G}^{a\mu \nu }}G_{\mu \nu }^{a}} \\
		& +\frac{{\tilde{\theta }_X}}{16{{\pi }^{2}}}\frac{g_{X}^{2}}{4}\int{{}^{\star }{{C}^{\mu \nu }}{{C}_{\mu \nu }}}
        +\frac{{{\theta }_{2H}}}{16{{\pi }^{2}}}\frac{g_{H}^{2}}{2}\int{{}^{\star }{{F}^{a\mu \nu }_H}F_{H\mu \nu }^{a}} \; , \\
	\end{aligned}
\end{equation}
where $g_i \, (i=1,2,3)$, $g_X$ and $g_H$ are the gauge coupling constants. Recall that the gauge field of the SM $U(1)_{Y}$ hypercharge is normalized such that ${Y}\in {\boldsymbol{Z}}/{6}\; $, and the gauge field of the dark $U(1)_{X}$ hypercharge is normalized such that ${X} \in {\boldsymbol{Z}}/{2}$.

After symmetry breakings, Eq.~(\ref{Stheta}) reduces to 
\begin{equation}
\label{SthetaSB}
	{{S}_{\theta }} =\frac{{{e}^{2}}}{16{{\pi }^{2}}}\frac{\tilde{\theta }_{Y}+18{{\theta }_{2L}}}{36}\int{\left( {}^{\star }F_{\rm em}^{\mu \nu }{{F}_{{\rm em}, \mu \nu }} \right)}+\cdots +\frac{{e_{D}^{2}}}{16{{\pi }^{2}}}\frac{\tilde{\theta }_X + 2{{\theta }_{2H}}}{4}\int{\left( ^{\star }F_{\rm dem}^{\mu \nu }{{F}_{{\rm dem}, \mu \nu }} \right)}+\cdots  \; .\\ 	
\end{equation}
Here, $e=g_1 g_2/\sqrt{g_1^2 + g_2^2}$ ($e_D = g_X g_H / \sqrt{ g_X^2 + g_H^2}$) denotes the (dark) electric charge and  $F_{{\rm em}, \mu \nu}$ ($F_{{\rm dem}), \mu \nu}$ is the (dark) electromagnetic field strength.

We observe that besides the electromagnetic $\theta$-angle, $\theta_{\rm em} = \frac{\tilde{\theta}_Y + 18 \theta_{2L}}{36}$, as first derived in~\cite{Tong:2017oea}, we also have the dark electromagnetic $\theta$-angle, $\theta_{\rm dem} = \frac{\tilde{\theta}_X + 2 \theta_{2H}}{4}$.
Both combinations are precisely those that cannot be rotated by chiral rotation as we demonstrated earlier.
Just like in the case of SM~\cite{Tong:2017oea}, this is of no coincidence: it follows from the 't Hooft anomaly matching conditions,
and the fact that both $H(\boldsymbol{2},\boldsymbol{2},\boldsymbol{1})_{3,-1}$ and $H(\boldsymbol{1},\boldsymbol{2},\boldsymbol{1})_{0,1}$ provide masses to all the SM and hidden fermions in minimal G2HDM. They are physical and cannot be transformed away.

The admissible ranges for ${\theta}_{\rm em}$ and ${\theta}_{\rm dem}$ resemble non-breaking $\theta$-angles, and both depend on the choices of $\Gamma$.

For electromagnetism, regardless of the choice of quotient group $G$, as long as $\Gamma$ contains either $\boldsymbol{1}$ or $\boldsymbol{Z}_{2L}$, the gauge group of electromagnetism is $U(1)_{\rm em}\times SU(3)_C$, with $\theta_{\rm em} \in \left[ 0,2\pi Q_{\rm em}^2 \right)$.
In contrast, when $\Gamma$ contains the element $\boldsymbol{Z}_{3}$ or $\boldsymbol{Z}_{2L}\times\boldsymbol{Z}_{3}$, the unbroken gauge group at low energy becomes $U(3)_C$, and $\theta_{\rm em} \in \left[ 0,18\pi \cdot Q_{\rm em}^2 \right)$. In particular, when $\Gamma$ contains $\boldsymbol{Z}_{2L}\times\boldsymbol{Z}_{3}$, the periodicity of $\theta_{\rm em}$ is $\left[0,2\pi\right)$, which coincides with the minimum fractional quark charge of $\frac{1}{3}$.

On the other hand, for dark electromagnetism, regardless of the choice of $G$, as long as $\Gamma$ contains either $\boldsymbol{1}$ or $\boldsymbol{Z}_{2H}$, the gauge group of dark electromagnetism is $U(1)_{D}$ and $\theta_{\rm dem} \in \left[ 0,2\pi \cdot Q^2_{D} \right)$, where $Q_{D}$ is the minimum dark charge.

These observations imply that when $\Gamma$ contains $\boldsymbol{1}, \boldsymbol{Z}_{2L}, \boldsymbol{Z}_{2H}$ or $\boldsymbol{Z}_{2L} \times \boldsymbol{Z}_{2H}$, the unbroken gauge group at low energy remains $U(1)_{\rm em}\times SU(3)_C$. However, when $\Gamma$ contains $\boldsymbol{Z}_{3}, \boldsymbol{Z}_{2L} \times \boldsymbol{Z}_{3}, \boldsymbol{Z}_{2H} \times \boldsymbol{Z}_{3}$ or $\boldsymbol{Z}_{2L} \times \boldsymbol{Z}_{2H} \times \boldsymbol{Z}_{3}$, the low energy gauge group is always $U(3)_C$. Meanwhile, the dark electromagnetism gauge group remains consistently $U(1)_D$. These results are derived from the symmetry-breaking patterns discussed above.

\section{Summary}\label{sec:summary}

In this work, we have extended the analysis of line operators in the SM to the minimal G2HDM. We enumerate possible quotient groups \( G = \tilde{G}/\Gamma \) for the minimal G2HDM, with the covering group \(\tilde{G}\) specified in Eq. (\ref{eq:G2HDMcoveringgroup}) and quotient patterns provided in Eq. (\ref{eq:quotientpatterns}). Notably, experimental validation is required to determine the correct gauge group in practice. As in the SM, the global structure of the G2HDM is influenced by the choice of the discrete group \(\Gamma\) used for quotienting. We determine the spectra of Wilson, 't Hooft, and dyonic line operators for several examples within the quotient patterns and their associated \(\Gamma\)s. With the inclusion of an additional hidden sector \( U(1)_X \times SU(2)_H \) — a dark replica of the SM electroweak gauge group — and a minimal matter content, the resulting spectra are significantly richer than those in the SM, as demonstrated in Section~\ref{sec:LO-G2HDM}.

The Witten effect was also incorporated by including the five \( CP \)-violating \(\theta\) terms for the G2HDM gauge group. Similar to the SM, the Witten effect transforms all 't Hooft lines into dyonic lines. Combined with non-perturbative and topological effects, we examined the impact of these terms on the periodicity of the \(\theta\)-angles and their specific values for \( CP \)-invariant theories in detail for the two quotient cases in Eqs.~(\ref{eq:CaseAex1X}) and (\ref{eq:CaseCex1}). The rich features of the periodicity and \( CP \)-invariant theories are summarized in Tables~\ref{tab:thetas4ex1} and~\ref{tab:thetas4ex2}.

As is well-known, after electroweak symmetry breaking in the SM, the \(\theta_{2L}\) term for \( SU(2)_L \) can be eliminated through by a chiral transformation, owing to the anomalous \( B+L \) global current. However, as noted in~\cite{Tong:2017oea}, a residual \(\theta_{\rm em}\) term remains for QED, which is a linear combination of \(\theta_{2L}\) and \(\tilde{\theta}_Y\), given by \(\theta_{\rm em} = (\tilde{\theta}_Y + 18 \theta_{2L})/36\). The physical implications of a nonzero QED $\theta_{\rm em}$ term at low energies were highlighted by Hsu~\cite{Hsu:2010jm,Hsu:2011sx}. Analogously, in minimal G2HDM, we find that the \(\theta_{2H}\) term for \( SU(2)_{H} \) can be similarly rotated away via a corresponding transformation. This leaves behind a residual \(\theta_{\rm dem} = (\tilde{\theta}_X + 2 \theta_{2H})/4\) term for the unbroken dark QED sector. These angular relations are consistent with considerations based on mixed anomalies of the global baryon and lepton number currents.

After the symmetry breaking in different quotient gauge groups, we found many minimal (dark) electric and (dark) magnetic charges. For any gauge group with $U(1)_Y$ and $U(1)_X$, when $\Gamma$ 
contains a factor of $\boldsymbol{Z}_3$, $\boldsymbol{Z}_{6L}$ or $\boldsymbol{Z}_{6H}$ (more precisely, when there is a $SU(3)_C$ center charge), the minimal electric and magnetic charges do {\it not} satisfy the electromagnetic DQC $Q_{\rm em} G_{\rm em} \in \boldsymbol{Z}$. The Dirac monopole is inconsistent with the fractional charge of the quarks. Therefore, the existence of colored magnetic charge is required to provide addition contribution to the pure electromagnetic DQC, which must be replaced by the GDQC like Eq.~(\ref{eq:DiracCondition}), depending on the quotient center~
\cite{Corrigan:1976wk}. 
However, for other cases of $\Gamma$, the minimal electric and magnetic charges do satisfy it. 
The minimal dark electric and magnetic charges do satisfy $Q_DG_D\in \boldsymbol{Z}$ for the hidden $U(1)_D$ for any $\Gamma$.
On the other hand, when the gauge group contains $U(1)_V$ and $U(1)_A$ where hypercharge and its dark counterpart are mixed, for $\Gamma$ contains either 
$\boldsymbol{Z}_{2L} \times \boldsymbol{Z}_{2H}$, $\boldsymbol{Z}_{6L} \times \boldsymbol{Z}_{2H}$,
or $\boldsymbol{Z}_{6H} \times \boldsymbol{Z}_{2L}$, 
the possible values of charge $h$ and $q$ are more diverse and flexible. As a result, when searching for the minimal charges, inconsistencies with both the electromagnetic DQC and its dark version may also arise. 
Tables~\ref{tab:minimaicharge1},
\ref{tab:minimaicharge2} and
\ref{tab:minimaicharge3} summarize the minimal charges and their Wilson and 't Hooft lines we found for the quotient groups 
Eqs.~(\ref{eq:SBex1}), (\ref{eq:SBex2}) and (\ref{eq:SBex3}) respectively. 

While we have explored the global properties and distinctions of all potential theories in the minimal G2HDM in detail, these discussions remain theoretical and do not imply the experimental observability of these differences, for instance, at the LHC.

Nevertheless, some general observations in parallel to those in the SM~\cite{Tong:2017oea} can be made.

\begin{table}
	\begin{center}
		\caption{
			The $\Gamma$ required for neutral quarks (electrically neutral colored Wilson lines) in the various quotient gauge groups in case (A). 
		}
		\label{tab:neutralquarksA}
		\resizebox{\textwidth}{!}{
			\begin{tabular} {|c|c|c|c|c|c|c|}
				
				\hline
				\multirow{2}*{\quad} & \multicolumn{2}{|c|}{$Q_{\rm em}=0,Q_D\neq0$} & \multicolumn{2}{|c|}{$Q_D=0,Q_{\rm em}\neq0$} & \multicolumn{2}{|c|}{$Q_{\rm em}=0,Q_D=0$} \\
				\cline{2-7}& \text{Quark} & $\Gamma$ & \text{Quark} & $\Gamma$ & \text{Quark} & $\Gamma$  \\
				
				\hline
				$\frac{U{{\left( 1 \right)}_{Y}}\times SU{{\left( 2 \right)}_{L}}\times SU{{\left( 3 \right)}_{C}}}{\Gamma_p}\times \frac{U{{\left( 1 \right)}_{X}} \times SU{{\left( 2 \right)}_{H}}}{\Gamma_m}\;$ &
				$W(\boldsymbol{1}\,\boldsymbol{2},\boldsymbol{3})_{0,h}$ &
				$\makecell[c] {\boldsymbol{1}\times\boldsymbol{1}, \boldsymbol{Z}_{2L}\times\boldsymbol{1}, \\ \boldsymbol{1}\times\boldsymbol{Z}_{2H}, \\ \boldsymbol{Z}_{2L}\times \boldsymbol{Z}_{2H}}$ & 
				$W(\boldsymbol{n},\boldsymbol{1},\boldsymbol{3})_{q,0}$ &
				$\text{All}$ & 
				$W(\boldsymbol{1},\boldsymbol{1},\boldsymbol{3})_{0,0}$ &
				$\makecell[c] {\boldsymbol{1}\times\boldsymbol{1}, \boldsymbol{Z}_{2L}\times\boldsymbol{1}, \\ \boldsymbol{1}\times\boldsymbol{Z}_{2H}, \\ \boldsymbol{Z}_{2L}\times \boldsymbol{Z}_{2H}}$ \\
				\hline
				
				$\frac{U{{\left( 1 \right)}_{Y}}\times SU{{\left( 2 \right)}_{H}}\times SU{{\left( 3 \right)}_{C}}}{\Gamma_p}\times \frac{U{{\left( 1 \right)}_{X}} \times SU{{\left( 2 \right)}_{L}}}{\Gamma_m}\;$ & 
				$W(\boldsymbol{1},\boldsymbol{2},\boldsymbol{3})_{0,0}$ &
				$\boldsymbol{1}\times\boldsymbol{1}, \boldsymbol{1}\times\boldsymbol{Z}_{2L}$ & 
				$W(\boldsymbol{n},\boldsymbol{1},\boldsymbol{3})_{q,0}$ &
				$\makecell[c]{\text{All except \;} \\ \boldsymbol{1}\times\boldsymbol{Z}_{2L}, \\  \boldsymbol{Z}_{2H}\times \boldsymbol{Z}_{2L}}$ & 
				$W(\boldsymbol{1},\boldsymbol{1},\boldsymbol{3})_{0,0}$ &
				$\makecell[c] {\boldsymbol{1}\times\boldsymbol{1}, \boldsymbol{1}\times\boldsymbol{Z}_{2L}, \\ \boldsymbol{Z}_{2H}\times\boldsymbol{1}, \\ \boldsymbol{Z}_{2H}\times \boldsymbol{Z}_{2L}}$ \\
				\hline
				
				$\frac{U{{\left( 1 \right)}_{Y}}\times SU{{\left( 2 \right)}_{H}}}{\Gamma_p}\times \frac{U{{\left( 1 \right)}_{X}} \times SU{{\left( 2 \right)}_{L}}\times SU{{\left( 3 \right)}_{C}}}{\Gamma_m}\;$ & 
				$W(\boldsymbol{1},\boldsymbol{n},\boldsymbol{3})_{0,h}$ &
				$\makecell[c]{\text{All except \;} \\ \boldsymbol{Z}_{2H}\times\boldsymbol{1}, \\  \boldsymbol{Z}_{2H}\times \boldsymbol{Z}_{2L}}$ & 
				$W(\boldsymbol{2},\boldsymbol{1},\boldsymbol{3})_{0,0}$ &
				$\boldsymbol{1}\times\boldsymbol{1}, \boldsymbol{Z}_{2H}\times\boldsymbol{1}$ & 
				$W(\boldsymbol{1},\boldsymbol{1},\boldsymbol{3})_{0,0}$ &
				$\makecell[c] {\boldsymbol{1}\times\boldsymbol{1}, \boldsymbol{1}\times\boldsymbol{Z}_{2L}, \\ \boldsymbol{Z}_{2H}\times\boldsymbol{1}, \\ \boldsymbol{Z}_{2H}\times \boldsymbol{Z}_{2L}}$ \\
				\hline
				
				$\frac{U{{\left( 1 \right)}_{Y}}\times SU{{\left( 2 \right)}_{L}}}{\Gamma_p}\times \frac{U{{\left( 1 \right)}_{X}} \times SU{{\left( 2 \right)}_{H}}\times SU{{\left( 3 \right)}_{C}}}{\Gamma_m}\;$ & 
				$W(\boldsymbol{1},\boldsymbol{n},\boldsymbol{3})_{0,h}$ &
				$\text{All}$ & 
				$W(\boldsymbol{2},\boldsymbol{1},\boldsymbol{3})_{q,0}$ &
				$\makecell[c] {\boldsymbol{1}\times\boldsymbol{1}, \boldsymbol{Z}_{2L}\times\boldsymbol{1}, \\ \boldsymbol{1}\times\boldsymbol{Z}_{2H}, \\ \boldsymbol{Z}_{2L}\times \boldsymbol{Z}_{2H}}$ & 
				$W(\boldsymbol{1},\boldsymbol{1},\boldsymbol{3})_{0,0}$ &
				$\makecell[c] {\boldsymbol{1}\times\boldsymbol{1}, \boldsymbol{Z}_{2L}\times\boldsymbol{1}, \\ \boldsymbol{1}\times\boldsymbol{Z}_{2H}, \\ \boldsymbol{Z}_{2L}\times \boldsymbol{Z}_{2H}}$ \\
				\hline

				$\frac{U(1)_V \times SU(2)_L \times SU(3)_C}{\Gamma_p} \times \frac{U{{\left( 1 \right)}_{A}} \times SU(2)_H}{\Gamma_m}$ &
				\sout{\phantom{\quad}} &
				\sout{\phantom{\quad}} &
				\sout{\phantom{\quad}} &
				\sout{\phantom{\quad}} &
				$W(\boldsymbol{1},\boldsymbol{1},\boldsymbol{3})_{0,0}$ &
				$\makecell[c] {\boldsymbol{1}\times\boldsymbol{1}, \boldsymbol{Z}_{2L}\times\boldsymbol{1}, \\ \boldsymbol{1}\times\boldsymbol{Z}_{2H}, \\ \boldsymbol{Z}_{2L}\times \boldsymbol{Z}_{2H}}$ \\
				\hline

				$\frac{U{{\left( 1 \right)}_{V}}\times SU{{\left( 2 \right)}_{H}}\times SU{{\left( 3 \right)}_{C}}}{\Gamma_p}\times \frac{U{{\left( 1 \right)}_{A}} \times SU{{\left( 2 \right)}_{L}}}{\Gamma_m}\;$ & 
				\sout{\phantom{\quad}} &
				\sout{\phantom{\quad}} &
				\sout{\phantom{\quad}} &
				\sout{\phantom{\quad}} &
				$W(\boldsymbol{1},\boldsymbol{1},\boldsymbol{3})_{0,0}$ &
				$\makecell[c] {\boldsymbol{1}\times\boldsymbol{1}, \boldsymbol{Z}_{2L}\times\boldsymbol{1}, \\ \boldsymbol{1}\times\boldsymbol{Z}_{2H}, \\ \boldsymbol{Z}_{2L}\times \boldsymbol{Z}_{2H}}$ \\
				\hline
				
				$\frac{U{{\left( 1 \right)}_{V}}\times SU{{\left( 2 \right)}_{H}}}{\Gamma_p}\times \frac{U{{\left( 1 \right)}_{A}} \times SU{{\left( 2 \right)}_{L}}\times SU{{\left( 3 \right)}_{C}}}{\Gamma_m}\;$ & 
				\sout{\phantom{\quad}} &
				\sout{\phantom{\quad}} &
				\sout{\phantom{\quad}} &
				\sout{\phantom{\quad}} &
				$W(\boldsymbol{1},\boldsymbol{1},\boldsymbol{3})_{0,0}$ &
				$\makecell[c] {\boldsymbol{1}\times\boldsymbol{1}, \boldsymbol{Z}_{2L}\times\boldsymbol{1}, \\ \boldsymbol{1}\times\boldsymbol{Z}_{2H}, \\ \boldsymbol{Z}_{2L}\times \boldsymbol{Z}_{2H}}$ \\
				\hline
				
				$\frac{U{{\left( 1 \right)}_{V}}\times SU{{\left( 2 \right)}_{L}}}{\Gamma_p}\times \frac{U{{\left( 1 \right)}_{A}} \times SU{{\left( 2 \right)}_{H}}\times SU{{\left( 3 \right)}_{C}}}{\Gamma_m}\;$ & 
				\sout{\phantom{\quad}} &
				\sout{\phantom{\quad}} &
				\sout{\phantom{\quad}} &
				\sout{\phantom{\quad}} &
				$W(\boldsymbol{1},\boldsymbol{1},\boldsymbol{3})_{0,0}$ &
				$\makecell[c] {\boldsymbol{1}\times\boldsymbol{1}, \boldsymbol{Z}_{2L}\times\boldsymbol{1}, \\ \boldsymbol{1}\times\boldsymbol{Z}_{2H}, \\ \boldsymbol{Z}_{2L}\times \boldsymbol{Z}_{2H}}$ \\
				\hline

		\end{tabular} }
	\end{center}
\end{table}

\begin{table}
	\begin{center}
		\caption{
			The $\Gamma$ required for neutral quarks (electrically neutral colored Wilson lines) in the various quotient gauge groups in case (B). 
		}
		\label{tab:neutralquarksB}
		\resizebox{\textwidth}{!}{
			\begin{tabular} {|c|c|c|c|c|c|c|}
				
				\hline
				\multirow{2}*{\quad} & \multicolumn{2}{|c|}{$Q_{\rm em}=0,Q_D\neq0$} & \multicolumn{2}{|c|}{$Q_D=0,Q_{\rm em}\neq0$} & \multicolumn{2}{|c|}{$Q_{\rm em}=0,Q_D=0$} \\
				\cline{2-7}& \text{Quark} & $\Gamma$ & \text{Quark} & $\Gamma$ & \text{Quark} & $\Gamma$  \\
				
				\hline

				$\frac{U{{\left( 1 \right)}_{Y}}\times SU{{\left( 3 \right)}_{C}}}{{\Gamma_{p}}}\times \frac{U{{\left( 1 \right)}_{X}} \times SU{{\left( 2 \right)}_{H}}}{{\Gamma_{m}}}\times \frac{SU{{\left( 2 \right)}_{L}}}{{\Gamma_{n}}}$ & 
				$W(\boldsymbol{1},\boldsymbol{2},\boldsymbol{3})_{0,h}$ &
				$\makecell[c] {\boldsymbol{1}\times\boldsymbol{1}\times\boldsymbol{1},  \boldsymbol{1}\times\boldsymbol{1}\times\boldsymbol{Z}_{2L}, \\ \boldsymbol{1}\times\boldsymbol{Z}_{2H}\times\boldsymbol{1}, \\ \boldsymbol{1}\times\boldsymbol{Z}_{2H}\times \boldsymbol{Z}_{2L}}$ & 
				$W(\boldsymbol{n},\boldsymbol{1},\boldsymbol{3})_{q,0}$ &
				$\makecell[c]{\text{All except \;} \\ \boldsymbol{1}\times\boldsymbol{1}\times\boldsymbol{Z}_{2L}, \\  \boldsymbol{1}\times\boldsymbol{Z}_{2H}\times \boldsymbol{Z}_{2L}}$ & 
				$W(\boldsymbol{1},\boldsymbol{1},\boldsymbol{3})_{0,0}$ &
				$\makecell[c] {\boldsymbol{1}\times\boldsymbol{1}\times\boldsymbol{1},  \boldsymbol{1}\times\boldsymbol{1}\times\boldsymbol{Z}_{2L}, \\ \boldsymbol{1}\times\boldsymbol{Z}_{2H}\times\boldsymbol{1}, \\ \boldsymbol{1}\times\boldsymbol{Z}_{2H}\times \boldsymbol{Z}_{2L}}$ \\
				\hline

				$\frac{U{{\left( 1 \right)}_{Y}}\times SU{{\left( 3 \right)}_{C}}}{{\Gamma_{p}}}\times \frac{U{{\left( 1 \right)}_{X}} \times SU{{\left( 2 \right)}_{L}}}{{\Gamma_{m}}}\times \frac{SU{{\left( 2 \right)}_{H}}}{{\Gamma_{n}}}$ & 
				$W(\boldsymbol{1},\boldsymbol{2},\boldsymbol{3})_{0,0}$ &
				$\makecell[c]{\boldsymbol{1}\times\boldsymbol{1}\times\boldsymbol{1}, \\ \boldsymbol{1}\times\boldsymbol{Z}_{2L}\times\boldsymbol{1}}$ & 
				$W(\boldsymbol{n},\boldsymbol{1},\boldsymbol{3})_{q,0}$ &
				$\makecell[c]{\text{All except \;} \\ \boldsymbol{1}\times\boldsymbol{Z}_{2L}\times\boldsymbol{1}, \\  \boldsymbol{1}\times\boldsymbol{Z}_{2L}\times \boldsymbol{Z}_{2H}}$ & 
				$W(\boldsymbol{1},\boldsymbol{1},\boldsymbol{3})_{0,0}$ &
				$\makecell[c] {\boldsymbol{1}\times\boldsymbol{1}\times\boldsymbol{1}, \\ \boldsymbol{1}\times\boldsymbol{Z}_{2L}\times\boldsymbol{1}, \\ \boldsymbol{1}\times\boldsymbol{1}\times\boldsymbol{Z}_{2H}, \\ \boldsymbol{1}\times\boldsymbol{Z}_{2L}\times \boldsymbol{Z}_{2H}}$ \\
				\hline
				
				$\frac{U{{\left( 1 \right)}_{Y}}\times SU{{\left( 2 \right)}_{H}}}{{\Gamma_{p}}}\times \frac{U{{\left( 1 \right)}_{X}} \times SU{{\left( 3 \right)}_{C}}}{{\Gamma_{m}}}\times \frac{SU{{\left( 2 \right)}_{L}}}{{\Gamma_{n}}}$ & 
				$W(\boldsymbol{1},\boldsymbol{n},\boldsymbol{3})_{0,h}$ &
				$\makecell[c]{\text{All except \;} \\ \boldsymbol{Z}_{2H}\times\boldsymbol{1}\times\boldsymbol{1}, \\  \boldsymbol{Z}_{2H}\times\boldsymbol{1}\times\boldsymbol{Z}_{2L}}$ & 
				$W(\boldsymbol{2},\boldsymbol{1},\boldsymbol{3})_{0,0}$ &
				$\makecell[c]{ \boldsymbol{1}\times\boldsymbol{1}\times\boldsymbol{1}, \\  \boldsymbol{Z}_{2H}\times\boldsymbol{1}\times\boldsymbol{1}}$ & 
				$W(\boldsymbol{1},\boldsymbol{1},\boldsymbol{3})_{0,0}$ &
				$\makecell[c] {\boldsymbol{1}\times\boldsymbol{1}\times\boldsymbol{1}, \\ \boldsymbol{1}\times\boldsymbol{1}\times\boldsymbol{Z}_{2L}, \\ \boldsymbol{Z}_{2H}\times\boldsymbol{1}\times\boldsymbol{1}, \\ \boldsymbol{Z}_{2H}\times\boldsymbol{1}\times\boldsymbol{Z}_{2L}}$ \\
				\hline
				
				$\frac{U{{\left( 1 \right)}_{Y}}\times SU{{\left( 2 \right)}_{L}}}{{\Gamma_{p}}}\times \frac{U{{\left( 1 \right)}_{X}} \times SU{{\left( 3 \right)}_{C}}}{{\Gamma_{m}}}\times \frac{SU{{\left( 2 \right)}_{H}}}{{\Gamma_{n}}}$ & 
				$W(\boldsymbol{1},\boldsymbol{n},\boldsymbol{3})_{0,h}$ &
				$\makecell[c]{\text{All except \;} \\ \boldsymbol{1}\times\boldsymbol{1}\times\boldsymbol{Z}_{2H}, \\  \boldsymbol{Z}_{2L}\times\boldsymbol{1}\times\boldsymbol{Z}_{2H}}$ & 
				$W(\boldsymbol{2},\boldsymbol{1},\boldsymbol{3})_{q,0}$ &
				$\makecell[c] {\boldsymbol{1}\times\boldsymbol{1}\times\boldsymbol{1}, \\ \boldsymbol{Z}_{2L}\times\boldsymbol{1}\times\boldsymbol{1}, \\ \boldsymbol{1}\times\boldsymbol{1}\times\boldsymbol{Z}_{2H}, \\ \boldsymbol{Z}_{2L}\times\boldsymbol{1}\times\boldsymbol{Z}_{2H}}$ & 
				$W(\boldsymbol{1},\boldsymbol{1},\boldsymbol{3})_{0,0}$ &
				$\makecell[c] {\boldsymbol{1}\times\boldsymbol{1}\times\boldsymbol{1}, \\ \boldsymbol{Z}_{2L}\times\boldsymbol{1}\times\boldsymbol{1}, \\ \boldsymbol{1}\times\boldsymbol{1}\times\boldsymbol{Z}_{2H}, \\ \boldsymbol{Z}_{2L}\times\boldsymbol{1}\times\boldsymbol{Z}_{2H}}$ \\
				\hline

				$\frac{U{{\left( 1 \right)}_{V}}\times SU{{\left( 3 \right)}_{C}}}{{\Gamma_{p}}}\times \frac{U{{\left( 1 \right)}_{A}} \times SU{{\left( 2 \right)}_{H}}}{{\Gamma_{m}}}\times \frac{SU{{\left( 2 \right)}_{L}}}{{\Gamma_{n}}}$ & 
				\sout{\phantom{\quad}} &
				\sout{\phantom{\quad}} &
				\sout{\phantom{\quad}} &
				\sout{\phantom{\quad}} &
				$W(\boldsymbol{1},\boldsymbol{1},\boldsymbol{3})_{0,0}$ &
				$\makecell[c] {\boldsymbol{1}\times\boldsymbol{1}, \boldsymbol{Z}_{2L}\times\boldsymbol{1}, \\ \boldsymbol{1}\times\boldsymbol{Z}_{2H}, \\ \boldsymbol{Z}_{2L}\times \boldsymbol{Z}_{2H}}$ \\
				\hline

				$\frac{U{{\left( 1 \right)}_{V}}\times SU{{\left( 3 \right)}_{C}}}{{\Gamma_{p}}}\times \frac{U{{\left( 1 \right)}_{A}} \times SU{{\left( 2 \right)}_{L}}}{{\Gamma_{m}}}\times \frac{SU{{\left( 2 \right)}_{H}}}{{\Gamma_{n}}}$ & 
				\sout{\phantom{\quad}} &
				\sout{\phantom{\quad}} &
				\sout{\phantom{\quad}} &
				\sout{\phantom{\quad}} &
				$W(\boldsymbol{1},\boldsymbol{1},\boldsymbol{3})_{0,0}$ &
				$\makecell[c] {\boldsymbol{1}\times\boldsymbol{1}, \boldsymbol{Z}_{2L}\times\boldsymbol{1}, \\ \boldsymbol{1}\times\boldsymbol{Z}_{2H}, \\ \boldsymbol{Z}_{2L}\times \boldsymbol{Z}_{2H}}$ \\
				\hline
				
				$\frac{U{{\left( 1 \right)}_{V}}\times SU{{\left( 2 \right)}_{H}}}{{\Gamma_{p}}}\times \frac{U{{\left( 1 \right)}_{A}} \times SU{{\left( 3 \right)}_{C}}}{{\Gamma_{m}}}\times \frac{SU{{\left( 2 \right)}_{L}}}{{\Gamma_{n}}}$ & 
				\sout{\phantom{\quad}} &
				\sout{\phantom{\quad}} &
				\sout{\phantom{\quad}} &
				\sout{\phantom{\quad}} &
				$W(\boldsymbol{1},\boldsymbol{1},\boldsymbol{3})_{0,0}$ &
				$\makecell[c] {\boldsymbol{1}\times\boldsymbol{1}, \boldsymbol{Z}_{2L}\times\boldsymbol{1}, \\ \boldsymbol{1}\times\boldsymbol{Z}_{2H}, \\ \boldsymbol{Z}_{2L}\times \boldsymbol{Z}_{2H}}$ \\
				\hline
				
				$\frac{U{{\left( 1 \right)}_{V}}\times SU{{\left( 2 \right)}_{L}}}{{\Gamma_{p}}}\times \frac{U{{\left( 1 \right)}_{A}} \times SU{{\left( 3 \right)}_{C}}}{{\Gamma_{m}}}\times \frac{SU{{\left( 2 \right)}_{H}}}{{\Gamma_{n}}}$ & 
				\sout{\phantom{\quad}} &
				\sout{\phantom{\quad}} &
				\sout{\phantom{\quad}} &
				\sout{\phantom{\quad}} &
				$W(\boldsymbol{1},\boldsymbol{1},\boldsymbol{3})_{0,0}$ &
				$\makecell[c] {\boldsymbol{1}\times\boldsymbol{1}, \boldsymbol{Z}_{2L}\times\boldsymbol{1}, \\ \boldsymbol{1}\times\boldsymbol{Z}_{2H}, \\ \boldsymbol{Z}_{2L}\times \boldsymbol{Z}_{2H}}$ \\
				\hline

		\end{tabular} }
	\end{center}
\end{table}

\begin{table}
	\begin{center}
		\caption{
			The $\Gamma$ required for neutral quarks (electrically neutral colored Wilson lines) in the various quotient gauge groups in case (C).
		}
		\label{tab:neutralquarksC}
		\resizebox{\textwidth}{!}{
			\begin{tabular} {|c|c|c|c|c|c|c|}
				
				\hline
				\multirow{2}*{\quad} & \multicolumn{2}{|c|}{$Q_{\rm em}=0,Q_D\neq0$} & \multicolumn{2}{|c|}{$Q_D=0,Q_{\rm em}\neq0$} & \multicolumn{2}{|c|}{$Q_{\rm em}=0,Q_D=0$} \\
				\cline{2-7}& \text{Quark} & $\Gamma$ & \text{Quark} & $\Gamma$ & \text{Quark} & $\Gamma$  \\
				
				\hline
			
				$\frac{U{{\left( 1 \right)}_{Y}}\times SU{{\left( 2 \right)}_{L}}}{{\Gamma_{p}}}\times \frac{U{{\left( 1 \right)}_{X}} \times SU{{\left( 2 \right)}_{H}}}{{\Gamma_{m}}}\times \frac{SU{{\left( 3 \right)}_{C}}}{{\Gamma_{n}}}$ & 
				$W(\boldsymbol{1},\boldsymbol{2},\boldsymbol{3})_{0,h}$ &
				$\makecell[c] {\boldsymbol{1}\times\boldsymbol{1}\times\boldsymbol{1}, \\ \boldsymbol{Z}_{2L}\times\boldsymbol{1}\times\boldsymbol{1}, \\ \boldsymbol{1}\times\boldsymbol{Z}_{2H}\times\boldsymbol{1}, \\ \boldsymbol{Z}_{2L}\times \boldsymbol{Z}_{2H}\times\boldsymbol{1}}$ & 
				$W(\boldsymbol{2},\boldsymbol{1},\boldsymbol{3})_{q,0}$ &
				$\makecell[c] {\boldsymbol{1}\times\boldsymbol{1}\times\boldsymbol{1}, \\ \boldsymbol{Z}_{2L}\times\boldsymbol{1}\times\boldsymbol{1}, \\ \boldsymbol{1}\times\boldsymbol{Z}_{2H}\times\boldsymbol{1}, \\ \boldsymbol{Z}_{2L}\times\boldsymbol{Z}_{2H}\times\boldsymbol{1}}$ & 
				$W(\boldsymbol{1},\boldsymbol{1},\boldsymbol{3})_{0,0}$ &
				$\makecell[c] {\boldsymbol{1}\times\boldsymbol{1}\times\boldsymbol{1}, \\ \boldsymbol{Z}_{2L}\times\boldsymbol{1}\times\boldsymbol{1}, \\ \boldsymbol{1}\times\boldsymbol{Z}_{2H}\times\boldsymbol{1}, \\ \boldsymbol{Z}_{2L}\times\boldsymbol{Z}_{2H}\times\boldsymbol{1}}$ \\
				\hline

				$\frac{U{{\left( 1 \right)}_{Y}}\times SU{{\left( 2 \right)}_{H}}}{{\Gamma_{p}}}\times \frac{U{{\left( 1 \right)}_{X}} \times SU{{\left( 2 \right)}_{L}}}{{\Gamma_{m}}}\times \frac{SU{{\left( 3 \right)}_{C}}}{{\Gamma_{n}}}$ & 
				$W(\boldsymbol{1},\boldsymbol{2},\boldsymbol{3})_{0,0}$ &
				$\makecell[c]{ \boldsymbol{1}\times\boldsymbol{1}\times\boldsymbol{1}, \\ \boldsymbol{1}\times\boldsymbol{Z}_{2L}\times\boldsymbol{1}}$ & 
				$W(\boldsymbol{2},\boldsymbol{1},\boldsymbol{3})_{0,0}$ &
				$\makecell[c]{ \boldsymbol{1}\times\boldsymbol{1}\times\boldsymbol{1}, \\ \boldsymbol{Z}_{2H}\times\boldsymbol{1}\times\boldsymbol{1}}$ & 
				$W(\boldsymbol{1},\boldsymbol{1},\boldsymbol{3})_{0,0}$ &
				$\makecell[c] {\boldsymbol{1}\times\boldsymbol{1}\times\boldsymbol{1}, \\ \boldsymbol{1}\times\boldsymbol{Z}_{2L}\times\boldsymbol{1}, \\ \boldsymbol{Z}_{2H}\times\boldsymbol{1}\times\boldsymbol{1}, \\ \boldsymbol{Z}_{2H}\times \boldsymbol{Z}_{2L}\times\boldsymbol{1}}$ \\
				\hline

				$\frac{U{{\left( 1 \right)}_{V}}\times SU{{\left( 2 \right)}_{L}}}{{\Gamma_{p}}}\times \frac{U{{\left( 1 \right)}_{A}} \times SU{{\left( 2 \right)}_{H}}}{{\Gamma_{m}}}\times \frac{SU{{\left( 3 \right)}_{C}}}{{\Gamma_{n}}}$ & 
				\sout{\phantom{\quad}} &
				\sout{\phantom{\quad}} &
				\sout{\phantom{\quad}} &
				\sout{\phantom{\quad}} &
				$W(\boldsymbol{1},\boldsymbol{1},\boldsymbol{3})_{0,0}$ &
				$\makecell[c] {\boldsymbol{1}\times\boldsymbol{1}, \boldsymbol{Z}_{2L}\times\boldsymbol{1}, \\ \boldsymbol{1}\times\boldsymbol{Z}_{2H}, \\ \boldsymbol{Z}_{2L}\times \boldsymbol{Z}_{2H}}$ \\
				\hline

				$\frac{U{{\left( 1 \right)}_{V}}\times SU{{\left( 2 \right)}_{H}}}{{\Gamma_{p}}}\times \frac{U{{\left( 1 \right)}_{A}} \times SU{{\left( 2 \right)}_{L}}}{{\Gamma_{m}}}\times \frac{SU{{\left( 3 \right)}_{C}}}{{\Gamma_{n}}}$ & 
				\sout{\phantom{\quad}} &
				\sout{\phantom{\quad}} &
				\sout{\phantom{\quad}} &
				\sout{\phantom{\quad}} &
				$W(\boldsymbol{1},\boldsymbol{1},\boldsymbol{3})_{0,0}$ &
				$\makecell[c] {\boldsymbol{1}\times\boldsymbol{1}, \boldsymbol{Z}_{2L}\times\boldsymbol{1}, \\ \boldsymbol{1}\times\boldsymbol{Z}_{2H}, \\ \boldsymbol{Z}_{2L}\times \boldsymbol{Z}_{2H}}$ \\
				\hline
				
		\end{tabular} }
	\end{center}
\end{table}

\begin{itemize}

\item
If SM neutral quark~\footnote{This is actually electrically neutral colored Wilson line. We follow the nomenclature used in Ref.~\cite{Tong:2017oea}.} \( W(\boldsymbol{1},\boldsymbol{2},\boldsymbol{3})_{0,0} \) (or more generally \( W(\boldsymbol{1},\boldsymbol{2},\boldsymbol{3})_{0,h} \) with $h \neq 0$) exists in Nature and 
is part of the representation of quotient group \( G = \frac{U(1)_Y \times SU(2)_L \times SU(3)_C}{\Gamma_p} \times \frac{U(1)_X \times SU(2)_H}{\Gamma_m} \), then \(\Gamma\) must be \(\boldsymbol{1} \times \boldsymbol{1} \), \(\boldsymbol{Z}_{2L} \times \boldsymbol{1}\), \(\boldsymbol{1} \times \boldsymbol{Z}_{2H}\), or \(\boldsymbol{Z}_{2L} \times \boldsymbol{Z}_{2H}\). 
Other representations of neutral quarks are allowed in this quotient group.
For instance, neutral quark \( W(\boldsymbol{1}, \boldsymbol{1}, \boldsymbol{3})_{0,0} \) 
with $Q_{\rm em} = Q_D = 0$
also require $\Gamma$ to be in the above four choices only. On the other hand, the 
$\Gamma$ required for the neutral quark \( W(\boldsymbol{n}, \boldsymbol{1}, \boldsymbol{3})_{q,0} \) with $Q_{\rm em}\ne 0 , Q_D=0$ can be of any choice.

\item
In the case of the quotient group \( G = \frac{U(1)_Y \times SU(2)_H \times SU(3)_C}{\Gamma_p} \times \frac{U(1)_X \times SU(2)_L}{\Gamma_m} \), the SM neutral quark \( W(\boldsymbol{1},\boldsymbol{2},\boldsymbol{3})_{0,0} \) is allowed provided that \(\Gamma\) to be $\boldsymbol{1} \times \boldsymbol{1}$ or $\boldsymbol{1} \times \boldsymbol{Z}_{2L}$. The neutral quark \( W(\boldsymbol{n}, \boldsymbol{1}, \boldsymbol{3})_{q,0} \) 
requires \(\Gamma\) to be any of its options except $\boldsymbol{1} \times \boldsymbol{Z}_{2L}, \boldsymbol{Z}_{2H}\times \boldsymbol{Z}_{2L}$, while the neutral quark \( W(\boldsymbol{1},\boldsymbol{1},\boldsymbol{3})_{0,0} \) 
requires $\Gamma$ to take \( \boldsymbol{1} \times \boldsymbol{1}\), \( \boldsymbol{1} \times \boldsymbol{Z}_{2L}\), \(\boldsymbol{Z}_{2H} \times \boldsymbol{1}\), or \(\boldsymbol{Z}_{2H} \times \boldsymbol{Z}_{2L}\).

\item
In the case of the quotient group \( G = \frac{U(1)_V \times SU(2)_L \times SU(3)_C}{\Gamma_p} \times \frac{U(1)_V \times SU(2)_H}{\Gamma_m} \), there is only one type of neutral quark, namely $W(\boldsymbol{1},\boldsymbol{1},\boldsymbol{3})_{0,0}$, and it requires that $\Gamma$ can be $\boldsymbol{1}\times\boldsymbol{1}, \boldsymbol{Z}_{2L}\times\boldsymbol{1}, \boldsymbol{1}\times\boldsymbol{Z}_{2H}$, or $\boldsymbol{Z}_{2L}\times \boldsymbol{Z}_{2H}$. 

\item 
For the above three and other 17 quotient gauge groups corresponding to the three cases (A), (B) and (C) discussed in Section~\ref{sec:G2HDM}, the possible $\Gamma$ for all neutral quarks are tabulated in Tables~\ref{tab:neutralquarksA}, ~\ref{tab:neutralquarksB} and ~\ref{tab:neutralquarksC} respectively. 

\item
For comparison, suppose a magnetic monopole is discovered that satisfies the minimal electromagnetic DQC for charged leptons ({\it i.e.}, colorless dyonic line operators) but {\it not} for quarks. In this case, \(\Gamma\) must contain either the factor \( \boldsymbol{Z}_{6L} \) or \( \boldsymbol{Z}_{6H} \), depending on whether the monopole charge resides in the SM or the dark sector. This follows from the fact that nontrivial magnetic monopoles satisfying the minimal electromagnetic DQC for charged leptons require \(\Gamma = \boldsymbol{Z}_{2L}\) or \(\Gamma = \boldsymbol{Z}_{2H}\). Furthermore, to ensure color neutrality, the condition \(z_3^e = 0\) must hold in the GDQC. As a result, \(\Gamma\) must necessarily include \( \boldsymbol{Z}_{6L} \) or \( \boldsymbol{Z}_{6H} \).

\end{itemize}

Finally, we direct interested readers to a similar analysis of line operators in~\cite{Guo:2025fkl} within the context of the left-right symmetric model~\cite{Mohapatra:1979ia,Mohapatra:1980yp}. Moreover, extending these investigations to the Pati-Salam model~\cite{Pati:1974yy}, which is based on the semi-simple gauge group 
\( SU(2)_L \times SU(2)_R \times SU(4)_C \), is an intriguing direction for future research. We leave this exploration for future work.

%\vfill 

\section*{Acknowledgment}
This work is supported by the
Natural Science Foundation of Chongqing, China, Grant No. CSTB2022NSCQ-MSX0351. TCY expresses his gratitude to Professor Khiem Hong Phan for hosting him at the Duy Tan University, Danang, Vietnam, where significant progress of this research was made.

\newpage

\appendix 

\section{Witten Effect for the Abelian Lines}\label{sec:appendix}

\begin{figure}[htbp!]

\subfigure[$\;z_2^e=0,z_2^m=0,\tilde{\theta}=0,4\pi$]{
\begin{tikzpicture}[scale=0.60]

	\draw[thick,->] (-0.75,0) -- (4.5,0) node[right] {$q$};
	\draw[thick,->] (0,-1.25) -- (0,2.5) node[above] {$g$};
	
	\filldraw[opacity=0.25] [fill=blue,draw=black](-0.25,-0.5) rectangle (1.75,0.5);
	
	\foreach \x in {-0.5,0, 0.5, 1, 1.5, 2, 2.5, 3, 3.5, 4} {
		\foreach \y in {-1,0,1,2} {
			\filldraw[draw=black,fill=white] (\x,\y) circle(0.2); 
		}
	}
		
	\foreach \x in {0,2,4} {
		\foreach \y in {-1,0,1,2} {
			\filldraw[draw=black,fill=green] (\x,\y) circle(0.2);	}}
	
	\node at (-0.5,-1) [below=10pt] {$-\frac{1}{2}\;$};
	\node at (0,-1.15) [below=10pt] {$0$};
	\node at (0.5,-1) [below=10pt] {$\frac{1}{2}$};
	\node at (1,-1.15) [below=10pt] {$1$};
	\node at (1.5,-1) [below=10pt] {$\frac{3}{2}$};
	\node at (2,-1.15) [below=10pt] {$2$};
	\node at (2.5,-1) [below=10pt] {$\frac{5}{2}$};
	\node at (3,-1.15) [below=10pt] {$3$};
	\node at (3.5,-1) [below=10pt] {$\frac{7}{2}$};
	\node at (4,-1.15) [below=10pt] {$4$};
	
	\node at (-0.5,2) [left=10pt] {$2$};
	\node at (-0.5,1) [left=10pt] {$1$};
	\node at (-0.5,0) [left=10pt] {$0$};
	\node at (-0.5,-1) [left=10pt] {$-1$};
	
\end{tikzpicture}
}

\subfigure[$\;z_2^e=0,z_2^m=0,\tilde{\theta}=2\pi,6\pi$]{
\begin{tikzpicture}[scale=0.60]
	
	\draw[thick,->] (-0.75,0) -- (4.5,0) node[right] {$q$};
	\draw[thick,->] (0,-1.25) -- (0,2.5) node[above] {$g$};
	
	\filldraw[opacity=0.25] [fill=blue,draw=black](-0.25,-0.5) rectangle (1.75,1.5);

	\foreach \x in {-0.5,0, 0.5, 1, 1.5, 2, 2.5, 3, 3.5, 4} {
		\foreach \y in {-1,0,1,2} {
			\filldraw[draw=black,fill=white] (\x,\y) circle(0.2); 
		}
	}
		
	\foreach \x in {0,2,4} {
		\foreach \y in {0,2} {
			\filldraw[draw=black,fill=green] (\x,\y) circle(0.2);	}}
		
	\foreach \x in {1,3} {
		\foreach \y in {-1,1} {
			\filldraw[draw=black,fill=green] (\x,\y) circle(0.2);	}}
	
	\node at (-0.5,-1) [below=10pt] {$-\frac{1}{2}\;$};
	\node at (0,-1.15) [below=10pt] {$0$};
	\node at (0.5,-1) [below=10pt] {$\frac{1}{2}$};
	\node at (1,-1.15) [below=10pt] {$1$};
	\node at (1.5,-1) [below=10pt] {$\frac{3}{2}$};
	\node at (2,-1.15) [below=10pt] {$2$};
	\node at (2.5,-1) [below=10pt] {$\frac{5}{2}$};
	\node at (3,-1.15) [below=10pt] {$3$};
	\node at (3.5,-1) [below=10pt] {$\frac{7}{2}$};
	\node at (4,-1.15) [below=10pt] {$4$};
	
	\node at (-0.5,2) [left=10pt] {$2$};
	\node at (-0.5,1) [left=10pt] {$1$};
	\node at (-0.5,0) [left=10pt] {$0$};
	\node at (-0.5,-1) [left=10pt] {$-1$};
		
\end{tikzpicture}
}

\subfigure[$\;z_2^e=1,z_2^m=0,\tilde{\theta}=0,4\pi$]{
\begin{tikzpicture}[scale=0.60]
	
	\draw[thick,->] (-0.75,0) -- (4.5,0) node[right] {$q$};
	\draw[thick,->] (0,-1.25) -- (0,2.5) node[above] {$g$};
	
	\filldraw[opacity=0.25] [fill=blue,draw=black](-0.25,-0.5) rectangle (1.75,0.5);

	\foreach \x in {-0.5,0, 0.5, 1, 1.5, 2, 2.5, 3, 3.5, 4} {
		\foreach \y in {-1,0,1,2} {
			\filldraw[draw=black,fill=white] (\x,\y) circle(0.2); 
		}
	}
		
	\foreach \x in {0,2,4} {
		\foreach \y in {-1,0,1,2} {
			\filldraw[draw=black,fill=green] (\x,\y) circle(0.2);	}}
	
	\node at (-0.5,-1) [below=10pt] {$\frac{1}{2}$};
	\node at (0,-1.15) [below=10pt] {$1$};
	\node at (0.5,-1) [below=10pt] {$\frac{3}{2}$};
	\node at (1,-1.15) [below=10pt] {$2$};
	\node at (1.5,-1) [below=10pt] {$\frac{5}{2}$};
	\node at (2,-1.15) [below=10pt] {$3$};
	\node at (2.5,-1) [below=10pt] {$\frac{7}{2}$};
	\node at (3,-1.15) [below=10pt] {$4$};
	\node at (3.5,-1) [below=10pt] {$\frac{9}{2}$};
	\node at (4,-1.15) [below=10pt] {$5$};
	
	\node at (-0.5,2) [left=10pt] {$2$};
	\node at (-0.5,1) [left=10pt] {$1$};
	\node at (-0.5,0) [left=10pt] {$0$};
	\node at (-0.5,-1) [left=10pt] {$-1$};
		
\end{tikzpicture}
}

\subfigure[$\;z_2^e=1,z_2^m=0,\tilde{\theta}=2\pi,6\pi$]{
\begin{tikzpicture}[scale=0.60]
	
	\draw[thick,->] (-0.75,0) -- (4.5,0) node[right] {$q$};
	\draw[thick,->] (0,-1.25) -- (0,2.5) node[above] {$g$};
	
	\filldraw[opacity=0.25] [fill=blue,draw=black](-0.25,-0.5) rectangle (1.75,1.5);
	
	\foreach \x in {-0.5,0, 0.5, 1, 1.5, 2, 2.5, 3, 3.5, 4} {
		\foreach \y in {-1,0,1,2} {
			\filldraw[draw=black,fill=white] (\x,\y) circle(0.2); 
		}
	}	
	
	\foreach \x in {0,2,4} {
		\foreach \y in {0,2} {
			\filldraw[draw=black,fill=green] (\x,\y) circle(0.2);	}}
	
	\foreach \x in {1,3} {
		\foreach \y in {-1,1} {
			\filldraw[draw=black,fill=green] (\x,\y) circle(0.2);	}}
	
	\node at (-0.5,-1) [below=10pt] {$\frac{1}{2}$};
	\node at (0,-1.15) [below=10pt] {$1$};
	\node at (0.5,-1) [below=10pt] {$\frac{3}{2}$};
	\node at (1,-1.15) [below=10pt] {$2$};
	\node at (1.5,-1) [below=10pt] {$\frac{5}{2}$};
	\node at (2,-1.15) [below=10pt] {$3$};
	\node at (2.5,-1) [below=10pt] {$\frac{7}{2}$};
	\node at (3,-1.15) [below=10pt] {$4$};
	\node at (3.5,-1) [below=10pt] {$\frac{9}{2}$};
	\node at (4,-1.15) [below=10pt] {$5$};
	
	\node at (-0.5,2) [left=10pt] {$2$};
	\node at (-0.5,1) [left=10pt] {$1$};
	\node at (-0.5,0) [left=10pt] {$0$};
	\node at (-0.5,-1) [left=10pt] {$-1$};	
	
\end{tikzpicture}
}
\caption{Witten effect for the Abelian lines in $\frac{U(1)\times SU(2)}{\boldsymbol{Z}_2}$ where $\tilde{\theta} = [0, 8 \pi)$ -- (1/3).}
\label{app:WittenEffectsZ2a}
\end{figure}

\vfill
\newpage

\begin{figure}[htbp!]

\subfigure[ $\;z_2^e=0,z_2^m=1,\tilde{\theta}=0$]{
\begin{tikzpicture}[scale=0.6]
	
	\draw[thick,->] (-0.75,0) -- (4.5,0) node[right] {$q$};
	\draw[thick,->] (0,-1.25) -- (0,2.5) node[above] {$g$};
	
	\filldraw[opacity=0.25] [fill=blue,draw=black](-0.25,-0.5) rectangle (1.75,0.5);

	\foreach \x in {-0.5,0, 0.5, 1, 1.5, 2, 2.5, 3, 3.5, 4} {
		\foreach \y in {-1,0,1,2} {
			\filldraw[draw=black,fill=white] (\x,\y) circle(0.2); 
		}
	}
		
	\foreach \x in {0,2,4} {
		\foreach \y in {-1,0,1,2} {
			\filldraw[draw=black,fill=green] (\x,\y) circle(0.2);	}}
	
	\node at (-0.5,-1) [below=10pt] {$-\frac{1}{2}\;$};
	\node at (0,-1.15) [below=10pt] {$0$};
	\node at (0.5,-1) [below=10pt] {$\frac{1}{2}$};
	\node at (1,-1.15) [below=10pt] {$1$};
	\node at (1.5,-1) [below=10pt] {$\frac{3}{2}$};
	\node at (2,-1.15) [below=10pt] {$2$};
	\node at (2.5,-1) [below=10pt] {$\frac{5}{2}$};
	\node at (3,-1.15) [below=10pt] {$3$};
	\node at (3.5,-1) [below=10pt] {$\frac{7}{2}$};
	\node at (4,-1.15) [below=10pt] {$4$};
	
	\node at (-0.5,2) [left=10pt] {$\frac{5}{2}$};
	\node at (-0.5,1) [left=10pt] {$\frac{3}{2}$};
	\node at (-0.5,0) [left=10pt] {$\frac{1}{2}$};
	\node at (-0.5,-1) [left=10pt] {$-\frac{1}{2}$};
	
\end{tikzpicture}
}

\subfigure[$\;z_2^e=0,z_2^m=1,\tilde{\theta}=2\pi$]{
\begin{tikzpicture}[scale=0.6]
	
	\draw[thick,->] (-0.75,0) -- (4.5,0) node[right] {$q$};
	\draw[thick,->] (0,-1.25) -- (0,2.5) node[above] {$g$};
	
	\filldraw[opacity=0.25] [fill=blue,draw=black](-0.25,-0.5) rectangle (1.75,1.5);
	
	\foreach \x in {-0.5,0, 0.5, 1, 1.5, 2, 2.5, 3, 3.5, 4} {
		\foreach \y in {-1,0,1,2} {
			\filldraw[draw=black,fill=white] (\x,\y) circle(0.2); 
		}
	}	
	
	\foreach \x in {0.5,2.5} {
		\foreach \y in {0,2} {
			\filldraw[draw=black,fill=green] (\x,\y) circle(0.2);	}}
	
	\foreach \x in {-0.5,1.5,3.5} {
		\foreach \y in {-1,1} {
			\filldraw[draw=black,fill=green] (\x,\y) circle(0.2);	}}
			
	\node at (-0.5,-1) [below=10pt] {$-\frac{1}{2}\;$};
	\node at (0,-1.15) [below=10pt] {$0$};
	\node at (0.5,-1) [below=10pt] {$\frac{1}{2}$};
	\node at (1,-1.15) [below=10pt] {$1$};
	\node at (1.5,-1) [below=10pt] {$\frac{3}{2}$};
	\node at (2,-1.15) [below=10pt] {$2$};
	\node at (2.5,-1) [below=10pt] {$\frac{5}{2}$};
	\node at (3,-1.15) [below=10pt] {$3$};
	\node at (3.5,-1) [below=10pt] {$\frac{7}{2}$};
	\node at (4,-1.15) [below=10pt] {$4$};
	
	\node at (-0.5,2) [left=10pt] {$\frac{5}{2}$};
	\node at (-0.5,1) [left=10pt] {$\frac{3}{2}$};
	\node at (-0.5,0) [left=10pt] {$\frac{1}{2}$};
	\node at (-0.5,-1) [left=10pt] {$-\frac{1}{2}$};	
	
\end{tikzpicture}
}

\subfigure[$\; z_2^e=0,z_2^m=1,\tilde{\theta}=4\pi$]{
\begin{tikzpicture}[scale=0.6]
	
	\draw[thick,->] (-0.75,0) -- (4.5,0) node[right] {$q$};
	\draw[thick,->] (0,-1.25) -- (0,2.5) node[above] {$g$};
	
	\filldraw[opacity=0.25] [fill=blue,draw=black](-0.25,-0.5) rectangle (1.75,0.5);
	
	\foreach \x in {-0.5,0, 0.5, 1, 1.5, 2, 2.5, 3, 3.5, 4} {
		\foreach \y in {-1,0,1,2} {
			\filldraw[draw=black,fill=white] (\x,\y) circle(0.2); 
		}
	}	
	
	\foreach \x in {1,3} {
		\foreach \y in {-1,0,1,2} {
			\filldraw[draw=black,fill=green] (\x,\y) circle(0.2);	}}	
	
	\node at (-0.5,-1) [below=10pt] {$-\frac{1}{2}\;$};
	\node at (0,-1.15) [below=10pt] {$0$};
	\node at (0.5,-1) [below=10pt] {$\frac{1}{2}$};
	\node at (1,-1.15) [below=10pt] {$1$};
	\node at (1.5,-1) [below=10pt] {$\frac{3}{2}$};
	\node at (2,-1.15) [below=10pt] {$2$};
	\node at (2.5,-1) [below=10pt] {$\frac{5}{2}$};
	\node at (3,-1.15) [below=10pt] {$3$};
	\node at (3.5,-1) [below=10pt] {$\frac{7}{2}$};
	\node at (4,-1.15) [below=10pt] {$4$};
	
	\node at (-0.5,2) [left=10pt] {$\frac{5}{2}$};
	\node at (-0.5,1) [left=10pt] {$\frac{3}{2}$};
	\node at (-0.5,0) [left=10pt] {$\frac{1}{2}$};
	\node at (-0.5,-1) [left=10pt] {$-\frac{1}{2}$};
		
\end{tikzpicture}
}

\subfigure[$\; z_2^e=0,z_2^m=1,\tilde{\theta}=6\pi$]{
\begin{tikzpicture}[scale=0.6]
	
	\draw[thick,->] (-0.75,0) -- (4.5,0) node[right] {$q$};
	\draw[thick,->] (0,-1.25) -- (0,2.5) node[above] {$g$};
	
	\filldraw[opacity=0.25] [fill=blue,draw=black](-0.25,-0.5) rectangle (1.75,1.5);
	
	\foreach \x in {-0.5,0, 0.5, 1, 1.5, 2, 2.5, 3, 3.5, 4} {
		\foreach \y in {-1,0,1,2} {
			\filldraw[draw=black,fill=white] (\x,\y) circle(0.2); 
		}
	}	
	
	\foreach \x in {-0.5,1.5,3.5} {
		\foreach \y in {0,2} {
			\filldraw[draw=black,fill=green] (\x,\y) circle(0.2);	}}
	
	\foreach \x in {0.5,2.5} {
		\foreach \y in {-1,1} {
			\filldraw[draw=black,fill=green] (\x,\y) circle(0.2);	}}
		
	\node at (-0.5,-1) [below=10pt] {$-\frac{1}{2}\;$};
	\node at (0,-1.15) [below=10pt] {$0$};
	\node at (0.5,-1) [below=10pt] {$\frac{1}{2}$};
	\node at (1,-1.15) [below=10pt] {$1$};
	\node at (1.5,-1) [below=10pt] {$\frac{3}{2}$};
	\node at (2,-1.15) [below=10pt] {$2$};
	\node at (2.5,-1) [below=10pt] {$\frac{5}{2}$};
	\node at (3,-1.15) [below=10pt] {$3$};
	\node at (3.5,-1) [below=10pt] {$\frac{7}{2}$};
	\node at (4,-1.15) [below=10pt] {$4$};
	
	\node at (-0.5,2) [left=10pt] {$\frac{5}{2}$};
	\node at (-0.5,1) [left=10pt] {$\frac{3}{2}$};
	\node at (-0.5,0) [left=10pt] {$\frac{1}{2}$};
	\node at (-0.5,-1) [left=10pt] {$-\frac{1}{2}$};	
	
\end{tikzpicture}
}
\caption{Witten effect for the Abelian lines in $\frac{U(1)\times SU(2)}{\boldsymbol{Z}_2}$ where $\tilde{\theta} = [0, 8 \pi)$ -- (2/3).}
\label{app:WittenEffectsZ2b}
\end{figure}

%%%
\vfill
\newpage

\begin{figure}[htbp!]

\subfigure[$\;z_2^e=1,z_2^m=1,\tilde{\theta}=0$]{
\begin{tikzpicture}[scale=0.6]
	
	\draw[thick,->] (-0.75,0) -- (4.5,0) node[right] {$q$};
	\draw[thick,->] (0,-1.25) -- (0,2.5) node[above] {$g$};
	
	\filldraw[opacity=0.25] [fill=blue,draw=black](-0.25,-0.5) rectangle (1.75,0.5);
	
	\foreach \x in {-0.5,0, 0.5, 1, 1.5, 2, 2.5, 3, 3.5, 4} {
		\foreach \y in {-1,0,1,2} {
			\filldraw[draw=black,fill=white] (\x,\y) circle(0.2); 
		}
	}	
	
	\foreach \x in {0,2,4} {
		\foreach \y in {-1,0,1,2} {
			\filldraw[draw=black,fill=green] (\x,\y) circle(0.2);	}}
	
	\node at (-0.5,-1) [below=10pt] {$\frac{1}{2}$};
	\node at (0,-1.15) [below=10pt] {$1$};
	\node at (0.5,-1) [below=10pt] {$\frac{3}{2}$};
	\node at (1,-1.15) [below=10pt] {$2$};
	\node at (1.5,-1) [below=10pt] {$\frac{5}{2}$};
	\node at (2,-1.15) [below=10pt] {$3$};
	\node at (2.5,-1) [below=10pt] {$\frac{7}{2}$};
	\node at (3,-1.15) [below=10pt] {$4$};
	\node at (3.5,-1) [below=10pt] {$\frac{9}{2}$};
	\node at (4,-1.15) [below=10pt] {$5$};
	
	\node at (-0.5,2) [left=10pt] {$\frac{5}{2}$};
	\node at (-0.5,1) [left=10pt] {$\frac{3}{2}$};
	\node at (-0.5,0) [left=10pt] {$\frac{1}{2}$};
	\node at (-0.5,-1) [left=10pt] {$-\frac{1}{2}$};
		
\end{tikzpicture}
}

\subfigure[\;$z_2^e=1,z_2^m=1,\tilde{\theta}=2\pi$]{
\begin{tikzpicture}[scale=0.6]
	
	\draw[thick,->] (-0.75,0) -- (4.5,0) node[right] {$q$};
	\draw[thick,->] (0,-1.25) -- (0,2.5) node[above] {$g$};
	
	\filldraw[opacity=0.25] [fill=blue,draw=black](-0.25,-0.5) rectangle (1.75,1.5);
	
	\foreach \x in {-0.5,0, 0.5, 1, 1.5, 2, 2.5, 3, 3.5, 4} {
		\foreach \y in {-1,0,1,2} {
			\filldraw[draw=black,fill=white] (\x,\y) circle(0.2); 
		}
	}	
	
	\foreach \x in {0.5,2.5} {
		\foreach \y in {0,2} {
			\filldraw[draw=black,fill=green] (\x,\y) circle(0.2);	}}
	
	\foreach \x in {-0.5,1.5,3.5} {
		\foreach \y in {-1,1} {
			\filldraw[draw=black,fill=green] (\x,\y) circle(0.2);	}}
	
	\node at (-0.5,-1) [below=10pt] {$\frac{1}{2}$};
	\node at (0,-1.15) [below=10pt] {$1$};
	\node at (0.5,-1) [below=10pt] {$\frac{3}{2}$};
	\node at (1,-1.15) [below=10pt] {$2$};
	\node at (1.5,-1) [below=10pt] {$\frac{5}{2}$};
	\node at (2,-1.15) [below=10pt] {$3$};
	\node at (2.5,-1) [below=10pt] {$\frac{7}{2}$};
	\node at (3,-1.15) [below=10pt] {$4$};
	\node at (3.5,-1) [below=10pt] {$\frac{9}{2}$};
	\node at (4,-1.15) [below=10pt] {$5$};
	
	\node at (-0.5,2) [left=10pt] {$\frac{5}{2}$};
	\node at (-0.5,1) [left=10pt] {$\frac{3}{2}$};
	\node at (-0.5,0) [left=10pt] {$\frac{1}{2}$};
	\node at (-0.5,-1) [left=10pt] {$-\frac{1}{2}$};
		
\end{tikzpicture}
}

\subfigure[$\; z_2^e=1,z_2^m=1,\tilde{\theta}=4\pi$]{
\begin{tikzpicture}[scale=0.6]
	
	\draw[thick,->] (-0.75,0) -- (4.5,0) node[right] {$q$};
	\draw[thick,->] (0,-1.25) -- (0,2.5) node[above] {$g$};
	
	\filldraw[opacity=0.25] [fill=blue,draw=black](-0.25,-0.5) rectangle (1.75,0.5);
	
	\foreach \x in {-0.5,0, 0.5, 1, 1.5, 2, 2.5, 3, 3.5, 4} {
		\foreach \y in {-1,0,1,2} {
			\filldraw[draw=black,fill=white] (\x,\y) circle(0.2); 
		}
	}
		
	\foreach \x in {1,3} {
		\foreach \y in {-1,0,1,2} {
			\filldraw[draw=black,fill=green] (\x,\y) circle(0.2);	}}

	\node at (-0.5,-1) [below=10pt] {$\frac{1}{2}$};
	\node at (0,-1.15) [below=10pt] {$1$};
	\node at (0.5,-1) [below=10pt] {$\frac{3}{2}$};
	\node at (1,-1.15) [below=10pt] {$2$};
	\node at (1.5,-1) [below=10pt] {$\frac{5}{2}$};
	\node at (2,-1.15) [below=10pt] {$3$};
	\node at (2.5,-1) [below=10pt] {$\frac{7}{2}$};
	\node at (3,-1.15) [below=10pt] {$4$};
	\node at (3.5,-1) [below=10pt] {$\frac{9}{2}$};
	\node at (4,-1.15) [below=10pt] {$5$};
	
	\node at (-0.5,2) [left=10pt] {$\frac{5}{2}$};
	\node at (-0.5,1) [left=10pt] {$\frac{3}{2}$};
	\node at (-0.5,0) [left=10pt] {$\frac{1}{2}$};
	\node at (-0.5,-1) [left=10pt] {$-\frac{1}{2}$};	
	
\end{tikzpicture}
}

\subfigure[$\; z_2^e=1,z_2^m=1,\tilde{\theta}=6\pi$]{
\begin{tikzpicture}[scale=0.6]
	
	\draw[thick,->] (-0.75,0) -- (4.5,0) node[right] {$q$};
	\draw[thick,->] (0,-1.25) -- (0,2.5) node[above] {$g$};
	
	\filldraw[opacity=0.25] [fill=blue,draw=black](-0.25,-0.5) rectangle (1.75,1.5);
	
	\foreach \x in {-0.5,0, 0.5, 1, 1.5, 2, 2.5, 3, 3.5, 4} {
		\foreach \y in {-1,0,1,2} {
			\filldraw[draw=black,fill=white] (\x,\y) circle(0.2); 
		}
	}
		
	\foreach \x in {-0.5,1.5,3.5} {
		\foreach \y in {0,2} {
			\filldraw[draw=black,fill=green] (\x,\y) circle(0.2);	}}
	
	\foreach \x in {0.5,2.5} {
		\foreach \y in {-1,1} {
			\filldraw[draw=black,fill=green] (\x,\y) circle(0.2);	}}	
	
	\node at (-0.5,-1) [below=10pt] {$\frac{1}{2}$};
	\node at (0,-1.15) [below=10pt] {$1$};
	\node at (0.5,-1) [below=10pt] {$\frac{3}{2}$};
	\node at (1,-1.15) [below=10pt] {$2$};
	\node at (1.5,-1) [below=10pt] {$\frac{5}{2}$};
	\node at (2,-1.15) [below=10pt] {$3$};
	\node at (2.5,-1) [below=10pt] {$\frac{7}{2}$};
	\node at (3,-1.15) [below=10pt] {$4$};
	\node at (3.5,-1) [below=10pt] {$\frac{9}{2}$};
	\node at (4,-1.15) [below=10pt] {$5$};
	
	\node at (-0.5,2) [left=10pt] {$\frac{5}{2}$};
	\node at (-0.5,1) [left=10pt] {$\frac{3}{2}$};
	\node at (-0.5,0) [left=10pt] {$\frac{1}{2}$};
	\node at (-0.5,-1) [left=10pt] {$-\frac{1}{2}$};
		
\end{tikzpicture}
}
\caption{Witten effect for the Abelian lines in $\frac{U(1)\times SU(2)}{\boldsymbol{Z}_2}$ 
where $\tilde{\theta} = [0, 8 \pi)$ -- (3/3).}
\label{app:WittenEffectsZ2c}
\end{figure}

\vfill
\newpage

\begin{figure}[htbp!]
\subfigure[$\;z_3^e=1,z_3^m=1,\tilde{\theta}=2\pi$]
{
\begin{tikzpicture}
	
	\draw[thick,->] (1/3,1/3) -- (22/3,1/3) node[right] {$q$};
	\draw[thick,->] (1,-3/3) -- (1,9/3) node[above] {$g$};
	
	\filldraw[opacity=0.25] [fill=blue,draw=black](7/6,0) rectangle (39/6,8/3);
	
	\foreach \x in {2/3,3/3,4/3,5/3,6/3,7/3,8/3,9/3,10/3,11/3,12/3,13/3,14/3,15/3,16/3,17/3,18/3,19/3,20/3,21/3} {
		\foreach \y in {-2/3,1/3,4/3,7/3} {
			\filldraw[draw=black,fill=white] (\x,\y) circle(0.15); 
		}
	}	
	
	\foreach \x in {4/3,13/3} {
		\filldraw[draw=black,fill=green] (\x,1/3) circle(0.15);	}
		
	\foreach \x in {10/3,19/3} {
		\foreach \y in {7/3,-2/3} {
			\filldraw[draw=black,fill=green] (\x,\y) circle(0.15);	}}
	
	\foreach \x in {7/3,16/3} {
		\filldraw[draw=black,fill=green] (\x,4/3) circle(0.15);	}	
	
	\node at (2/3,-3/3) [below] {$\frac{2}{3}$};
	\node at (3/3,-3/3) [below] {$\frac{3}{3}$};
	\node at (4/3,-3/3) [below] {$\frac{4}{3}$};
	\node at (5/3,-3/3) [below] {$\frac{5}{3}$};
	\node at (6/3,-3/3) [below] {$\frac{6}{3}$};
	\node at (7/3,-3/3) [below] {$\frac{7}{3}$};
	\node at (8/3,-3/3) [below] {$\frac{8}{3}$};
	\node at (9/3,-3/3) [below] {$\frac{9}{3}$};
	\node at (10/3,-3/3) [below] {$\frac{10}{3}$};
	\node at (11/3,-3/3) [below] {$\frac{11}{3}$};
	\node at (12/3,-3/3) [below] {$\frac{12}{3}$};
	\node at (13/3,-3/3) [below] {$\frac{13}{3}$};
	\node at (14/3,-3/3) [below] {$\frac{14}{3}$};
	\node at (15/3,-3/3) [below] {$\frac{15}{3}$};
	\node at (16/3,-3/3) [below] {$\frac{16}{3}$};
	\node at (17/3,-3/3) [below] {$\frac{17}{3}$};
	\node at (18/3,-3/3) [below] {$\frac{18}{3}$};
	\node at (19/3,-3/3) [below] {$\frac{19}{3}$};
	\node at (20/3,-3/3) [below] {$\frac{20}{3}$};
	\node at (21/3,-3/3) [below] {$\frac{21}{3}$};
	
	\node at (1/3,-2/3) [left] {$-\frac{2}{3}$};
	\node at (1/3,1/3) [left] {$\frac{1}{3}$};
	\node at (1/3,4/3) [left] {$\frac{4}{3}$};
	\node at (1/3,7/3) [left] {$\frac{7}{3}$};
	
\end{tikzpicture}
}

\hspace{5pt}

\subfigure[ $\;z_3^e=1,z_3^m=1,\tilde{\theta}=16\pi$]
{\begin{tikzpicture}
	
	\draw[thick,->] (1/3,1/3) -- (22/3,1/3) node[right] {$q$};
	\draw[thick,->] (1,-3/3) -- (1,9/3) node[above] {$g$};
	
	\filldraw[opacity=0.25] [fill=blue,draw=black](9/6,0) rectangle (41/6,8/3);
	
	\foreach \x in {2/3,3/3,4/3,5/3,6/3,7/3,8/3,9/3,10/3,11/3,12/3,13/3,14/3,15/3,16/3,17/3,18/3,19/3,20/3,21/3} {
		\foreach \y in {-2/3,1/3,4/3,7/3} {
			\filldraw[draw=black,fill=white] (\x,\y) circle(0.15); 
		}
	}	
	
	\foreach \x in {11/3,20/3} {
		\filldraw[draw=black,fill=green] (\x,1/3) circle(0.15);	}
	
	\foreach \x in {8/3,17/3} {
		\filldraw[draw=black,fill=green] (\x,4/3) circle(0.15);	}
	
	\foreach \x in {5/3,14/3} {
		\foreach \y in {7/3,-2/3} {
			\filldraw[draw=black,fill=green] (\x,\y) circle(0.15);	}  }
	
	\node at (2/3,-3/3) [below] {$\frac{2}{3}$};
	\node at (3/3,-3/3) [below] {$\frac{3}{3}$};
	\node at (4/3,-3/3) [below] {$\frac{4}{3}$};
	\node at (5/3,-3/3) [below] {$\frac{5}{3}$};
	\node at (6/3,-3/3) [below] {$\frac{6}{3}$};
	\node at (7/3,-3/3) [below] {$\frac{7}{3}$};
	\node at (8/3,-3/3) [below] {$\frac{8}{3}$};
	\node at (9/3,-3/3) [below] {$\frac{9}{3}$};
	\node at (10/3,-3/3) [below] {$\frac{10}{3}$};
	\node at (11/3,-3/3) [below] {$\frac{11}{3}$};
	\node at (12/3,-3/3) [below] {$\frac{12}{3}$};
	\node at (13/3,-3/3) [below] {$\frac{13}{3}$};
	\node at (14/3,-3/3) [below] {$\frac{14}{3}$};
	\node at (15/3,-3/3) [below] {$\frac{15}{3}$};
	\node at (16/3,-3/3) [below] {$\frac{16}{3}$};
	\node at (17/3,-3/3) [below] {$\frac{17}{3}$};
	\node at (18/3,-3/3) [below] {$\frac{18}{3}$};
	\node at (19/3,-3/3) [below] {$\frac{19}{3}$};
	\node at (20/3,-3/3) [below] {$\frac{20}{3}$};
	\node at (21/3,-3/3) [below] {$\frac{21}{3}$};
		
	\node at (1/3,-2/3) [left] {$-\frac{2}{3}$};
	\node at (1/3,1/3) [left] {$\frac{1}{3}$};
	\node at (1/3,4/3) [left] {$\frac{4}{3}$};
	\node at (1/3,7/3) [left] {$\frac{7}{3}$};
	
\end{tikzpicture}
}
\caption{Witten effect for the Abelian lines in $\frac{U(1)\times SU(3)}{\boldsymbol{Z}_3}$
where $\tilde{\theta} = [0, 18 \pi)$.}
\label{app:WittenEffectsZ3}
\end{figure}

\vfill
\newpage

\begin{figure}[htbp!]
\subfigure[$\;z_2^e=1,z_2^m=1;z_3^e=1,z_3^m=1,\tilde{\theta}=2\pi$]
{
\begin{tikzpicture}[scale=1.25]
	
	\draw[thick,->] (3/6,5/6) -- (44/6,5/6) node[right] {$q$};
	\draw[thick,->] (6/6,-3/6) -- (6/6,37/6) node[above] {$g$};
	
	\filldraw[opacity=0.25] [fill=blue,draw=black](4/6,-2/6) rectangle (81/12,30/6);
	
	\draw[thick, dashed] (5/6,35/6) -- (5/6,-1/6);
	\draw[thick, dashed] (11/6,35/6) -- (11/6,-1/6);
	\draw[thick, dashed] (17/6,35/6) -- (17/6,-1/6);
	\draw[thick, dashed] (23/6,35/6) -- (23/6,-1/6);
	\draw[thick, dashed] (29/6,35/6) -- (29/6,-1/6);
	\draw[thick, dashed] (35/6,35/6) -- (35/6,-1/6);
	\draw[thick, dashed] (41/6,35/6) -- (41/6,-1/6);
	
	\foreach \x in {5/6, 6/6, 7/6, 8/6, 9/6, 10/6, 11/6, 12/6, 13/6, 14/6, 15/6, 16/6, 17/6, 18/6, 19/6, 20/6, 21/6, 22/6, 23/6, 24/6, 25/6, 26/6, 27/6, 28/6, 29/6, 30/6, 31/6, 32/6, 33/6, 34/6, 35/6, 36/6, 37/6, 38/6, 39/6, 40/6, 41/6, 42/6} {
		\foreach \y in {-1/6,5/6,11/6,17/6,23/6,29/6,35/6} {
			\filldraw[draw=black,fill=white] (\x,\y) circle(0.07); 
		}
	}
	
	\foreach \x in {5/6,41/6} {
		\foreach \y in {-1/6,35/6}{
			\filldraw[draw=black,fill=green] (\x,\y) circle(0.07);	
		}
	}
	
		\filldraw[draw=black,fill=green] (11/6,5/6) circle(0.07);	

		\filldraw[draw=black,fill=green] (17/6,11/6) circle(0.07);	

		\filldraw[draw=black,fill=green] (23/6,17/6) circle(0.07);		

		\filldraw[draw=black,fill=green] (29/6,23/6) circle(0.07);	

		\filldraw[draw=black,fill=green] (36/6,29/6) circle(0.07);	
	
	\node at (5/6,-2/6) [below] {$\frac{5}{6}$};
	\node at (11/6,-2/6) [below] {$\frac{11}{6}$};
	\node at (17/6,-2/6) [below] {$\frac{17}{6}$};
	\node at (23/6,-2/6) [below] {$\frac{23}{6}$};
	\node at (29/6,-2/6) [below] {$\frac{29}{6}$};
	\node at (35/6,-2/6) [below] {$\frac{35}{6}$};
	\node at (41/6,-2/6) [below] {$\frac{41}{6}$};	
	
	\node at (3/6,35/6) [left] {$\frac{35}{6}$};
	\node at (3/6,29/6) [left] {$\frac{29}{6}$};
	\node at (3/6,23/6) [left] {$\frac{23}{6}$};
	\node at (3/6,17/6) [left] {$\frac{17}{6}$};
	\node at (3/6,11/6) [left] {$\frac{11}{6}$};
	\node at (3/6,5/6) [left] {$\frac{5}{6}$};
	\node at (3/6,-1/6) [left] {$-\frac{1}{6}$};
	
\end{tikzpicture}
}

\hspace{5pt}

\subfigure[$\;z_2^e=1,z_2^m=1;z_3^e=1,z_3^m=1,\tilde{\theta}=70\pi$]
{
\begin{tikzpicture}[scale=1.25]
	
	\draw[thick,->] (3/6,5/6) -- (46/6,5/6) node[right] {$q$};
	\draw[thick,->] (6/6,-3/6) -- (6/6,37/6) node[above] {$g$};
	
	\filldraw[opacity=0.25] [fill=blue,draw=black](13/12,-2/6) rectangle (85/12,30/6);
	
	\draw[thick, dashed] (7/6,35/6) -- (7/6,-1/6);
	\draw[thick, dashed] (13/6,35/6) -- (13/6,-1/6);
	\draw[thick, dashed] (19/6,35/6) -- (19/6,-1/6);
	\draw[thick, dashed] (25/6,35/6) -- (25/6,-1/6);
	\draw[thick, dashed] (31/6,35/6) -- (31/6,-1/6);
	\draw[thick, dashed] (37/6,35/6) -- (37/6,-1/6);
	\draw[thick, dashed] (43/6,35/6) -- (43/6,-1/6);
	
	\foreach \x in {5/6, 6/6, 7/6, 8/6, 9/6, 10/6, 11/6, 12/6, 13/6, 14/6, 15/6, 16/6, 17/6, 18/6, 19/6, 20/6, 21/6, 22/6, 23/6, 24/6, 25/6, 26/6, 27/6, 28/6, 29/6, 30/6, 31/6, 32/6, 33/6, 34/6, 35/6, 36/6, 37/6, 38/6, 39/6, 40/6, 41/6, 42/6, 43/6, 44/6} {
		\foreach \y in {-1/6,5/6,11/6,17/6,23/6,29/6,35/6} {
			\filldraw[draw=black,fill=white] (\x,\y) circle(0.07); 
		}
	}
	
	\foreach \x in {7/6,43/6} {
		\foreach \y in {-1/6,35/6}{
			\filldraw[draw=black,fill=green] (\x,\y) circle(0.07);	
		}
	}
	
		\filldraw[draw=black,fill=green] (13/6,5/6) circle(0.07);		

		\filldraw[draw=black,fill=green] (19/6,11/6) circle(0.07);	
	
		\filldraw[draw=black,fill=green] (25/6,17/6) circle(0.07);		

		\filldraw[draw=black,fill=green] (31/6,23/6) circle(0.07);		

		\filldraw[draw=black,fill=green] (37/6,29/6) circle(0.07);	
		
	\node at (7/6,-2/6) [below] {$\frac{7}{6}$};
	\node at (13/6,-2/6) [below] {$\frac{13}{6}$};
	\node at (19/6,-2/6) [below] {$\frac{19}{6}$};
	\node at (25/6,-2/6) [below] {$\frac{25}{6}$};
	\node at (31/6,-2/6) [below] {$\frac{31}{6}$};
	\node at (37/6,-2/6) [below] {$\frac{37}{6}$};
	\node at (43/6,-2/6) [below] {$\frac{43}{6}$};
		
	\node at (3/6,35/6) [left] {$\frac{35}{6}$};
	\node at (3/6,29/6) [left] {$\frac{29}{6}$};
	\node at (3/6,23/6) [left] {$\frac{23}{6}$};
	\node at (3/6,17/6) [left] {$\frac{17}{6}$};
	\node at (3/6,11/6) [left] {$\frac{11}{6}$};
	\node at (3/6,5/6) [left] {$\frac{5}{6}$};
	\node at (3/6,-1/6) [left] {$-\frac{1}{6}$};
	
\end{tikzpicture}
}
\caption{Witten effect for the Abelian lines in 
$\frac{U(1)\times SU(2)\times SU(3)}{\boldsymbol{Z}_6}$ where $\tilde{\theta} = [0, 72 \pi)$.}
\label{app:WittenEffectsZ6}
\end{figure}

\vfill
\newpage

\bibliographystyle{unsrt}
\bibliography{Refs}

\end{document}